\documentclass[12pt]{amsart}
\usepackage{mathrsfs}
\usepackage{epsfig}
\begin{document}
 \topmargin=5mm
 \oddsidemargin=6mm
 \evensidemargin=6mm
\baselineskip=20pt
\parindent 20pt
{\flushleft{\Large\bf  On Negative Order KdV Equations}}\\[4pt]
{\large\bf  Zhijun Qiao\footnote{E-mail address: \  qiao@utpa.edu}
}\\
 {\small  Department of Mathematics, The University of Texas-Pan
American,
 1201 W University Drive, Edinburg, TX 78539, USA} \\[4pt]
{\large\bf  Engui Fan}\footnote{ E-mail
address: \  faneg@fudan.edu.cn}\\
{\small School of Mathematical Sciences, Institute of Mathematics
and Key Laboratory of Mathematics for Nonlinear Science, Fudan
University, Shanghai, 200433, P.R.  China}\vspace{4mm}
\begin{center}
\begin{minipage}{6.6in}

\baselineskip=19pt { \small {\bf Abstract.}

In this paper, %based on the regular KdV system,
we study
negative order KdV (NKdV) equations and give their Hamiltonian
structures, Lax pairs, infinitely many conservation laws, and
explicit multi-soliton and multi-kink wave solutions thorough
bilinear B\"{a}cklund transformations. The NKdV equations
studied in the paper are differential and can be derived from the
first member in the negative order KdV hierarchy.
 The NKdV equations are not only gauge-equivalent to
 the Camassa-Holm equation through some hodograph transformations,
 but also closely related to  the Ermakov-Pinney
 systems and the Kupershmidt deformation.
The bi-Hamiltonian structures and a Darboux transformation of the
NKdV equations are constructed with the aid of trace identity and
their Lax pairs, respectively. The 1- and 2- kink wave and
soliton solutions are given in an explicit formula through the
Darboux transformation. The 1-kink wave solution is expressed in the
form of $tanh$ while the 1-bell soliton is in the form of $sech$,
and both forms are very standard. The collisions of 2-kink-wave and
2-bell-soliton solutions, are analyzed in details, and this singular
interaction is a big difference from the regular KdV equation.
Multi-dimensional binary Bell polynomials  are employed to find
bilinear formulation and B\"{a}cklund transformations, which produce
$N$-soliton solutions. A direct and unifying scheme is proposed for
explicitly building up quasi-periodic wave solutions
 of the NKdV equations.
 Furthermore, the
relations between quasi-periodic wave solutions and soliton
solutions are clearly described. Finally,  we show the
quasi-periodic wave solution convergent to the soliton solution
under some limit conditions. }
\end{minipage}
\end{center}
\begin{center}
\begin{minipage}{6.6in}
\baselineskip=17pt {
\\[4pt]}
\end{minipage}
\end{center}
%%%%%%%%%%%%%%%%%%%%%%%%%%%%%%%%%%%%%%%%%%%%%%%%%%%%%%%%%%%%%%%%%%%%%%%%%%%%%%%%%%%%%%%%%%%%%%%%%%%%
%%%%%%%%%%%%%%%%%%%%%%%%%%%%%%%%%%%%%%%%%%%%%%%%%%%%%%%%%%%%%%%%%%%%%%%%%%%%%%%%%%%%%%%%%%%%%%%%%%%%
{\bf Table of Contents }\\
1. Introduction\\
2. Hamiltonian structures of the NKdV hierarchy

2.1.  The NKdV hierarchy

2.2. Bi-Hamiltonian structures\\
3.   Relations %of NKdV equations
     to  other important  equations

3.1. Kupershmidt deformation

3.2. NKdV hierarchy with self-consistent sources

3.3. Reduction of NKdV equation

3.4. Ermakov-Pinney  systems\\
4. Darboux transformation of NKdV equations

4.1.  Darboux transformation

4.2.  Reduction of Darboux transformation\\
 5.  Applications of Darboux transformation

 5.1. Kink waves and interaction

5.2. Bell solitons and interaction\\
 6.   Bilinearization of NKdV
equations

6.1.  Binary Bell polynomials

6.2. Bilinear formulation

6.3. $N$-soliton solutions\\
7.  Bilinear B\"{a}cklund transformation of NKdV equations

7.1.  Bilinear B\"{a}cklund transformation

7.2. Inverse scattering formulation\\
8.  Darboux covariant Lax pair of NKdV equations\\
9.  Conservation laws of NKdV equations\\
10. Quasi-periodic solutions of NKdV equations

10.1. Theta functions

10.2.  Bilinear formulas

 10.3. One-periodic wave solutions and  long wave limit

 10.4. Two-periodic wave solutions  and  long wave limit

 10.5.   Multi-periodic wave solutions\\[16pt]
 %%%%%%%%%%%%%%%%%%%%%%%%%%%%%%%%%%%%%%%%%%%%%%%%%%%%%%%%%%%%%%%%%%%%%%%%%%%%%%%%%%%
 %%%%%%%%%%%%%%%%%%%%%%%%%%%%%%%%%%%%%%%%%%%%%%%%%%%%%%%%%%%%%%%%%%%%%%%%%%%%%%%%%
{\bf\large 1. Introduction}

The Korteweg-de Vries (KdV) equation
$$u_t+6uu_x+u_{xxx}=0$$
was proposed by Korteweg and de Vries  in fluid dynamics
\cite{Korteweg}, starting from the observation and subsequent
experiments by Russell \cite{Russell}.
 There are many excellent sources for the highly interesting background and
historical development of the KdV equation, which brings it to the
forefront of modern mathematical physics.  In 1967,  Gardner,
Greener, Kruskal and Miura found the inverse scattering
transformation method to solve  the Cauchy problem of  the KdV
equation with sufficiently decaying initial data
\cite{Gardner1}. Soon thereafter, Lax  explained the magical
isospectral property of the time dependent family of Schrodinger
operators which is now called the Lax pair, and introduced the
KdV hierarchy through a recursive procedure
 \cite{Lax1}. In the same year a sequence of infinitely many
polynomial conservation laws were obtained with the help of
Miura's transformation \cite{Miura1, Miura2}.

There are tools to view the KdV equation as a completely
integrable system by Gardner, and Zakharov and Faddeev
\cite{Gardner2, Zakharov}. The bilinear derivative method was
developed by Hirota to find $N$-soliton solutions of the KdV
equation \cite{Hirota}. The KdV hierarchy was constructed  by Lax
\cite{Lax} through a recursive approach, and further studied by
Gel'fand and Dikii  \cite{Dikii}. On the base of the
inverse spectral theory and algebro-geometric methods, the inverse
scattering method was extended to periodic initial data by %was developed
by Novikov, Dubrovin,  Lax, Its,  Matveev  et al \cite{Du, Its,
Mc,Nov}. For more recent reviews on the KdV equation one may refer to
%for instance, to 
literature
\cite{Ab1,Ba1,Belo,Bullough,Che1,Fritz1,Lax3,Palais,Tu1,Tu2}.

All the work done in the above mentioned publications dealt with the
positive order KdV hierarchy, which includes the KdV equation as a
special member. However, there was only little work on  the NKdV
hierarchy.     Verosky \cite{Verosky} studied symmetries and
negative powers of recursion operator and gave the following
negative order KdV equation (called the NKdV equation thereafter)
%NKdV equation
$$
\begin{array}{l}
v_t=w_x,\\[8pt]
w_{xxx}+4vw_x+2v_xw=0.
\end{array}\eqno(1.1)
$$
and Lou \cite{Lou} presented additional symmetries based on the
invertible recursion operator of the KdV system and particularly
provided the following NKdV equation (called the NKdV-1
equation thereafter)
$$v_t=2uu_x,\ \ u_{xx}+vu=0, \Longleftrightarrow \left(\frac{u_{xx}}{u}\right)_t+2uu_x=0, \eqno(1.2)$$
which can be reduced from the NKdV equation (1.1) under the following transformation
$$w=u^2, \ \ v=-\frac{u_{xx}}{u}. \eqno(1.3)$$
Moreover,   the second part of NKdV-1 equation (1.2) is a linear
Schr\"{o}dinger equation or Hill equation
$$u_{xx}+vu=0.$$
Fuchssteiner \cite{Fuchssteiner} pointed out the gauge-equivalent
relation between the NKdV equation (1.1) and the Camassa-Holm (CH)
equation \cite{Camassa}
$$m_t+m_xu+2mu_x=0,\ \ m=u-u_{xx}$$
 through some hodograph transformation, and
later on Hone proposed the associate CH equation, which is actually
equivalent to  the NKdV equation (1.1), and gave soliton solutions
through the KdV system \cite{H1}.  Zhou \cite{Zhou1} generalized the Kupershmidt
deformation and proposed  a kind of the mixed KdV hierarchy, which contains
the NKdV equation (1.1) as a special case.

Very recently,  Qiao and Li \cite{Qiao} gave a unifying formulation
of the Lax representations for both negative and positive order KdV
hierarchies, and furthermore studied all possible traveling wave
solutions, including soliton, kink wave, and periodic wave
solutions, of the integrable NKdV-1 equation (1.2) with % possesses
the following Lax pair
$$
\begin{array}{l}
L\psi\equiv\psi_{xx}+v\psi=\lambda\psi,\\ \\
\psi_t=\frac{1}{2}u^2\lambda^{-1}\psi_x-\frac{1}{2}uu_x\lambda^{-1}\psi.
\end{array}\eqno(1.4)
$$
The most interesting \cite{Qiao} is: the NKdV-1 equation has both soliton and kink solutions,
which is the first integrable example, within our knowledge, having such a property in soliton theory.%integrable systems.

Studying negative order integrable hierarchies plays an important
role in the theory of peaked soliton (peakon) and cusp soliton
(cuspon). For instance, the well-known CH peakon equation is
actually produced through its negative order hierarchy while its
positive order hierarchy includes the remarkable Harry-Dym type
equation \cite{Qiao1}. The Degasperis-Procesi (DP) peakon equation
\cite{DP[1999]} can also be generated through its negative order
hierarchy \cite{Qiao-AAM}. Both the CH equation and the DP equation
are typical integrable peakon and cupson systems with nonlinear
quadratic terms
\cite{Camassa,DHH2002,Lundmark-Szm,Qiao1,Zhang-qiao}. Recently, some
nonlinear cubic integrable equations have also been found to have
peakon and cupson solutions \cite{HW1, Novikov,Qiao2,Qiao-JMP2}.

In this paper, we study the NKdV hierarchy, in particular, focus on
the NKdV equation (1.1) and the NKdV-1 equation (1.2). Actually, as
per \cite{Lou,Qiao-book}, the NKdV equation (1.1) can embrace other
possible differential-integro forms according to the kernel of operator
$K=\frac{1}{4}\partial_x^3+\frac{1}{2}(v\partial_x+\partial_x v)$.
Here we just list the NKdV-1 equation (1.2) as it is differential and
%We find that the first negative order KdV equation
also equivalent to a nonlinear quartic integrable system:
$$uu_{xxt}-u_{xx}u_t-2u^3u_x=0. $$

 %with both classic soliton and kink wave solutions.

The purpose of this paper is  to   investigate  integrable
properties,  $N$-soliton and $N$-kink solutions of the NKdV equation
(1.1) and  NKdV-1 equation (1.2). In section 2, the trace identity
technique is employed to construct the bi-Hamiltonian structures of the
NKdV hierarchy.  In section 3, we show  that the NKdV equation (1.1) is
related to the Kupershmidt deformation and the Ermakov-Pinney systems,
and is also able to reduced to the NKdV-1 equation (1.2) under a transformation.
%The relation  between solution of  NKdV equation (1.1) and that of NKdV-1 equation (1.2) is given.
In section 4, a Darboux
transformation of the NKdV equation (1.1) is provided with the help of its
Lax pairs. In section 5, as a direct application of the Darboux
transformation, the kink-wave and bell soliton solutions are
explicitly given, and the collision of two soliton solutions is
analyzed in detail through two-solitons.   In section 6, an
extra auxiliary variable  is introduced to bilinearize the NKdV
equation (1.1) through binary Bell polynomials.  In section 7, the
 bilinear B\"{a}cklund transformations  are obtained and  Lax pairs are also recovered.
In section 8, we will give a kind of Darboux covariant Lax pair, and
in section 9, infinitely many conservation
laws of the NKdV equation (1.1) are presented through its Lax equation and
  a generalized Miura transformation.   All conserved densities and fluxes are
recursively given in an explicit formula.  In sections 10,
 a direct and unifying
scheme is proposed for building up quasi-periodic wave solutions
 of the NKdV equation (1.1) in an explicit formula.  Furthermore, the
relations between quasi-periodic wave solutions and soliton
solutions are clearly described. Finally,  we show the
quasi-periodic wave solution convergent to the soliton solution
under some limit conditions.% and the assumption of small amplitude.% limit.
%\end{document}
 \\[12pt]
%%%%%%%%%%%%%%%%%%%%%%%%%%%%%%%%%%%%%%%%%%%% %%%%%%%%%%%%%%%%%%%%%%%%%%%%
%%%%%%%%%%%%%%%%%%%%%%%%%%%%%%%%%%%%%%%%%%%%%%%%%%%%%%%%%%%%%%%%%%%%%%%%%%%%%%
{\bf\large  2.   Hamiltonian structures of the NKdV hierarchy   }\\

To find the Hamiltonian structures of the NKdV hierarchy, let us
re-derive the  NKdV hierarchy in matrix
form.\\[8pt]
%%%%%%%%%%%%%%%%%%%%%%%%%%%%%%%%%%%%%%%%%%%%%%%%%%%%%%%%%%%%%%%%%%%%%%%%%%%
{\bf 2.1. The NKdV hierarchy}

Consider the Schr\"{o}dinger-KdV spectral problem
$$
\begin{aligned}
\psi_{xx}+v\psi=\lambda\psi,
\end{aligned}\eqno(2.1)
$$
where $\lambda$ is an eigenvalue, $\psi$ is the eigenfunction
corresponding to the eigenvalue $\lambda$, and $v$ is a potential
function.

Let $\varphi_1=\psi,\ \ \varphi_2=\psi_x$, then the spectral problem
(2.1)   becomes
$$
\begin{aligned}
\varphi_x=U\varphi=\left(
\begin{matrix}0&1\cr
\lambda -v &0\end{matrix}\right)\varphi,
\end{aligned}\eqno(2.2)
$$
where $\varphi=(\varphi_1, \varphi_2)^T$ is a two-dimensional vector
of eigenfunctions.

The Gateaux derivative of spectral operator $U$ in direction $\xi$
at point $v$ is
$$U'[\xi]=\frac{d}{d\varepsilon}U(v+\varepsilon\xi)|_{\varepsilon=0}=\left(
\begin{matrix}0&0\cr
-\xi&0\end{matrix}\right),\eqno(2.3)$$ which is injective and linear
with respect to the variable $\xi$.

 The Lenard recursive sequence $\{ G_m\}$ of the spectral problem (2.1)
 is defined by
$$\begin{aligned}
&G_{-1}\in Ker K=\{G|KG=0\}, \ \ G_0\in Ker J=\{G|JG=0\}\\
&KG_{m-1}=JG_{m},\ \ m=0, -1, -2\cdots,
\end{aligned}\eqno(2.4)$$ which directly produces the NKdV
hierarchy:
$$v_t=KG_{m-1}=JG_{m},\ \ m= -1, -2\cdots\eqno(2.5)$$
where
$$
\begin{aligned}
&K=\frac{1}{4}\partial_x^3+\frac{1}{2}(v\partial_x+\partial_x v),\ \
\ J=\partial_x,
\end{aligned}\eqno(2.6)
$$
and  $K$  is exactly  a recursion operator of the well-known KdV
hierarchy
$$v_t=K^nv_x,
\ \ n=0, 1, 2, \cdots.$$

The first  equation ($m=0$)  in the NKdV hierarchy (2.5) is trivial
equation
$$v_t=JG_0=0, \ \ JG_0=KG_{-1}=0.$$

The second equation ($m=-1$)  in the NKdV hierarchy (2.5) takes
$$v_t=G_{-1,x}, \ \ \ KG_{-1}=0,$$
which is exactly the NKdV equation (1.1) by replacing $G_{-1}=w$.

In a similar way to the paper \cite{Qiao},  we construct zero
curvature representation for NKdV hierarchy.

 {\bf Proposition  1.} Let
$U$ be the spectral matrix defined  in (2.2), then for an
arbitrarily smooth function $G\in C^{\infty}(\mathbb{R})$, the
following operator equation
$$
\begin{aligned}
V_x-[U,V]=U'[KG]-\lambda U'[JG]
\end{aligned}\eqno(2.7)
$$
admits a  matrix solution
$$V=V(G)=\left(
\begin{matrix}\displaystyle{-\frac{1}{4}G_x}&\displaystyle{\frac{1}{2}G}\cr\cr
\displaystyle{-\frac{1}{4}G_{xx}-\frac{1}{2}vG+\frac{1}{2}\lambda
G}&\displaystyle{\frac{1}{4}G_x}\end{matrix}\right)\lambda^{-1},
$$ which is a linear function with respect to $G$,  and
Gateaux derivative is defined by (2.3).

{\bf Theorem 1.} Suppose that  $\{G_j, \ \ j=-1,-2, \cdots\}$ is the
first Lenard sequence defined by (2.4), and $V_j=V(G_j)$ is a
corresponding solution to  the operator equation (2.7) for $G=G_j$.
With  $V_j$ being its  coefficients, a $m$th matrix polynomial in
$\lambda$ is constructed as follows
$$W_m=\sum_{j=1}^{m}V_{j}\lambda^{-m+j}.$$
Then we conclude that the NKdV hierarchy (2.5) admits zero curvature
representation
$$U_t-W_{m,x}+[U,W_m]=0,$$
which is equivalent to
$$
\begin{aligned}
&\varphi_x=U\varphi=\left(
\begin{matrix}0&1\cr
\displaystyle{\lambda-v}&0\end{matrix}\right)\varphi,\\ \\
 &\varphi_t=W_m\varphi=\sum_{j=1}^m
 \left(
\begin{matrix}-{\frac{1}{4}G_{j,x}}&\frac{1}{2}G_{j}\cr\cr
{-\frac{1}{4}G_{j,xx}-\frac{1}{2}vG_{j}+\frac{1}{2}\lambda
G_{j}}&{\frac{1}{4}G_{j,x}}\end{matrix}\right)\lambda^{-m+j-1}\varphi.
\end{aligned}\eqno(2.8)
$$
This theorem  actually provides  an unified formula of the Lax pairs
for the whole NKdV hierarchy (2.5).

According to theorem 1,  the  NKdV equation (1.1)  admits  Lax pair
with parameter $\lambda$
$$
\begin{array}{l}
L\psi\equiv\psi_{xx}+v\psi=\lambda\psi,\\ \\
\psi_t=\frac{1}{2}w\lambda^{-1}\psi_x-\frac{1}{4}w_x\lambda^{-1}\psi,
\end{array}
$$
or equivalently,
$$
\begin{array}{l}
L\psi=(\partial_{x}^2+v)\psi=\lambda\psi,\\[8pt]
M\psi=(4\partial_x^2\partial_t+4v\partial_t+2w\partial_x+3w_{x})\psi=0.
\end{array}\eqno(2.9)
$$

The NKdV equation (1.1) also possesses Lax pair without parameter
$$
\begin{array}{l}
L\psi=(\partial_{x}^2+v)\psi=0,\\[8pt]
M\psi=(4\partial_x^2\partial_t+4v\partial_t+2w\partial_x+3w_{x})\psi=0.
\end{array}\eqno(2.10)
$$

 Especially,  taking the constraint   $v=-{u_{xx}}/{u}$ and $w=u^2\in
Ker K$,  we then further get the NKdV equation (1.2) and its Lax
parir (1.4).
\\[8pt]
%%%%%%%%%%%%%%%%%%%%%%%%%%%%%%%%%%%%%%%%%%%%%%%%%%%%%%%%%%%%%%%%%%%%%%%%%%%%%%%%%%%%%%%%%%
{\bf 2.2. Hamiltonian structures}

{\bf Proposition  2.} \cite{Tu1} For the spectral problem (2.2),
assume that $V$ is a solution to the following stationary zero
curvature equation with the given homogeneous rank
$$V_x=[U, V].\eqno(2.11)$$ Then there exists a constant
$\beta$, such that %it holds that
$$\frac{\delta }{\delta v}\left\langle V, \frac{\partial U}{\partial
\lambda}\right\rangle=\left(\lambda^{-\beta}\frac{\partial}{\partial\lambda}\lambda^{\beta}\right)\left\langle
V, \frac{\partial U}{\partial v}\right\rangle,\eqno(2.12)$$ holds,
where $\langle \cdot, \cdot\rangle$ stands for the trace of the
product of two matrices.

Let $\{G_m, \ \ m=-1,-2 \cdots\}$ be
the negative order Lenard sequence recursively given through (2.4) and %. We construct
$$G_{\lambda}=\sum_{m=-\infty}^{-1}G_{m}\lambda^{-m},\eqno(2.13)$$
be a series with respect to $\lambda$. Assume that
$V_{\lambda}=V(G_{\lambda})$ is the matrix solution for the operator
equation (2.9) corresponding to $G=G_{\lambda}$. So, $V_{\lambda}$
can be written as
$$V_{\lambda}=\sum_{m=-\infty}^{-1}V_{m}\lambda^{-m}.$$

Then, we have the following proposition.

{\bf Proposition  3.} $V_{\lambda}$ satisfies the following Lax form
$$V_{\lambda, x}=[U,V_{\lambda}].$$

{\it Proof.}  By (2.4), we have

$$
\begin{aligned}
&(K-\lambda
J)G_{\lambda}=\sum_{m=-\infty}^{-1}KG_{m}\lambda^{-m}-\sum_{m=-\infty}^{-1}JG_{m}\lambda^{-m+1}
\\
&\ \ \ \ \ \ \ \ \ \ \ \ \ \ \ \ \ \ \
=KG_{-1}\lambda^{-1}+\sum_{m=-\infty}^{-1}(KG_{m-1}-JG_{m})\lambda^{-m}=0.
\end{aligned}
$$
Therefore, Proposition  1 implies
$$V_{\lambda,
x}-[U,V_{\lambda}]=U'[KG_{\lambda}]-\lambda
U'[JG_{\lambda}]=U'[KG_{\lambda}-\lambda JG_{\lambda}]=0.$$
 $\square$

Next, we discuss the Hamiltonian structures of
the hierarchy (2.5). It is crucial to find %show the existence of
infinitely many conserved densities.

{\bf Theorem 2.} \ \
\begin{enumerate}
\item  The hierarchy (2.5) possesses the
bi-Hamiltonian structures
$$v_t=K\frac{\delta H_{m-1}}{\delta
v}=J\frac{\delta H_{m}}{\delta v}, \ \ m=-1, -2\cdots,\eqno(2.14)$$
where the Hamiltonian functions $H_m$  are implicitly  given through
the following formulas
$$\
H_{-1}=G_{-1}\in KerK,\ \ H_{m}=\frac{G_{m}}{m}, \ \ \ m=-1,
-2\cdots.\eqno(2.15)$$
\item The hierarchy (2.5) is integrable in the Liouville sense.
\item
The Hamiltonian functions $\{H_m\}$ are conserved densities of the whole
hierarchy (2.5) and therefore they are in involution in pairs.
\end{enumerate}

{\it Proof.} A direction calculation leads to
$$\left\langle V_{\lambda}, \frac{\partial U}{\partial \lambda}\right\rangle=\frac{1}{2}G_{\lambda},\
\ \
 \left\langle V_{\lambda}, \frac{\partial U}{\partial v}\right\rangle=-\frac{1}{2}G_{\lambda}.$$
 By using the trace identity (2.12) and  the expansion (2.13), we obtain
$$
\begin{aligned}
&\frac{\delta}{\delta
v}\left(\sum_{m=-\infty}^{-1}G_{m}\lambda^{-m}\right)
=\sum_{m=-\infty}^{-1}(m-1-\beta)G_{m-1}\lambda^{-m}
+(-1-\beta)G_{-1}, \\[8pt]
 &  \ m=-1, -2\cdots. \end{aligned}\eqno(2.16)$$

If taking $G_{-1}\not=0$, form (2.16) we find $\beta=-1$ and
$$\frac{\delta H_m}{\delta v}=G_{m-1}, \ \ \ m=-1, -2\cdots,\eqno(2.17)$$
where $H_m$ are given by (2.15). Substituting (2.17) into (2.5)
yields the bi-Hamiltonian structures (2.14).

Next, we consider infinitely many conserved densities to guarantee
integrability of the hierarchy  (2.16).
 Since $J$ and $K$ are skew-symmetric operators,
we infer that
$$\mathcal{L}^*J=(J^{-1}K)^*J=-K^*=K=J\mathcal{L},$$
 which implies
$$
\begin{array}{l}
\displaystyle{\{H_n,H_m\}=\left(\frac{\delta H_n}{\delta v},
J\frac{\delta H_m}{\delta
v}\right)=(\mathcal{L}^nG_{-1},J\mathcal{L}^mG_{-1})=(\mathcal{L}^nG_{-1},
\mathcal{L}^*J\mathcal{L}^{m-1}G_{-1})}\\
\displaystyle{=(\mathcal{L}^{n+1}G_{-1},J\mathcal{L}^{m-1}G_0)=\{H_{n+1},H_{m-1}\}},
\ \ m, n\leq -1.
\end{array}
$$
Repeating the above argument gives
$$\{H_n,H_m\}=\{H_m,H_n\}=\{H_{m+n},H_{-1}\}.\eqno(2.18)$$
On the other hand, we find
$$\{H_m,H_n\}=(\mathcal{L}^mG_{-1},J\mathcal{L}^nG_{-1})=(J^*\mathcal{L}^mG_{-1},\mathcal{L}^nG_{-1})
=-\{H_n,H_m\}.\eqno(2.19)$$ Then combining (2.18) with  (2.19) leads
to
$$\{H_m,H_n\}=0,$$
which implies that $\{H_m\}$ are in involution, and therefore the
hierarchy (2.14)  are integrable in Liouville sense.

Especially, under the constraint (1.3),  we obtain  bi-Hamilton
structures of the NKdV equation (1.2)
$$v_t=K\frac{\delta H_{-1}}{\delta
v}=J\frac{\delta H_{0}}{\delta u},$$ where two Hamiltonian functions
are given by
$$H_0=\frac{1}{3}u^3,\ \  \ \ H_{-1}=-u^2,$$
which can also be written in a conserved density form in the sense
of equivalence class
 $$H_{0} \sim
 -\frac{1}{3}\int
u^3dx,\
 \ \ \ H_{-1} \sim
 -\int u^2dx.$$
\\[12pt]
%%%%%%%%%%%%%%%%%%%%%%%%%%%%%%%%%%%%%%%%%%%%%%%%%%%%%%%%%%%%%%%%%%%%%%%%%%%%%%%%%%%%%%%%%%%%%%%%%%%%%%%%%%%%%%%%%%%%%%%%%%%%%%%%%%%%%%
%%%%%%%%%%%%%%%%%%%%%%%%%%%%%%%%%%%%%%%%%%%%%%%%%%%%%%%%%%%%%%%%%%%%%%%%%%%%%%%%%%%%%%%%%%%%%%%%%%%%%%%%%%%%%%%%%%%%%%%%%%%%%%%%%%%%%
{\bf\large  3.  Relations to  other important equations}\\[12pt]
{\bf 3.1. Kupershmidt  deformation}\\

Recently a class of new integrable systems, known as the Kupershmidt
 deformation of soliton equations, have attracted much
attention.  This topic starts from Kupershmidt, Karasu-Kalkani' work
\cite{Guha,Kersten,Kupershmidt}.

For the Lenard operator pair (2.6), we define Lenard gradients
recursively by
$$KG_j=JG_{j+1}, \ KG_{-1}=JG_0=0, \ \ j=0, \pm 1, \pm 2, \cdots,$$
then  KdV hierarchy is
$$v_t=KG_{m-1}=JG_m, \ \ m=0,\pm 1,\pm 2,\cdots\eqno(3.1)$$
which contains both the NKdV hierarchy and the positive order KdV hierarchy.

The first  equation ($m=0$)  in the KdV hierarchy (3.1) is trivial system
$$v_t=JG_0=0, \ \ KG_{-1}=JG_0=0,\eqno(3.2)$$
which can be regarded as is a ``sharp threshold"  equation of the NKdV
hierarchy and positive order KdV hierarchy.

A Kupershmidt nonholonomic
deformation of the hierarchy (3.1)  takes
$$
\begin{array}{l}
v_t=JG_{m}+Jw,\ \  m=0, \pm 1, \pm 2,\cdots,\\[8pt]
 Kw=0,\end{array}
\eqno(3.3) $$ where two operators $K$ and $J$  are given by (1.4).
Then the first flow  ($m=0$) of the hierarchy (3.3) is exactly the
NKdV equation (1.1)
$$
\begin{array}{l}
v_t=w_x,\\[8pt]
w_{xxx}+4vw_x+2v_xw=0,
\end{array}
$$
which may be regarded as a Kupershmidt nonholonomic deformation of
the threshold equation (3.2)
\\[8pt]
%%%%%%%%%%%%%%%%%%%%%%%%%%%%%%%%%%%%%%%%%%%%%%%%%%%%%%%%%%%%%%%%%%%%%%%%%%%%%%%%%%%
{\bf 3.2.  NKdV hierarchy with self-consistent sources }\\

Soliton equations with self-consistent sources have important
physical applications, for example, the KdV equation with
self-consistent source describes the interaction of long and short
capillary-gravity waves \cite{Leon, Me1, Me2, Me3}.

For the $N$ distinct $\lambda_j$ of  the spectral problem (2.1), the
functional gradient of $\lambda_j$ with respect to $v$ is
$$\frac{\delta \lambda_j}{\delta v}=\psi_j^2.$$
Here we define the whole KdV hierarchy  with self-consistent sources 
as follows
%is given by
$$
\begin{array}{l}
\displaystyle{v_t=JG_m+\alpha J\frac{\delta \lambda}{\delta v}=JG_m+\alpha J\sum_{j=1}^N\psi_j^2,}\\[8pt]
\psi_{j,xx}+(v+\lambda_j)\psi_j=0,\\[8pt]
 m=0, \pm 1, \pm 2,\cdots; \
j=1, \cdots, N.
\end{array}\eqno(3.4)
$$

Taking $m=1$ in the hierarchy (3.4) leads to the KdV equation with
self-consistent sources
$$
\begin{array}{l}
\displaystyle{v_t=\frac{1}{4}(v_{xxx}+6vv_x)+\alpha\partial_x\sum_{j=1}^N\psi_j^2,}\\[8pt]
\psi_{j,xx}+(v+\lambda_j)\psi_j=0, \ j=1, \cdots, N,
\end{array}
$$
while choosing $m=-1$ in the hierarchy (3.4) gives the NKdV equation with
self-consistent sources
$$
\begin{array}{l}
\displaystyle{v_t=w_x+\alpha\partial_x\sum_{j=1}^N\psi_j^2,}\\[8pt]
w_{xxx}+4vw_x+2v_xw=0,\\[8pt]
\psi_{j,xx}+(v+\lambda_j)\psi_j=0, \ j=1, \cdots, N.
\end{array}
$$

 Obviously,
taking $N=1,\  m=0, \ \alpha=1,\  v\rightarrow v+\lambda_1$ in the
hierarchy (3.4),  then we get the NKdV equation (1.2)
$$
\begin{array}{l}
v_t=(\psi_1^2)_x,\ \ \psi_{1,xx}+v\psi_1=0,
\end{array}
$$
which may be regarded as the threshold equation (3.2)  with
self-consistent sources.
\\[8pt]
%%%%%%%%%%%%%%%%%%%%%%%%%%%%%%%%%%%%%%%%%%%%%%%%%%%%%%%%%%%%%%%%%%%%%%%%%%%%%%%%%%%
{\bf 3.3.  Reduction of  the NKdV equation (1.1)}\\

{\bf Theorem  3.}   $(u, v)$ is a solution of NKdV-1 equation
(1.2) if and only if %$\Longleftrightarrow $
$(w, v)$ with $w=u^2$  is a solution of NKdV equation (1.1) under
the transformation
$$ u_{xx}+vu=0,\eqno(3.5)$$
which is actually a linear Schr\"{o}dinger equation or Hill
equation.

{\it Proof.}  Let
 $$w=u^2,\eqno(3.6)$$
 then by %the first equation of the NKdV equation
 (1.1), we have % becomes
$$v_t=w_x=2uu_x,$$
which is the  first equation of %NKdV-1  equation
(1.2).   By %using
(3.6), the second equation of (1.1) leads to %can be rewritten as
$$3u_x(u_{xx}+vu)+u(u_{xx}+vu)_x=0,$$
or equivalently,
$$[u^3(u_{xx}+vu)]_x=0,\eqno(3.7)$$

Apparently, according to (3.7), if  $(u, v)$ is a solution of the
NKdV-1 equation (1.2), then  $(w, v)$ is a solution of the NKdV
equation (1.1) where $w=u^2$. Reversely, if $(w, v)$ is a solution
of the NKdV equation (1.1), then  $(u, v)$ is also  a solution of
the NKdV-1 equation (1.2) under the transformation (3.5).

For a given function $\phi$, let us define the following Baker-Akhiezer function
$$u=\exp\left(\int_0^x\phi dx\right),%\ \ \ w=\exp\left(2\int_0^x\phi dx\right),
\eqno(3.8)$$ then  (3.2) yields the following  Riccati equation
$$\phi_x+\phi^2+v=0.\eqno(3.9)$$
So, we have

 {\bf Theorem  4. }    $(u, v)$ is a solution of the NKdV-1 equation
(1.2) if and only if  $(w, v)$   is a solution of the NKdV equation
(1.1) as $\phi$  is a solution of the Riccati equation (3.9) while
$u$ is the Baker-Akhiezer function (3.8) and $w=u^2$.
\\[8pt]
%%%%%%%%%%%%%%%%%%%%%%%%%%%%%%%%%%%%%%%%%%%%%%%%%%%%%%%%%%%%%%%%%%%%%%%%%%%%%%%%%%%
 {\bf 3.4. Ermakov-Pinney equation}\\

 The Ermakov-Ray-Reid systems
$$\begin{aligned}
\psi_{xx}+\omega^2(x)\psi=\frac{1}{\psi^2\phi}F(\frac{\phi}{\psi}),\\
\phi_{xx}+\omega^2(x)\phi=\frac{1}{\psi\phi^2}G(\frac{\psi}{\phi}),
\end{aligned}$$
were originally  introduced by Ermakov \cite{VE, RR80}. Due to their
nice mathematical properties of Ermakov systems admitting a novel
integral
 of motion  together with a concomitant nonlinear superposition principle
 and extensively physical applications, there has been numerous an extensive
 literature devoting to the analysis
 of the Ermakov systems \cite{ath,Pinney,CHR,CBKA,CRHA,RS96}.  The
most simple
  case is equation
$$\psi_{xx}+\omega^2(x)\psi=\frac{c}{\psi^3},$$
which is called the Ermakov-Pinney equation. The Ermakov-Pinney
equation  is a quite famous example of a nonlinear ordinary
differential equation. Such an equation (and generalizations
thereof) have been shown to be relevant to a number of physical
contexts including quantum cosmology, quantum field theory,
nonlinear elasticity and nonlinear optics
\cite{Fine,Rosu,Shahinpoor}. A recent account of some of its
properties along with applications in cosmological settings can be
found in Ref. \cite{HL}.

{\bf Proposition  4. }   Suppose that  $(w, v)$  is a  solution of
the
 NKdV equation (1.1).   Let
 $$w=p_t=\psi^2, \ v=p_x,$$
    then $\psi$
satisfies a Ermakov-Pinney equation
$$\psi_{xx}+v\psi=\frac{\mu}{\psi^3},\eqno(3.9) $$
where $\mu$ is an integration constant.

Especially, if  $(u, v)$ is the solution of the NKdV-1 equation
(1.2), let
$$u=\phi\exp\left(i\int\mu\phi^{-2}dx\right),\eqno(3.10)$$
then $\phi$ satisfies the Ermakov-Pinney equation
$$\phi_{xx}+v\phi=\frac{\mu}{\phi^3}.\eqno(3.11)$$

{\it Proof.}  Substituting transformation $w=\psi^2$ into  the
second equation of the NKdV equation (1.1) yields
$$
\begin{array}{l}
w_{xxx}+4vw_x+2v_xw=2\psi(\psi_{xx}+v\psi)_x+6\psi_x(\psi_{xx}+v\psi)\\[8pt]
=\displaystyle{\frac{2}{\psi^2}}[(\psi_{xx}+v\psi)\psi^3]_x=0,
\end{array}
$$
which leads to (3.9).

 Substituting transformation (3.10)  into  the
second equation of the NKdV equation (1.2) yields
$$
\begin{aligned}
u_{x}+vu=(\phi_{xx}+v\phi-\frac{\mu}{\phi^3})\exp\left(i\int\mu\phi^{-2}dx\right)=0,
\end{aligned}
$$
which implies  (3.11).

{\bf Proposition 5. } For given function $v$, let $\psi_1, \psi_2$
are two solutions of linear Schr\"{o}dinger equation
$$u_{xx}+vu=0,\eqno(3.12)$$
 then equation
$$w_{xxx}+4w_xv+2v_xw=0\eqno(3.13)$$
admits a general solution
$$w=a\psi_1^2+2b\psi_1\psi_2+c\psi_2^2,\eqno(3.14) $$
where
 $$ac-b^2=\frac{\mu}{2W},\ \ W=\psi_1\psi_{2,x}-\psi_{1,x}\psi_2.$$

{\it Proof.}  Let $w=\psi^2$, by proposition 4, then $\psi$
satisfies a Ermakov-Pinney equation (3.9).  It is easy to check that
if $\psi_1$ and $ \psi_2$ are two solutions of equation (3.14), then
$$\psi=\sqrt{a\psi_1^2+2b\psi_1\psi_2+c\psi_2^2}$$
is  a solution of equation (3.9).  So (3.14) is a general solution
of the equation (3.13).
\\[12pt]
%%%%%%%%%%%%%%%%%%%%%%%%%%%%%%%%%%%%%%%%%%%%%%%%%%%%%%%%%%%%%%%%%%%%%%%%%%%%%%%%%%%%%%%%%%%%%%%%%%%%%%%%%%%%%%%%%%%%%%%%%%%%%%%%%%%%%%
%%%%%%%%%%%%%%%%%%%%%%%%%%%%%%%%%%%%%%%%%%%%%%%%%%%%%%%%%%%%%%%%%%%%%%%%%%%%%%%%%%%%%%%%%%%%%%%%%%%%%%%%%%%%%%%%%%%%%%%%%%%%%%%%%%%%%%
{\bf\large  4.  Darboux transformation of NKdV equations  }\\

In this section,  we shall construct a Darboux transformation for
 general  NKdV equation (1.1), and then reduce it  to the NKdV-1 equation
 (1.2).\\[8pt]
 {\bf 4.1. Darboux transformation}\\

 A Darboux transformation is actually a special gauge
transformation
$$\tilde{\psi}=T\psi\eqno(4.1)$$
of solutions of the Lax pair  (2.9), here $T$ is a differential
operator (For the Lax pair (2.10), the Darboux transformation with
$\lambda=0$ can be obtained). It requires that $\tilde{\psi}$ also
satisfies the same Lax pair (2.9)  with some $\tilde{L}$ and
$\tilde{M}$, i. e.
$$
\begin{array}{l}
\tilde{L}\tilde{\psi}=\lambda\tilde{\psi},\ \ \ \
\tilde{L}=TLT^{-1},\\[8pt]
\tilde{M}\tilde{\psi}=0,\ \ \ \ \tilde{M} =TMT^{-1}
\end{array}\eqno(4.2)$$
Apparently, we have
$$[\tilde{L},\tilde{M}]=
T[L,M]T^{-1},$$
 which implies that $\tilde{L}$ and $\tilde{M}$
are required to have the same forms as $L$ and  $M$, respectively,
in  order to make  system (2.9)  invariant under the gauge
transformation (3.4). At the same time the old potentials $u$ and
$v$  in $L$, $M$ will be mapped into new potentials $\tilde{u}$ and
$\tilde{v}$ in $\tilde{L}$, $\tilde{M}$. This process can be done
continually  and usually it may yield a series of multi-soliton
solutions.

Let us now set up a Darboux transformation for  the system (2.9).
Let $\psi_0=\psi_0(x,t)$  be a basic solution of Lax pair (2.9)  for
$\lambda_0$, and use it  to define the following gauge
transformation
$$\tilde{\psi}=T\psi,\eqno(4.3)$$
where $$ T=\partial_x-\sigma, \ \ \sigma=\partial_{x}\ln
\psi_0.\eqno(4.4)$$
From  (2.9) and (4.4),  one can see %we can show
that $\sigma$ satisfies
$$\sigma_{x}+\sigma^2+v-\lambda=0\eqno(4.5)$$
$$4\sigma_{xxt}+12\sigma_x\sigma_t+4v\sigma_t+2w\sigma_x+6\sigma\sigma_{xt}+3w_{xx}=0.\eqno(4.6)$$

{\bf Proposition 6.}  The operator  $\tilde{L}$ determined by (4.2)
has the same form as $L$, that is,
$$\tilde{L}=\partial_x^2+\tilde{v},$$
where the transformation between  $v$ and  $\tilde{v}$ is given by
$$  \tilde{v}=v+2\sigma_x. \eqno(4.7)$$
The transformation: $(\psi,v)\rightarrow (\tilde{\psi}, \tilde{v})$
is called a Darboux transformation of the first spectral problem of
Lax pair (2.9).

{\it Proof.}      According to (4.2), we just prove
$$\tilde{L}T=TL,$$
that is,
$$(\partial_{x}^2+\tilde{v})(\partial_x-\sigma)=(\partial_x-\sigma)(\partial_{x}^2+v),$$
which is true through  (4.5) and (4.7).

{\bf Proposition 7.}  Under the transformation (4.3), the operator
$\tilde{M}$ determined by (4.2) has the same form as $M$, that is,
$$\tilde{M}=4\partial_x^2\partial_t+4\tilde{v}\partial_t-2\tilde{w}\partial_x-3\tilde{w}_{x},\eqno(4.8)$$
where the transformations between $w$, $v$ and $\tilde{w}$,
$\tilde{v}$ are given by
$$\tilde{w}=w+2\sigma_t,\ \ \ \tilde{v}=v+2\sigma_x. \eqno(4.9)$$
The transformation: $(\psi,w, v)\rightarrow (\tilde{\psi},
\tilde{w},\tilde{v})$ is  Darboux transformation of the second
spectral problem of Lax pair (2.9).

{\it Proof.}   To see  that $\tilde{M}$ has the form (4.8) same as
$M$, we just prove
$$\tilde{M}T=TM,\eqno(4.10) $$
where
$$\tilde{M}=4\partial_x^2\partial_t+f\partial_t+g\partial_x+h,\eqno(4.11)$$
with  three functions $f, g$, and $h$ to be determined. Substituting
$\tilde{M}, \ M, \ L$ into (4.10) and  comparing the coefficients of
all distinct operators  lead to:

coefficient of operator $\partial_x\partial_t$
$$f=4v+8\sigma_x=4\tilde{v},$$
which holds by using (4.9).

coefficient of operator  $\partial_x^2$
$$g=2w+4\sigma_t=2\tilde{w},$$
which implies  from  (4.9).

coefficient of operator $\partial_x$
$$
\begin{array}{l}
h=8\sigma_{xt}+5w_x-2\sigma w+g\sigma
=6\sigma_{xt}+3w_x+2(\sigma_x+\sigma^2+v)_t\\[4pt]
=6\sigma_{xt}+3w_x=3\tilde{w}_x,
\end{array}
$$
here we have used equation  (4.5) and (4.9).

 coefficient of  operator $\partial_t$
$$-4\sigma_{xx}-f\sigma=4v_x-4v\sigma,$$
that is,
 $$\sigma_{xx}+2\sigma\sigma_x+v_x=0.$$
which holds by using (4.5).

 coefficient of  non-operator:
$$4\sigma_{xxt}+f\sigma_t+g\sigma_x+\sigma h+3w_{xx}-3\sigma w_x=0,$$
that is,
 $$4\sigma_{xxt}+12\sigma_x\sigma_t+4v\sigma_t+2w\sigma_x+6\sigma\sigma_{xt}+3w_{xx}=0,$$
which is  the equation  (4.6). We complete the proof. $\square$

Propositions 4 and 5 tell us that the  transformations (4.3) and
(4.9) send  the Lax pair   (2.9) to another Lax pair (4.2)
 in the same type.  Therefore, both of the Lax pairs lead to the
 same NKdV  equation (1.1).   So, we  call the transformation
 $(\psi,w,v)\rightarrow(\tilde{\psi},\tilde{w},\tilde{v})$
  a Darboux transformation of   the NKdV equation  (1.1).
  In summary, we arrive at the following theorem.

  {\bf Theorem 5.}\ \   A solution $w, \ v$ of the
   NKdV  equation (1.1)  is  mapped into its  new solution $\tilde{w}, \ \tilde{v}$
   under the Darboux transformations (4.3) and (4.9).\\[8pt]
 %%%%%%%%%%%%%%%%%%%%%%%%%%%%%%%%%%%%%%%%%%%%%%%%%%%%%%%%%%%%%%%%%%%%%%%%%%%%%%%%%%%
{\bf 4.2. Reduction of Darboux transformations}

To get  Darboux transformations for the NKdV-1 equation (1.2),  we
consider two reductions of the Darboux transformations (4.3) and (4.9).

{\bf Corollary 1.} Let $\lambda=k^2>0$, then under the constraints
$w=u^2, \ v=-u_{xx}/u$, the Darboux transformations (4.3) and (4.9)
are  reduced to a Darboux transformation of  the NKdV-1 equation
(1.2) 
 $(\psi,v,u)\rightarrow(\tilde{\psi},\tilde{v},\tilde{u})$,
  where
$$
\begin{array}{l}
\tilde{\psi}=T\psi, \ \ \tilde{v}=v+2\sigma_x, \ \
\tilde{u}=k^{-1}(u_x-\sigma u)=k^{-1}Tu.
\end{array}\eqno(4.13)$$

{\it Proof.}  For $\lambda>0$, suppose that   $(v,u)$ is a solution
of the NKdV-1 equation and $\psi$ is an eigenfunction of the Lax pair (1.4),
then we have
$$\lambda^{-1}(u\psi_x-u_x\psi)=\partial_x^{-1}(u\psi). $$
Therefore, the Lax pair (1.4) can be rewritten as
 $$\begin{aligned}
&\psi_{xx}+v\psi=\lambda\psi,\\
&\psi_t=\frac{1}{2}u\lambda^{-1}(u\psi_x-u_x\psi)=\frac{1}{2}u\partial_x^{-1}(u\psi)=N(u,\lambda)\psi,
\end{aligned}\eqno(4.12)$$
where $N=N(u,\lambda)=\frac{1}{2}u\partial_x^{-1}u$.

According to Proposition 6,  the spectral problem of Lax pair
(4.12) is covariant under the transformation (4.13), that is, 
$$\tilde{\psi}_{xx}+\tilde{v}\tilde{\psi}=\lambda\tilde{\psi}.$$
So, we only need to prove %that
$$\tilde{\psi}_t=N(\tilde{u},\lambda)\tilde{\psi}\eqno(4.14)$$

Substituting (4.13) into the left  hand side of (4.14) yields %gives
 $$\begin{aligned}
&\tilde{\psi}_t=(\psi_{t})_x-(\sigma\psi)_t=(N\psi)_x-\sigma
N\psi-(\psi_0^{-1}N\psi_0)_x\psi,\\
& =\frac{1}{2}[(u_x-\sigma
 u)\partial_x^{-1}(u\psi)-\psi_0^{-1}\psi(u_x-\sigma
 u)\partial_x^{-1}(u\psi_0)]\\
 &=\frac{1}{2}k\tilde{u}[\partial_x^{-1}(u\psi)+k^{-2}(u_x-\sigma u)\psi].
\end{aligned}\eqno(4.15)$$
In the same way, substituting (4.13) into the right hand side of (4.14)
produces
 $$\begin{aligned}
&N(\tilde{u},\lambda)\tilde{\psi}=\frac{1}{2}\tilde{u}\partial_x^{-1}[k^{-1}(u_x-\sigma
u)(\psi_x-\sigma\psi)]\\
&=\frac{1}{2}k^{-1}\tilde{u}[u_x\psi-\partial_x^{-1}(u_{xx}\psi)-\sigma
u\psi+\partial_x^{-1}(\psi_0^{-1}\psi_{0,xx}u\psi)]\\
 &=\frac{1}{2}k^{-1}\tilde{u}[k^2\partial_x^{-1}(u\psi)+(u_x-\sigma u)\psi].
\end{aligned}\eqno(4.16)$$
Combining (4.15) with (4.16) implies that  (4.14) holds.
$\square$

In a similar way, we also have the following result.
 
{\bf Corollary 2.}  Let $\lambda=0$, then under the constraints
$w=u^2, \ v=-u_{xx}/u$, the Darboux transformations (4.3) and (4.9)
are  reduced to  another Darboux transformation of the NKdV-1 equation
(1.2) 
 $(\psi,v,u)\rightarrow(\tilde{\psi},\tilde{v},\tilde{u})$,
  where %in which
 $$\begin{aligned}
&\tilde{v}=v+2\sigma_x, \ \ \tilde{\psi}=\psi-\psi_0^{-1}\sigma\partial_x^{-1}(\psi_0\psi),\\
&\tilde{u}=\left\{\begin{matrix}\displaystyle{\psi_0^{-1}\sigma},&u=0,\cr\cr
\displaystyle{u-\psi_0^{-1}\sigma\partial_x^{-1}(\psi_0u)},&u\not=0.\end{matrix}\right.
\end{aligned}\eqno(4.17)$$
with $\sigma=\partial_x\ln (1+\partial_x^{-1}\psi_0^2)$. \\\\[12pt]
%%%%%%%%%%%%%%%%%%%%%%%%%%%%%%%%%%%%%%%%%%%%%%%%%%%%%%%%%%%%%%%%%%%%%%%%%%%%%%%%%%%%%%%%%%%%%%%%%%%%%%%%%%%%%%%%%%%%
  {\large\bf 5. Applications of the Darboux transformation}\\

   In this section, we shall apply the Darboux
transformations (4.3) and (4.9)  to
obtain  kink-type  and bell-type of  explicit  solutions for the NKdV equation (1.1). \\[12pt]
{\bf 5.1.   The kink-wave solutions  }

For the case of $\lambda=k^2>0$,  we  substitute  $v=0, w=1$ into
the Lax pair (2.9) and choose  the following basic solution
$$\psi=e^{\xi}+e^{-\xi}=2\cosh\xi, \ \ \xi=kx-\frac{1}{2k}t+\gamma, \eqno(5.1)$$
where $\gamma$ and $k$ are  two arbitrary constants.

 Taking $\lambda=k_1^2$,  then  (4.4) and (5.1) lead to
$$\sigma_1=\partial_x\ln\psi=k_1\tanh\xi_1, \ \ \xi_1=k_1x-\frac{1}{2k_1}t+\gamma_1.$$
The  Darboux transformation (4.9)  gives  bell-type solution for the
NKdV equation (1.1)
$$
\begin{array}{l}
\tilde{v}^{I}=2\sigma_{1,x}=2k_1^2{\rm sech}^2\xi_1,\\[8pt]
\tilde{w}^{I}=1-2\sigma_{1,t}=\tanh^2\xi_1.
\end{array}\eqno(5.2)
$$
By using Darboux trasformation (4.13), we get  a kink-type wave
solution for the NKdV equation (1.2)
$$\tilde{u}^{I}=k_1^{-1}(u_x-\sigma u)=-\tanh\xi_1, \ \ \xi_1=k_1x-\frac{1}{2k_1}t+\gamma_1.\eqno(5.3)$$

{\bf Remark 1.}  There  is much  difference between  traveling waves of the
NKdV equation (1.2) and  of the classical KdV equation. For the
NKdV equation (1.2), its one-wave solution is a
negative-moving (i.e. from right to left) kink-wave with
velocity $-{1}/{2k_1^2}$, amplitude $\pm 1$ and width ${1}/{k_1}.$ %, respectively.
Its amplitude is  independent of velocity, and width is directly
proportional to the velocity.
For the KdV equation
$$u_t+6uu_x+u_{xxx}=0,\eqno(5.4)$$
one-soliton solution is
$$u=\frac{k^2}{2}{\rm sech}^2\frac{k(x-k^2t)}{2},\eqno(5.5)$$
%The one-soiton solution (5.5)
which is a bell-type  positive-moving wave with velocity $k^2$,
amplitude $k^2/2$ and width $1/k$, respectively. Its  amplitude is
directly proportional to velocity, and width is inversely
proportional to the velocity.

Let us now construct two-kink %soliton
solutions to see the interaction of two
kink %soliton
solutions.   According to (4.4),
$$\tilde{\psi}=T\psi=(\partial_x-\sigma_1) (e^{\xi}+e^{-\xi})\eqno(5.6)$$
is also an eigenfunction of Lax pair (2.9).  Taking $\lambda=k_2^2$,
we have
$$\sigma_2=-k_1\tanh\xi_1+\frac{k_1^2-k_2^2}{k_1\tanh\xi_1-k_2\tanh\xi_2}.\eqno(5.7)$$

Repeating the Darboux transformation (4.9) one more time, we get two
soliton solution for the the NKdV equation (1.1)
$$\tilde{v}^{II}=\tilde{v}^I+2\sigma_{2,x}=\frac{(k_1^2-k_2^2)(k_2^2{\rm sech}^2\xi_2-k_1^2{\rm sech}^2\xi_1)}
{(k_1\tanh\xi_1-k_2\tanh\xi_2)^2},$$
$$\tilde{w}^{II}=\tilde{w}^I-2\sigma_{2,t}=\left(\frac{k_1\tanh\xi_2-k_2\tanh\xi_1}
{k_1\tanh\xi_1-k_2\tanh\xi_2}\right)^2.$$ Therefore, we obtain a two-kink wave solution
of the NKdV equation (1.2)
$$\tilde{\tilde{u}}=\frac{k_2\tanh\xi_1-k_1\tanh\xi_2}
{k_1\tanh\xi_1-k_2\tanh\xi_2}.\eqno(5.8)$$

\input epsf
\begin{figure}
\footnotesize $\textbf{(a)}$ \ \ \ \ \ \  \ \ \ \ \ \ \ \ \   \ \ \
\ \ \ \ \ \ \ \ \ \  \ \ \ \ \ \ \  \ \  \ \ \ \ \ \ \ \ \ \ \ \ \ \
\ \ \ \ \ \ \ \ \ \ \footnotesize $\textbf{(b)}$ \ \ \ \
 \begin{center}
\includegraphics[scale=0.55]{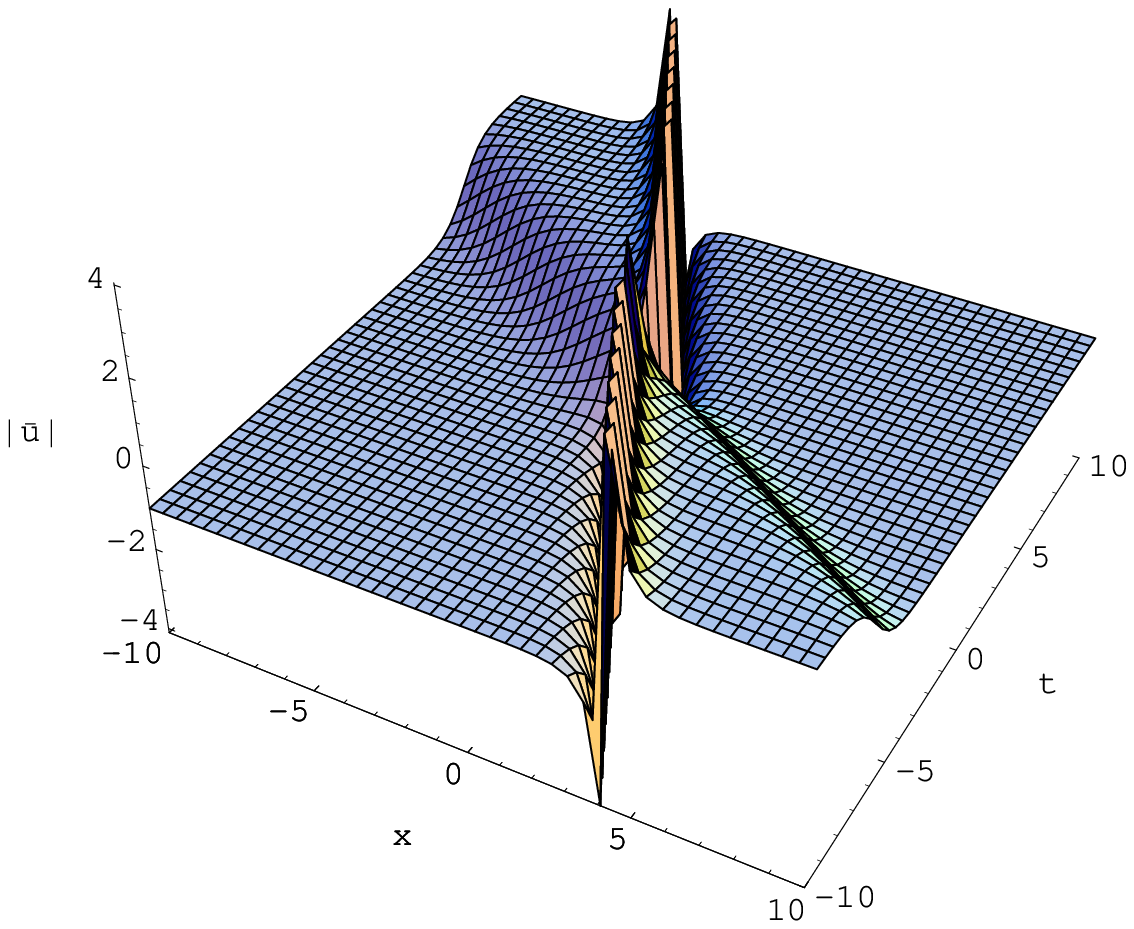}\ \ \ \ \ \ \ \ \
\includegraphics[scale=0.55]{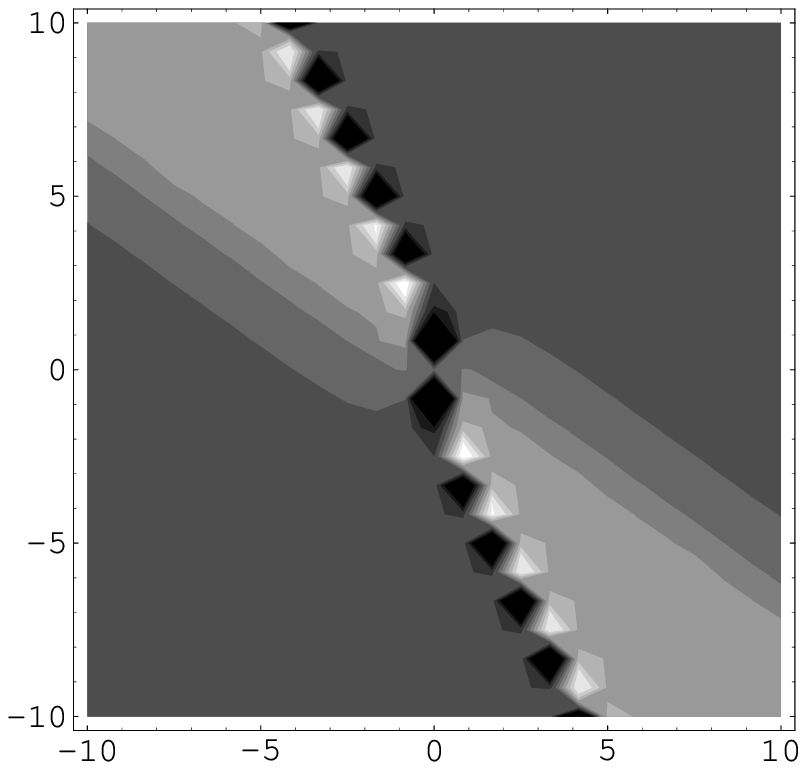}
\end{center}

\caption{\footnotesize\baselineskip=15pt The two-kink wave solution $u(x,t)$
with parameters: $k_1=1, $  $k_2=0.6$.  (a) Perspective view of the
wave. (b) Overhead view of the wave, with contour plot shown. The
bright lines are crests and the dark lines are troughs.}
\end{figure}

\input epsf
\begin{figure}
\ \ \ \ \ \ \ \ \ \  \ \ \ \ \ \ \ \ \footnotesize $\textbf{(a)}$ \
\ \ \ \ \  \ \ \ \ \ \ \ \ \   \ \ \ \ \ \ \ \ \ \ \ \ \
\footnotesize $\textbf{(b)}$ \ \ \ \ \ \ \  \ \ \ \ \ \ \ \  \ \ \ \
 \begin{center}
\includegraphics[scale=0.50]{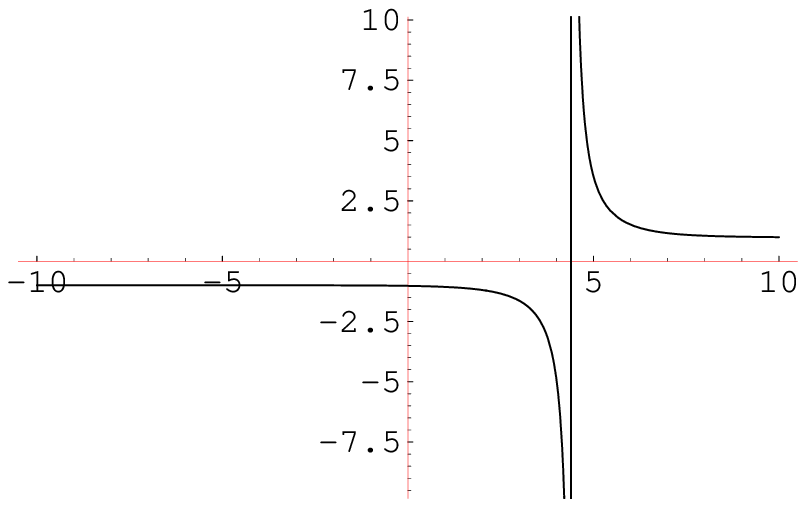}\ \
\includegraphics[scale=0.50]{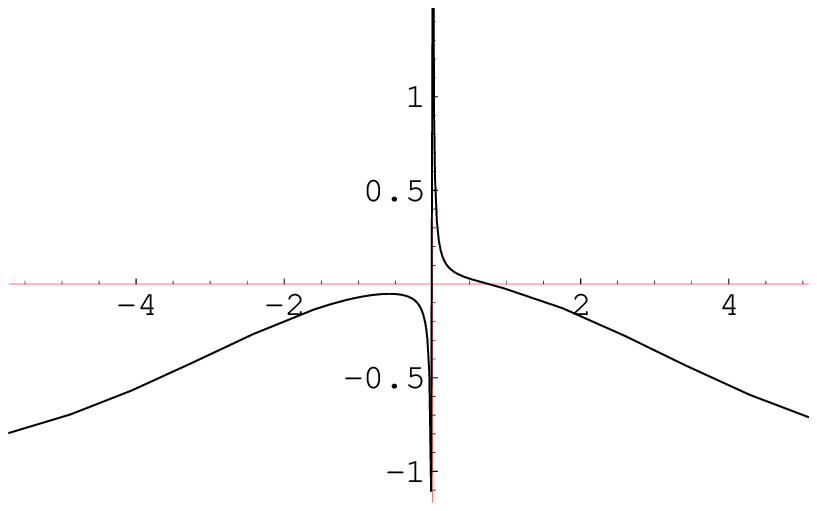}
\\[12pt]
\footnotesize $\textbf{(c)}$ \ \ \ \ \ \  \ \ \ \ \ \ \ \ \   \ \ \
\ \ \ \ \ \ \ \ \ \  \footnotesize $\textbf{(d)}$ \ \ \ \ \ \ \  \ \
\ \ \ \ \ \ \ \ \ \ \ \ \ \ \ \ \ \ \ \ \ \ \ \ \footnotesize
$\textbf{(e)}$ \ \ \ \
\\[8pt]
\includegraphics[scale=0.43]{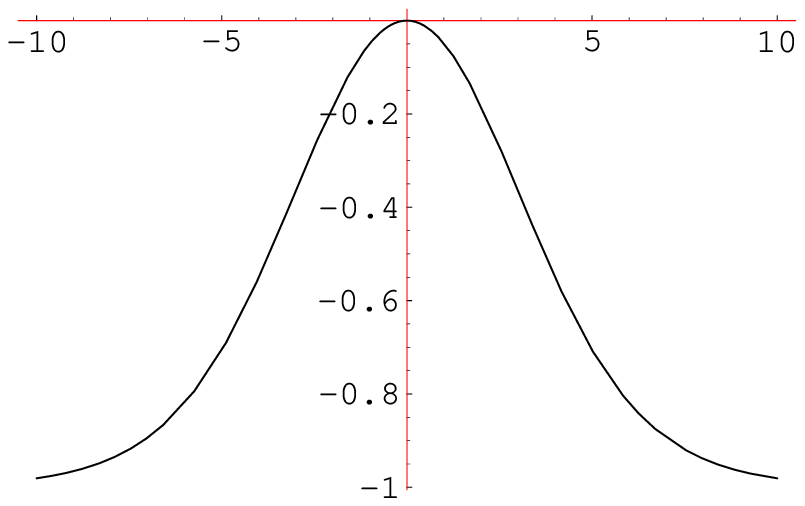}
\includegraphics[scale=0.456]{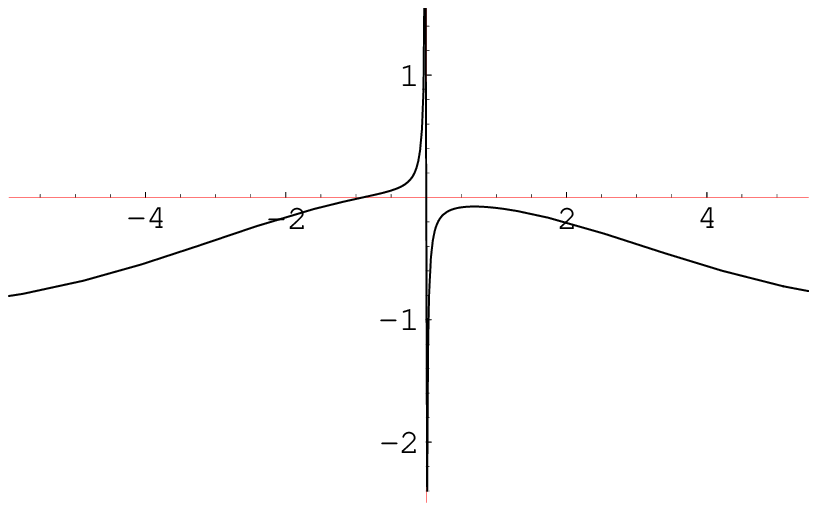}
\includegraphics[scale=0.46]{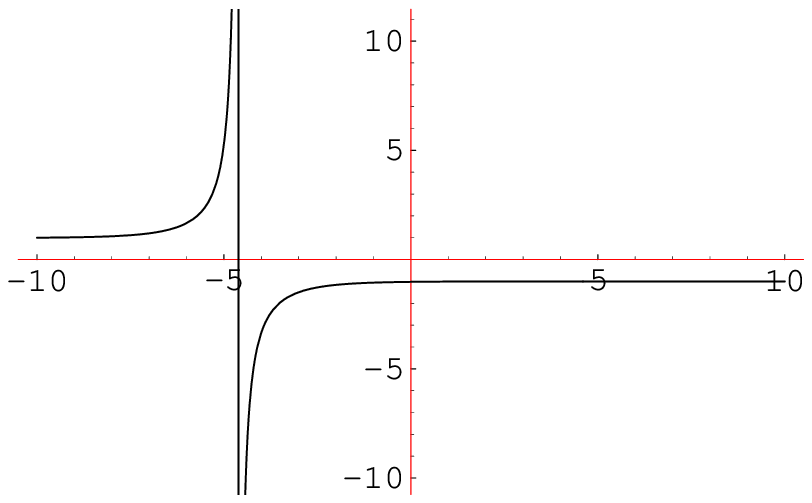}
\end{center}

\caption{\footnotesize\baselineskip=15pt Interaction between
singular soliton   ${\rm csch}\xi_1$ and smooth soliton ${\rm
sech}\xi_2$ with parameters: (a) $t=-3$,  (b) $ t=-0.05$, (c) $
t=0$, (d) $t=0.05$, (e) $ t=3$.}
\end{figure}

Let us use the two-kink wave solution (5.8) to analyze interaction
of the two one-soliton solutions.  Without loss of generality,  we
suppose $k_1>k_2>0$, then we have
$$\xi_2=\frac{k_2}{k_1}\left[\xi_1-\frac{k_1}{2}(\frac{1}{k_2^2}-\frac{1}{k_1^2})t\right].$$
Therefore,  on the fixed line $\xi_1=$constant,  we get
$$\tanh\xi_2\sim -1,\ \ t\rightarrow
+\infty,
$$
and it follows (5.8) that
$$\tilde{\tilde{u}}\sim \frac{k_2\tanh\xi_1+k_1}
{k_1\tanh\xi_1+k_2}=
\coth\left(\xi_1-\frac{1}{2}\ln\frac{k_1-k_2}{k_1+k_2}\right), \
t\rightarrow +\infty.\eqno(5.9)
$$

In a similar way, one can get
$$\tanh\xi_2\sim 1 \ \ {\rm as} \ t\rightarrow -\infty,$$
 which are main parts compared with terms $1$ and $e^{2\xi_1}$,
 and it follows (3.19) that
 $$\tilde{\tilde{u}}\sim\frac{k_2e^{2\xi_1}-k_1}
{k_1e^{2\xi_1}-k_2}=
\coth\left(\xi_1+\frac{1}{2}\ln\frac{k_1-k_2}{k_1+k_2}\right), \
t\rightarrow -\infty.\eqno(5.10)$$

 In a similar way,
on the line $\xi_2=$constant, we will arrive at
$$\tilde{\tilde{u}}\sim
\tanh\left(\xi_2+\frac{1}{2}\ln\frac{k_1-k_2}{k_1+k_2}\right),\ \
{\rm as}\ \ t\rightarrow +\infty, \eqno(5.11)$$
$$\tilde{\tilde{u}}\sim
\tanh\left(\xi_2-\frac{1}{2}\ln\frac{k_1-k_2}{k_1+k_2}\right),\ \
{\rm as}\ \ t\rightarrow -\infty.\eqno(5.12)$$

 {\bf Remark 2.} From expressions (5.9)-(5.12), we see that the
two-kink wave solution (5.8) is a singular solution, which is able
to be decomposed into a singular kink-type solution and a smooth
kink wave solutions.  The expressions (5.10) and (5.12) show that
the wave $\tanh\xi_2$
 is on the left of the wave $\coth\xi_1$ before their interaction, while
the expressions (5.9) and (5.11) show that the wave $\coth\xi_1$
 is on the left of the wave $\tanh\xi_2$  after their interaction.
 The shapes of the two kink waves $\coth\xi_1$
and $\tanh\xi_2$ don't change except their phases.
 Their phases of the two waves $\coth\xi_1$ and $\tanh\xi_2$
 are
  $\ln\frac{k_1-k_2}{k_1+k_2}>0$ and
  $-\ln\frac{k_1-k_2}{k_1+k_2}<0$, respectively
  as the wave is negatively going
 along the $x-$axis.   Very interesting case is particular at $t=0$:
 collision of such two kink waves forms a
 smooth  bell-type soliton  and its singularity disappears (See Figure
 2).

After their interaction It can be seen that the the two kink waves
resume their original shapes. At the right moment of interaction,
the two kink waves are fused into a smooth bell-type soliton. The
two-kink wave interactions possess the regular elastic-collision
features and pass through each other, and their shapes keep
unchanged with a phase shift after the interaction. Here, we also
demonstrate a fact that the large-amplitude kink wave with faster
velocity overtakes the small-amplitude one, after collision, the
smaller one is left behind.
\\[12pt]
%%%%%%%%%%%%%%%%%%%%%%%%%%%%%%%%%%%%%%%%%%%%%%%%%%%%%%%%%%%%%%%%%%%%%%%%%%%%%%%%%%%%%%%%%%%%%%%%%%%%%%%%%%%%%%%%%%%%
{\bf 5.2. The bell-type soliton solutions}

(i)  For the case of $\lambda=0$ (i.e. without parameter $\lambda$),
we substitute $v=-k^2, \ w=0$ into the Lax pair  (2.10), and choose
the following basic solution as
$$\psi=e^{\xi}+e^{-\xi}, \ \ \xi=kx+\frac{1}{2k}t,$$
where $k$ is  an arbitrary constant.

 Taking $k=k_1$, (4.4) gives
$$\sigma=\sigma_1=\partial_x\ln\psi=k_1\tanh\xi_1, \ \ \xi_1=k_1x+\frac{1}{2k_1}t.\eqno(5.13)$$
Using  the Darboux transformation (4.9), we have  one-soliton
solution for the NKdV equation (1.1)
$$\begin{aligned}
&\tilde{v}=v+2\sigma_{1,x}=2k_1^2{\rm sech}^2\xi_1-k_1^2,\\
&\tilde{w}=-2\sigma_{1,t}={\rm sech}^2\xi_1.
\end{aligned}\eqno(5.14)$$

 So,  we get a one-soliton solution for the NKdV-1
equation (1.2) by using Darboux transformation (4.17)
$$\tilde{u}= {\rm sech}\xi_1, \ \ \xi_1=k_1x+\frac{1}{2k_1}t.\eqno(5.15)$$

{\bf Remark 3.}    For the negative order  KdV equation (1.2),  its
one-soiton solution (5.15) is a smooth bell-type negative-moving
wave, whose velocity, amplitude and width are ${1}/{2k_1^2}$, $\pm
1$ and ${1}/{k_1}$, respectively. Its amplitude is independent of
velocity, and width is directly proportional to the velocity.

(ii)  For the  case of $\lambda=-k^2$,
 we take a seed solution of  $v=-2k^2, w=1$  in  the
Lax pair  (2.9), and choose the following basic solution  as
$$\psi=e^{\xi}+e^{-\xi}, \ \ \xi=kx-\frac{1}{2k}t+\gamma,$$
where $k$ is  an arbitrary constant.

Taking $k=k_1$ sends  (3.7) to
$$\sigma=\sigma_1=\partial_x\ln\psi=k_1\tanh\xi_1, \ \ \xi_1=k_1x-\frac{1}{2k_1}t+\gamma_1.\eqno(5.13)$$
Using the Darboux transformation (3.12), we then get one-soliton
solution
$$\begin{aligned}
&\tilde{v}^{I}=v+2\sigma_{1,x}=-2k_1^2{\rm tanh}^2\xi_1+\gamma_1,\\
&\tilde{w}^{I}=1-2\sigma_{1,t}=1+{\rm sech}^2\xi_1,
\end{aligned}\eqno(5.14)$$
 which cannot satisfies the constraint (3.3), so  $\sqrt{\tilde{w}^{I}}$ is not   soliton for
the NKdV equation (1.2).

{\bf Remark 4.}  For the NKdV equation (1.1),  its
one-soiton solution (5.14) is a smooth bell-type positive-moving %going movement
wave, whose velocity, amplitude and width are ${1}/{2k_1^2}$, $\pm
1$ and ${1}/{k_1}$, respectively. Its amplitude is  independent of
velocity, and width is directly proportional to the velocity.

Let's construct a two-soliton solution of the NKdV equation (1.1).
According to the gauge transformation (4.4),
$$\tilde{\psi}=T\psi=(\partial_x-\sigma_1) (e^{\xi}+e^{-\xi})$$
is also an eigenfunction of Lax (2.9). We have
$$\sigma_2=-k_1\tanh\xi_1+\frac{k_1^2-k_2^2}{k_1\tanh\xi_1-k_2\tanh\xi_2}.$$

Repeating the Darboux transformation (4.9) one more time, we obtain
$$\tilde{v}^{II}=\tilde{v}^I+2\sigma_{2,x}=\frac{(k_1^2-k_2^2)(k_2^2{\rm sech}^2\xi_2-k_1^2{\rm sech}^2\xi_1)}
{(k_1\tanh\xi_1-k_2\tanh\xi_2)^2},$$
$$\tilde{w}^{II}=\tilde{w}^I-2\sigma_{2,t}=\left(\frac{k_1\tanh\xi_2-k_2\tanh\xi_1}
{k_1\tanh\xi_1-k_2\tanh\xi_2}\right)^2,$$ which is the same for
NKdV-1 equation (1.2). So we get  two-soliton solution  with (5.8)
$$\tilde{\tilde{u}}=\pm\frac{k_1\tanh\xi_2-k_2\tanh\xi_1}
{k_1\tanh\xi_1-k_2\tanh\xi_2},$$ but here $\xi_j=k_jx
-\frac{1}{2k_j}t, \ j=1,  2$.
\\[12pt]
 %%%%%%%%%%%%%%%%%%%%%%%%%%%%%%%%%%%%%%%%%%%%%%%%%%%%%%%%%%%%%%%%%%%%%%%%%%%%%%%%%%%%%%%%%%%%
 %%%%%%%%%%%%%%%%%%%%%%%%%%%%%%%%%%%%%%%%%%%%%%%%%%%%%%%%%%%%%%%%%%%%%%%%%%%%%%%%%%%%%%%%%%%%%%
{\bf\large  6.  Bilinearization of the NKdV
 equation}

The bilinear derivative method, developed by Hirota \cite{Hirota},
has become a powerful approach to construct exact solutions of
nonlinear equations.  Once a nonlinear equation is written in  a
bilinear form by using some transformation, then multi-solitary wave
solutions or  quasi-periodic wave solutions  can usually be obtained %constructed
\cite{Hirota1, Hirota2,Hu,Hu2,Liu4,Zhang,Fan1}. However,
unfortunately, this method is not as direct as many people might
wish because the original equation is reduced to two or more
bilinear equations under new variables called Hirota's variables.
Since no a general rule to select Hirota's variables,  there is no
rule to choose some essential formulas (such as exchange formulas),
either. Especially the construction of bilinear B\"{a}cklund
transformation  relies on a particular skill and appropriate
exchange formulas.  On the other hand, in recent years Lambert and
his co-workers have found a kind of the generalized Bell polynomials
playing important role in seeking the characterization of
bilinearized equations. Based on the Bell polynomials, they
presented  an alternative procedure  to obtain parameter families of
a bilinear B\"{a}cklund transformation and Lax pairs for soliton
equations in a quick and short way
\cite{Gilson,Lambert1,Lambert2}. \\[8pt]
{\bf 6.1. Multi-dimensional binary  Bell polynomials }

The main tool we use here  is a class of generalized
multi-dimensional binary  Bell polynomials
\cite{Gilson}-\cite{Lambert2}.

{\bf Definition 1.}  Let $n_k\geq 0, \ k=1, \cdots, \ell$ denote
arbitrary integers, $f=f(x_1,\cdots, x_\ell)$ be a $C^{\infty}$
multi-variable function, then
$$Y_{n_1x_1,\cdots,n_\ell x_\ell}(f)\equiv \exp({-f})\partial_{x_1}^{n_1}\cdots\partial_{x_\ell}^{n_\ell}
\exp(f)\eqno(6.1)$$ is a  polynomial in the partial derivatives of
$f$ with respect to $x_1, \cdots, x_\ell$,  which we call
a multi-dimensional Bell polynomial (a generalized Bell polynomial or
$Y$-polynomial).

For the two dimensional %special
case, let  $f=f(x, t)$,  then the associated %two-dimensional
Bell polynomials through (6.1) can produce the following representatives:
$$\begin{aligned}
&{Y}_x(f) =f_x, \ {Y}_{2x}(f)=f_{2x}+f_x^2,\ \ \
{Y}_{3x}(f)=f_{3x}+3f_xf_{2x}+f_x^3,\\
&{Y}_{x,t}(f) =f_{x,t}+f_xf_t, \ \
{Y}_{2x,t}(f)=f_{2x,t}+f_{2x}f_t+2f_{x,t}f_x+f_x^2f_t, \cdots.
\end{aligned}$$

{\bf Definition 2. } Based on the use of above Bell polynomials
(6.21), the multi-dimensional binary  Bell polynomials (
$\mathcal{Y}$-polynomials)   are defined by %as follows
$$\mathcal{Y}_{n_1x_1,\cdots,n_{\ell}x_{\ell}}(g,h)=Y_{n_1x_1,\cdots,n_{\ell}x_\ell}
(f)\mid_{f_{r_1x_1,\cdots,r_{\ell}x_{\ell}}=\left\{\begin{matrix}g_{r_1x_1,
\cdots,r_{\ell}x_{\ell}},&r_1+\cdots+r_{\ell} \ \ {\rm is\ \
odd},\cr\cr h_{r_1x_1,\cdots,r_{\ell}x_{\ell}},&r_1+\cdots+r_{\ell}
\ \ {\rm is\ \ even},\end{matrix}\right.}$$
which is a multi-variable
polynomial with respect to all  partial derivatives
$g_{r_1x_1,\cdots,r_\ell x_\ell}$ ($ r_1+\cdots+r_{\ell} \ \ {\rm is \
odd}$)  and $h_{r_1x_1,\cdots,r_\ell x_\ell}$  ($
r_1+\cdots+r_{\ell} \ \ {\rm is \ even}$),  $ r_k=0,\cdots, n_k,\ k=0,
\cdots, \ell$.

The binary  Bell polynomials  also inherit %the easily recognizable
partial structures of the Bell polynomials.   The first few lower %st
order binary Bell Polynomials are
 $$\begin{aligned}
&\mathcal{Y}_x (g)=g_x, \ \mathcal{Y}_{2x}(g,h)=h_{2x}+g_x^2,\ \ \mathcal{Y}_{x,t}(g,h)=h_{xt}+g_xg_t.\\
&\mathcal{Y}_{3x}(g, h)=g_{3x}+3g_xh_{2x}+g_x^3, \cdots.
\end{aligned}\eqno(6.2)$$

{\bf Proposition 8.}  The link between binary  Bell polynomials
$\mathcal{Y}_{n_1x_1,\cdots,n_{\ell}x_{\ell}}(g, h)$ and the
standard Hirota bilinear expression $D_{x_1}^{n_1}\cdots
D_{x_{\ell}}^{n_{\ell}}F\cdot G$ can be given by an identity
$$\mathcal{Y}_{n_1x_1,\cdots,n_{\ell}x_{\ell}}(g=\ln F/G, h=\ln FG)=
(FG)^{-1}D_{x_1}^{n_1}\cdots D_{x_{\ell}}^{n_{\ell}} F\cdot
G,\eqno(6.3)$$
where% in which
${n_1}+n_2+\cdots+n_{\ell}\geq 1$, and
operators $D_{x_1},\cdots, D_{x_\ell}$  are classical Hirota's
bilinear operators defined by
 $$D_{x_1}^{n_1}\cdots D_{x_{\ell}}^{n_{\ell}} F\cdot
G=(\partial_{x_1}-\partial_{x_1'})^{n_1}\cdots
(\partial_{x_\ell}-\partial_{x_\ell'})^{n_\ell} F(x_1, \cdots,
x_\ell)G(x_1', \cdots, x_\ell')|_{x_1'=x_1,  \cdots,
x_\ell'=x_\ell}.$$

 In the special case of  $F=G$, the formula (6.4) becomes
 $$\begin{aligned}
&F^{-2}D_{x_1}^{n_1}\cdots D_{x_{\ell}}^{n_{\ell}} G\cdot
G=\mathcal{Y}_{n_1x_1,\cdots,n_{\ell}x_{\ell}}(0,q=2\ln
G)\\[6pt]
&=\left\{\begin{matrix}0,&n_1+\cdots+n_{\ell} \ \ {\rm is\ \
odd},\cr\cr
P_{n_1x_1,\cdots,n_{\ell}x_{\ell}}(q),&n_1+\cdots+n_{\ell} \ \ {\rm
is\ \ even}.\end{matrix}\right.\end{aligned}\eqno(6.4)$$  The first
few  $P$-polynomial are
 $$\begin{aligned}
&P_{2x}(q)=q_{2x}, \ P_{x,t}(q)=q_{xt},  \ P_{4x}(q)=q_{4x}+3q_{2x}^2,\\
&P_{6x}(q)=q_{6x}+15q_{2x}q_{4x}+15q_{2x}^3, \cdots.
\end{aligned}\eqno(6.5)$$
The formulas (6.4) and (6.5) will prove particularly useful in
connecting nonlinear equations to their corresponding bilinear
forms. This means that if a nonlinear equation is expressed%ible
by a linear combination of $P$-polynomials, then the nonlinear equation  can be
transformed into a linear equation.

{\bf Proposition 9.}  The binary Bell polynomials
$\mathcal{Y}_{n_1x_1,\cdots,n_{\ell}x_{\ell}}(v,w)$ can be separated
into $P$-polynomials and $Y$-polynomials
$$\begin{aligned}
&(FG)^{-1}D_{x_1}^{n_1}\cdots D_{x_{\ell}}^{n_{\ell}}F\cdot
G=\mathcal{Y}_{n_1x_1,\cdots,n_{\ell}x_{\ell}}(g, h)|_{g=\ln F/G, h=\ln FG}\\
&=\mathcal{Y}_{n_1x_1,\cdots,n_{\ell}x_{\ell}}(g, g+q,)|_{g=\ln F/G, q=2\ln G}\\
&=\sum_{n_1+\cdots+n_\ell=even}\sum_{r_1=0}^{n_1}\cdots\sum_{r_\ell=0}^{n_\ell}\prod_{i=1}^{\ell}
\left(\begin{matrix}n_i\cr
r_i\end{matrix}\right)P_{r_1x_1,\cdots,r_{\ell}x_{\ell}}(q)Y_{(n_1-r_1)x_1,\cdots,(n_\ell-r_\ell)x_\ell}(v).
\end{aligned}\eqno(6.6)$$
The key property of the multi-dimensional  Bell polynomials
$$Y_{n_1x_1,\cdots,n_{\ell}x_{\ell}}(g)|_{g=\ln\psi}={\psi_{n_1x_1,\cdots,n_{\ell}x_{\ell}}}/{\psi},\eqno(6.7)$$
implies that the binary Bell polynomials
$\mathcal{Y}_{n_1x_1,\cdots,n_{\ell}x_{\ell}}(g,h)$ can still be
linearized by means of the Hopf-Cole transformation $g=\ln \psi$,
that is, $\psi=F/G$.  The formulas (6.6) and (6.7) will then provide
the shortest way to the associated Lax system of nonlinear
equations.
\\[8pt]
%%%%%%%%%%%%%%%%%%%%%%%%%%%%%%%%%%%%%%%%%%%%%%%%%%%%%%%%%%%%%%%%%%%%%%%%%%%%%%%%%%%%%%%%%%%%%%%%%%%%%%%%%%%%%%%%%%%%%%%%%%%%%%%%%%%%%%
{\bf 6.2.   Bilinearization   }

{\bf Theorem 6.}   Under the transformation
$$v=v_0+2(\ln G)_{2x}, \ \ w=w_0+2(\ln G)_{xt},$$
 the  NKdV equation (1.1) can be bilinearized into
 $$\begin{aligned}
 &(D_{x}^4+12v_0 D_x^2-D_xD_y)G\cdot
G=0,\\
 &(2D_tD_x^3+6w_0 D_x^2+
 D_tD_y)G\cdot
G=0.
 \end{aligned}\eqno(6.8)$$
where $y$ ia an auxiliary variable, and  $u_0$, $v_0$ are two
constant solutions of the NKdV equation (1.1).

{\it Proof.}  The invariance of the NKdV equation (1.1) under the
scale transformation
$$x\rightarrow \lambda x, \  \ t\rightarrow \lambda^{\alpha} t, \ \ v\rightarrow \lambda^{-2}v,
\ \ w\rightarrow \lambda^{-\alpha-1}w$$
 shows that the dimensions of
the  fields  $v$  and $w$ are $-2$ and $-(\alpha+1)$,  respectively.
So we may  introduce a dimensionless  potential field $q$ by setting
$$v=v_0+q_{2x},\ \ \ w=w_0-q_{xt}.\eqno(6.9)$$
Substituting the transformation (6.9) into the equation (1.1), we
can write the resulting equation in the following form
$$\begin{aligned}
 &q_{4x,t}+4q_{2x}q_{2x,t}+2q_{3x}q_{xt}+4v_0 q_{2x,t}+2w_0 q_{3x}=0,
 \end{aligned}$$
which is regrouped as follows
$$\begin{aligned}
 &\frac{2}{3}q_{4x,t}+2(q_{2x}q_{2x,t}+q_{xt}q_{3x})+\frac{1}{3}q_{4x,t}+2q_{2x}q_{2x,t}+4v_0 q_{2x,t}+2w_0
 q_{3x}=0,
 \end{aligned}\eqno(6.10)$$
where we will see that Such an expression is necessary  to get a
bilinear form of the equation (1.1).  Further integrating the
equation (6.10)  with respect to $x$ yields
 $$\begin{aligned}
 &E(q) \equiv \frac{2}{3}(q_{3x,t}+3q_{2x}q_{xt}+3w_0 q_{2x})+
 \frac{1}{3}\partial_x^{-1}\partial_t(q_{4x}+3q_{2x}^2+12v_0 q_{2x})=0.
 \end{aligned}\eqno(6.11)$$

In order to  write the equation (6.11) in a local bilinear form, let us
first get rid of the integral operator $\partial_x^{-1}$. To
do so, we introduce an auxiliary variable $y$ and impose a
subsidiary constraint condition
$$q_{4x}+3q_{2x}^2+12v_0 q_{2x}-q_{xy}=0.\eqno(6.12)$$
Then, the equation (6.10) becomes
 $$\begin{aligned}
 &2(q_{3x,t}+3q_{2x}q_{xt}+3w_0 q_{2x})+
 q_{yt}=0.
 \end{aligned}\eqno(6.13)$$

According to the formula (6.5),  the  equations (6.12) and
(6.13) are then cast into
  a pair of equations in the form of   $P$-polynomials
 $$\begin{aligned}
 &P_{4x}(q)+12v_0 P_{2x}(q)-P_{xy}(q)=0,\\
 &2P_{3x,t}(q)+6w_0 P_{2x}(q)+P_{yt}(q)+3\gamma=0.
 \end{aligned}$$

Finally, by the property (6.4),  making  the  following variable
$$q=2\ln G\ \ \Longleftrightarrow \ \ v=v_0+2(\ln G)_{2x}, \ \ w=w_0+2(\ln G)_{xt},$$
change   above system  to the following bilinear forms of the NKdV
equation (1.1) as follows
 $$\begin{aligned}
 &(D_{x}^4+12v_0 D_x^2-D_xD_y)G\cdot
G=0,\\
 &(2D_tD_x^3+6w_0 D_x^2+
 D_tD_y)G\cdot
G=0,
 \end{aligned}\eqno(6.14)$$
 which is also simultaneously bilinear system  in $y$.  This system is easily solved with multi-soliton solutions by
using the Hirota's bilinear method. $\square$

Finally,  we show that the NKdV-1 equation (1.1) can be directly
bilinearized through a transformation, not Bell polynomials.  Making
dependent variable transformation
$$v=v_0+2(\ln F)_{xx}, \ \ u=G/F,\eqno(6.15)$$
we can change the equation (1.2) into
 $$\begin{aligned}
 &2(F_{xt}-F_xF_t)=G^2,\\
 &F_{xx}G-2F_xG_x+G_{xx}F+v_0FG=0,
 \end{aligned}$$
 which is equivalent to the bilinear form
  $$\begin{aligned}
 &D_xD_t F\cdot
F=G^2,\ \ (D_x^2+v_0)F\cdot G=0.
 \end{aligned}\eqno(6.16)$$
 It is obvious that the  bilinear form of the NKdV-1 (6.16)  is more simple than the  bilinear form of NKdV
 (6.15).
\\[8pt]
 %%%%%%%%%%%%%%%%%%%%%%%%%%%%%%%%%%%%%%%%%%%%%%%%%%%%%%%%%%%%%%%%%%%%%%%%%%%%%%%%%%%%%%%%%%
{\bf 6.3. N-soliton solutions}

As usual as %In the same procedure as
the normal perturbation method, let us expand
$G$ in the power series of a small parameter $\varepsilon$ as follows
$$G=1+\varepsilon g^{(1)}+ \varepsilon^2g^{(2)}+\varepsilon^3g^{(3)}+\cdots$$
Substituting the above equation into (6.7) and sorting %arranging
each order of $\varepsilon$, we have
 $$\begin{aligned}
 &\varepsilon:\ \   (D_{x}^4+12v_0 D_x^2-D_xD_y)g_1\cdot
1=0,\\
&\ \ \ \ (2D_tD_x^3+6w_0 D_x^2+
 D_tD_y)g^{(1)}\cdot
1=0,
 \end{aligned}\eqno(6.16)$$
$$\begin{aligned}
&\varepsilon^2:\ \ (D_{x}^4+12v_0 D_x^2-D_xD_y)(2g^{(2)}\cdot
1+g^{(1)}\cdot g^{(1)})=0,\\
&\ \ \ \ (2D_tD_x^3+6w_0 D_x^2+
 D_tD_y)(2g^{(2)}\cdot
1+g^{(1)}\cdot g^{(1)})=0,
 \end{aligned}\eqno(6.17)$$
$$\begin{aligned}
&\varepsilon^3:\ \ (D_{x}^4+12v_0 D_x^2-D_xD_y)(g^{(3)}\cdot
1+g^{(1)}\cdot
g^{(2)})=0, \\
&\ \ \ \ (2D_tD_x^3+6w_0 D_x^2+
 D_tD_y)(g^{(3)}\cdot 1+g^{(1)}\cdot
g^{(2)})=0, \\
&\cdots\cdots
 \end{aligned}\eqno(6.18)$$

By employing formulae mentioned above,  the system (6.16) is
equivalent to the following linear system
 $$\begin{aligned}
& g^{(1)}_{xxxx}+12v_0  g^{(1)}_{xx}- g^{(1)}_{xy}=0,\\
&2 g^{(1)}_{xxxt}+6w_0  g^{(1)}_{xx}+
  g^{(1)}_{yt}=0,
 \end{aligned}$$
which has solution
$$g^{(1)}=e^{\xi}, \ \ \ \xi=k x-\frac{2kw_0}{k^2+4v_0}t+(k^3+12v_0k)y+\sigma,\eqno(6.19)$$
where $k$ and $\sigma$ are two arbitrary parameters.

Substituting (6.12) into (6.10) and (6.11)  and choosing
$g^{(2)}=g^{(3)}=\cdots=0$, then the $G$'s expansion  is truncated
with a finite sum as
$$G=1+e^{\xi},$$
which gives regular one-soliton solution of  the NKdV equation (1.1)
 $$\begin{aligned}
&v=v_0+2\partial_x^2\ln(1+e^{\xi})=v_0+\frac{k^2}{2}{\rm sech}^2\xi/2,\\
&w=w_0+2\partial_t\partial_x\ln(1+e^{\xi})=
w_0+\frac{k^2w_0}{k^2+4v_0}{\rm sech}^2\xi/2, \\
&\xi=k x-\frac{2kw_0}{k^2+4v_0}t+\gamma,
\end{aligned}\eqno(6.20)$$
where $\gamma=(k^3+12v_0k)y+\sigma$, and  $k$, $v_0$, $w_0$  are
constants.

Let $w_0=1, \ v_0=0$,  then the solution (6.20) reads as a kink-type
 solution of the NKdV-I equation (1.2)
$$u=\pm\tanh\xi/2, \ \ \ \xi=k x-\frac{2}{k}t+\gamma.$$

In a similar way,  taking
$$g^{(1)}=e^{\xi_1}+e^{\xi_2}, \ \ \xi_j=k_j x-\frac{2k_jw_0}{k_j^2+4v_0}t+\gamma_j,\ \ j=1, 2,$$
we  get a two-soliton wave solution
$$\begin{aligned}
&v=v_0+ 2\partial_x^2\ln(1+e^{\xi_1}+e^{\xi_2}+e^{\xi_1+\xi_2+A_{12}})\\
&w=w_0-2\partial_t\partial_x\ln(1+e^{\xi_1}+e^{\xi_2}+e^{\xi_1+\xi_2+A_{12}}),\\
&A_{12}=\ln\left(\frac{k_1-k_2}{k_1+k_2}\right)^2.
\end{aligned}\eqno(6.21)$$

In general, we can get  a N-soliton solution of the NKdV equation
(1.1)
$$\begin{aligned}
&v=v_0+2\partial_x^2\ln\left(\sum_{\mu_j=0,1} \exp(\sum_{j=1}^{N} \mu_j\xi_j+\sum_{1\leq j\leq N}^{N}\mu_j\mu_lA_{jl}\right),\\
&w=w_0-\partial_t\partial_x\ln\left(\sum_{\mu_j=0,1} \exp(\sum_{j=1}^{N} \mu_j\xi_j+\sum_{1\leq j\leq N}^{N}\mu_j\mu_lA_{jl}\right),\\
&A_{jl}=\ln\left(\frac{k_j-k_l}{k_j+k_l}\right)^2.
\end{aligned}$$
 where  the notation  $\sum_{\mu_j=0,1}$
  represents  all possible  combinations  $\mu_j=0, 1$, and $\xi_j=k_j x-\frac{2k_jw_0}{k_j^2+4v_0}t+
  \gamma_j,\ \ j=1, 2,\cdots, N.$

  In the following, we discuss the soliton solutions for NKdV-1
  equation by using bilinear equation (6.16). Let us expand
$F$ and $G$ in the power series of a small parameter $\varepsilon$
as follows
$$F=1+ f^{(2)}\varepsilon^2+ f^{(4)}\varepsilon^4+f^{(6)}\varepsilon^6+\cdots$$
$$G= g^{(1)}\varepsilon+ g^{(3)}\varepsilon^3+g^{(5)}\varepsilon^5+\cdots$$
Substituting the above equation into (6.16) and arranging each order
of $\varepsilon$, we have
 $$\begin{aligned}
 &g^{(1)}_{xx}+v_0g^{(1)}=0,\\
&g^{(3)}_{xx}+v_0g^{(3)}=-(D_x^2+v_0)f^{(2)}\cdot g^{(1)},\\
&g^{(5)}_{xx}+v_0g^{(5)}=-(D_x^2+v_0)(f^{(2)}\cdot g^{(3)}+f^{(4)}\cdot g^{(1)}),\\
&\cdots\cdots
 \end{aligned}\eqno(6.23)$$
$$\begin{aligned}
&2f^{(2)}_{xt}=(g^{(1)})^2,\\
&2f^{(4)}_{xt}= 2g^{(1)}g^{(3)}-D_xD_tf^{(2)}\cdot f^{(2)},\\
&2f^{(6)}_{xt}= 2g^{(1)}g^{(5)}+2(g^{(3)})^2-2D_xD_tf^{(3)}\cdot f^{(3)},\\
&\cdots\cdots
 \end{aligned}\eqno(6.24)$$
Let $v_0=-k^2$,  it follows  from the first equation of (6.23) and
(6.24) that
$$g^{(1)}=e^{\xi}, \ \ f^{(2)}=\frac{1}{4}e^{2\xi}, \ \ \xi=kx+\frac{1}{2k}t+\gamma.\eqno(6.25)$$
 Substituting (6.25) into the second equation of
(6.23) leads to
$$g^{(3)}_{xx}-k^2g^{(3)}=0,$$
from which we may take $g^{(3)}=0$, further choose
$g^{(5)}=\cdots=0, \ f^{(4)}=\cdots=0.$
 So  $F$ and $G$ are  truncated
with a finite sum as
$$F=1+\frac{1}{4}e^{2\xi}, \ \ G=e^{\xi}.$$
Finally, the formula  (6.14)  gives one-soliton solution of the
NKdV-1 equation (1.2)
$$v=2k^2{\rm sech}^2\xi-k^2, \ \ \ u={\rm sech}\xi.$$
 \\[12pt]
%%%%%%%%%%%%%%%%%%%%%%%%%%%%%%%%%%%%%%%%%%%%%%%%%%%%%%%%%%%%%%%%%%%%%%%%%%%%%%%%%%%%%%%%%%%%%%%%%%%%%
{\bf 7.  Bilinear B\"{a}cklund transformation }\\

In this section,  we search for the bilinear B\"{a}cklund
transformation and Lax pair of the NKdV equation (1.1).
\\[8pt]
%%%%%%%%%%%%%%%%%%%%%%%%%%%%%%%%%%%%%%%%%%%%%%%%%%%%%%%%%%%%%%%%%%%%%%%
{\bf 7.1. Bilinear B\"{a}cklund transformation}

{\bf Theorem 7.} Suppose that  $F$  is  a  solution of the bilinear
equation (6.8), and if $ G$ satisfying
$$\begin{aligned}
&(D_x^2-\lambda)F\cdot G=0,\\
&[D_tD_x^2+2w_0 D_x+ (4v_0+3\lambda) D_t]F\cdot G=0,
\end{aligned}\eqno(7.1)$$
then G is another solution of the equation (6.8).

{\it Proof. }
  Let
$$q=2\ln G, \ \ \ \tilde{q}=2\ln F$$
 be  two different solutions of the
equation (6.10).    Introducing two new variables
$$h=(\tilde{q}+q)/2=\ln(FG), \ \ g=(\tilde{q}-q)/2=\ln(F/G),$$
makes the function $E$ invariant under the two fields $\tilde{q}$ and $q$: % condition
 $$\begin{aligned}
&E(\tilde{q})-E(q)=E(h+g)-E(h-g)\\
&=8v_0 g_{xt}+4w_0 g_{2x}+2g_{3x,t}+4h_{2x}g_{x,t}+4h_{x,t}g_{2x}+4\partial_x^{-1}(h_{2x}g_{2x,t}+h_{2x,t}g_{2x})\\
&=2\partial_x(\mathcal{Y}_{2x,t}(g,h)+4v_0\mathcal{Y}_t(g)-2w_0\mathcal{Y}_x(g))+R(g,h)=0,
\end{aligned}\eqno(7.2)$$
where
$$R(g,h)=-2\partial_x[(h_{2x}+g_x^2)g_t]+4h_{2x}g_{xt}-4h_{2x,t}g_{x}+4\partial_x^{-1}
(h_{2x}g_{2x,t}+h_{2x,t}g_{2x}).$$
This two-field invariant condition can be
regarded as a natural ansatz for a bilinear B\"{a}cklund
transformation and may produce some required transformations under
additional appropriate constraints.

In order to decouple  the two-field condition (7.2),% into a pair of constraints,
let us impose a constraint so as to express
$R(g,h)$ in the form of  $x$-derivative of
$\mathcal{Y}$-polynomials. The simple %possible
choice of the constraint may be
$$\mathcal{Y}_{2x}(g,h)=h_{2x}+g_x^2=\lambda,\eqno(7.3)$$
which directly leads to %thecomputing the $R(g,h)$, we find that
$$\begin{aligned}
R(g,h)=2\lambda &g_{xt}+4h_{2x}g_{xt}-4h_{2x,t}g_{x}-4g_x^2g_{xt}
=6\lambda g_{xt},
\end{aligned}\eqno(7.4)$$
where $h_{2x,t}=-2g_xg_{xt}$ and
$h_{2x}=\lambda-g_x^2$ are used.

Using the relations (7.2)-(7.4),  we derived  a coupled
 system of $\mathcal{Y}$-polynomials
$$\begin{aligned}
&\mathcal{Y}_{2x}(g,h)-\lambda=0,\\
&\mathcal{Y}_{2x,t}(g,h)+(4v_0+3\lambda)\mathcal{Y}_t(g)+2w_0\mathcal{Y}_x(g)=0,
\end{aligned}\eqno(7.5)$$
where we prefer the second equation to be expressed in the form of conserved quantity without
integration with respect to $x$. This is very useful to construct
conservation laws.  Apparently,  the identity (6.2) directly sends the
system (7.5) %immediately  leads
to   the following bilinear B\"{a}cklund
transformation
$$\begin{aligned}
&(D_x^2-\lambda)F\cdot G=0,\\
&[D_tD_x^2+2w_0 D_x+ (4v_0+3\lambda) D_t]F\cdot G=0,
\end{aligned}\eqno(7.6)$$
where we have integrated the  second equation in the system (7.5)
with respect to $x$, and  $w_0$ is the corresponding integration
constant. $\square$\\[8pt]
%%%%%%%%%%%%%%%%%%%%%%%%%%%%%%%%%%%%%%%%%%%%%%%%%%%%%%%%%%%%%%%%%%%%%%%
{\bf 7.2. Inverse scattering formulation}

 {\bf Theorem 8.}  The NKdV equation (1.1) admits a Lax pair
$$\begin{aligned}
&\psi_{2x}+v\psi=\lambda\psi,  \\
&4\psi_{2x,t}+4v\psi_t-2w\psi_x-3w_x\psi=0.
\end{aligned}\eqno(7.7)$$

{\it Proof.}  By the transformation  $v=\ln \psi$, it follows  from  the
formulas (6.5) and (6.6) that
$$\begin{aligned}
&\mathcal{Y}_t(g)=\psi_t/\psi, \ \ \mathcal{Y}_x(g)=\psi_x/\psi, \ \ \mathcal{Y}_{2x}(g,h)=q_{2x}+\psi_{2x}/\psi,\\
&\mathcal{Y}_{2x,t}(g,h)=2q_{xt}\psi_{x}/\psi+q_{2x}\psi_t/\psi+\psi_{2x,t}/\psi,\
\
\end{aligned}$$
which make the system (7.5) linearized into a Lax pair with
parameter  $\lambda$
$$\begin{aligned}
&L\psi\equiv(\partial_x^2+q_{2x})\psi=\lambda\psi,
\end{aligned}\eqno(7.8)$$
$$\begin{aligned}
&M\psi\equiv
[\partial_t\partial_x^2+(4v_0+q_{2x})\partial_t+2(q_{xt}+w_0)\partial_x+3\lambda\partial_t]\psi,
\end{aligned}\eqno(7.9)$$
or equivalently,
$$\begin{aligned}
&\psi_{2x}+v\psi=\lambda\psi,  \\
&4\psi_{2x,t}+4v\psi_t-2w\psi_x-3w_x\psi=0,
\end{aligned}$$
where the equation (7.8) is used to get the second equation. One can
easily verify from equations (7.8) and
(7.9) that %the integrability condition
$$[L, M]=q_{4x,t}+4(v_0+q_{2x})q_{2x,t}+2q_{3x}(q_{xt}+w_0)=0$$ exactly gives
the  NKdV equation (1.1) through replacing $v_0+q_{2x}$ and
$w_0+q_{xt}$ by $v$ and $w$, respectively. $\square$
\\[16pt]
%%%%%%%%%%%%%%%%%%%%%%%%%%%%%%%%%%%%%%%%%%%%%%%%%%%%%%%%%%%%%%%%%%%%%%%%%%%%%%%%%%%%%%%%%%%%%%%%%%%%%
{\bf 8. Darboux covariant Lax pair }\\

In this section,  we will give a kind of Darboux covariant Lax pair,
whose form is invariant under the  gauge transformation (4.3).

{\bf Theorem 9.}    The NKdV equation (1.1) possesses the following
Darboux covariant  Lax pair
$$\begin{aligned}
&L\psi=\lambda\psi,\\
 &{M}_{{\rm cov}}\psi=0, \ \
{M}_{{\rm cov}}=M+3\partial_xL,
\end{aligned}$$
under the gauge transformation $\tilde{\psi}=T\psi$. This is
actually equivalent to the Lax pair  (2.9).

{\it Proof. }
 In section 4, we have shown   that  the gauge transformation  (4.1)
 maps the operator $L(q)$  onto a similar operator
$$\tilde{{L}}(\tilde{q})=TL(q)T^{-1},$$
which satisfies the following covariance condition
$$\tilde{{L}}(\tilde{q})=L(q+\Delta q),\  \tilde{q}=q+\Delta q, \ {\rm
with} \ \ \Delta q=2\ln\phi.$$

 Next,  we want to find a third order operator ${M}_{{\rm cov}}(q)$ with
appropriate coefficients, such that $M_{{\rm cov}}(q)$ is mapped by
gauge transformation (8.1) onto a similar operator $\tilde{M}_{{\rm
{\rm cov}}}(\tilde{q})$, which satisfies the  covariance condition
$$\tilde{M}_{{\rm cov}}(\tilde{q})={M}_{{\rm cov}}(q+\Delta q), \ \tilde{q}=q+\Delta q.$$

Suppose that $\phi$ is a solution of the following Lax pair
$$\begin{aligned}
&L\psi=\lambda\psi,\\
 &{M}_{{\rm cov}}\psi=0, \ \
{M}_{{\rm
cov}}=4\partial_t\partial_x^2+b_1\partial_x+b_2\partial_t+b_3,
\end{aligned}\eqno(8.2)$$
where $b_1, \ b_2$ and $b_3$ are functions to be determined. Then,
%It suffice that we require that
the transformation $T$ is required to map the operator
$M_{{\rm cov}}$ to the similar one
$$TM_{{\rm cov}}T^{-1}=\tilde{M}_{{\rm cov}}, \ \
\tilde{L}_{2,{\rm
cov}}=4\partial_t\partial_x^2+\tilde{b}_1\partial_x+\tilde{b}_2\partial_t+\tilde{b}_3,\eqno(8.3)$$
where $\tilde{b}_1, \tilde{b}_2$ and $\tilde{b}_3$ satisfy the {\rm
cov}ariant condition
$$\tilde{b}_j={b}_j(q)+\Delta b_j={b}_j(q+\Delta q), \ \ j=1, 2, 3.\eqno(8.4)$$

It follows from (8.2) and (5.3) that
$$\begin{aligned}
&\Delta b_1=\tilde{b}_1-b_1=4\sigma_t, \ \ \Delta
b_2=\tilde{b}_2-b_2=8\sigma_x,
\end{aligned}\eqno(8.5)$$
$$\begin{aligned}
 &\Delta b_3=\tilde{b}_3-b_3=\sigma\Delta b_1+8\sigma_{xt}+b_{1,x},
\end{aligned}\eqno(8.6)$$
and $\sigma$ satisfy
$$4\sigma_{2x,t}+\tilde{b}_1\sigma_x+\tilde{b}_2\sigma_t+\sigma\Delta
b_3+b_{3,x}=0.\eqno(8.7)$$
According to the relation  (8.4), it
remains to determine $b_1,\  b_2$ and $b_3$ in the form of
polynomial expressions in terms of $q$'s derivatives %of $q$
$$b_j=F_j(q,q_{x}, q_{y}, q_{xy}, q_{2x}, q_{2y}, q_{2x,y}, \cdots), \ \ j= 1, 2, 3$$
such that
$$\Delta F_j=F_j(q+\Delta q, q_{x}+\Delta q_{x}, q_{t}+\Delta q_{t}, \cdots)
-F_j(q, q_{x},q_{t},  \cdots) =\Delta b_j,\eqno(8.8)$$ with $\Delta
q_{kx,lt}=2(\ln \phi)_{kx,lt}, \ k, l=1, 2, \cdots$, and  $\Delta
b_j$ being given through the relations (8.5)-(8.7).

Expanding the left hand of the equation (8.8), we obtain
$$\Delta b_1=\Delta F_1=F_{1,q}\Delta q+F_{1,q_{x}}\Delta q_{x}+
F_{1,q_{y}}\Delta q_{y}+F_{1,q_{xy}}\Delta q_{xt}+ \cdots
=4\sigma_t=2\Delta q_{xt},$$
 which implies that we can determine
$b_1$ up to a  arbitrary constant $c_1$, namely,
$$b_1=F_1(q_{xt})=2q_{xt}+c_1, \ c_1\ {\rm is \ an\ arbitrary\ constant} \eqno(8.9)$$

Proceeding in the same way  deduce the  function $b_2$  as follows
$$b_2=F_2(q_{2x})=4 q_{2x}+c_2,\eqno(8.10)$$
where $c_2$ is an arbitrary constant.

We see  from the relation (8.6) that  $\Delta b_3$ contains the term
$b_{1,x}=q_{2x,t}$,  which should be eliminated such that  $\Delta
b_3$ admits the form (8.8). By the Lax pair %means of the eigenvalue equation in
(8.2), we have the following relation
$$q_{2x,t}=-\sigma_{xt}-2\sigma\sigma_t.\eqno(8.11) $$
Substituting (8.9) and (8.11) into (8.6) yields
$$\Delta b_3=4\sigma\sigma_t+8\sigma_{xt}+2q_{2x,t}=6\sigma_{xt}=3\Delta q_{2x,t}.$$
If choosing
$$b_3=F_3(q_{2x,t})=3q_{2x,t}+c_3,\eqno(8.12)$$
the third condition
$$\Delta F_3=F_{3,q}\Delta q+F_{3,q_{x}}\Delta q_{x}+F_{3,q_{t}}\Delta q_{t} \cdots
=\Delta b_3$$
holds, %can be satisfied,
where $c_3$ is an arbitrary constant.

Letting $c_1=-2v_0, \ c_2=0, \ c_3=w_0$ in (8.9), (8.10) and (8.12),
then it follows from (8.2) that we have the following Darboux
covariant evolution equation
$$M_{{\rm cov}}\psi=0, \ \
M_{{\rm
cov}}=4\partial_t\partial_x^2+2q_{xt}\partial_x+4q_{2x}\partial_t+3q_{2x,t},$$
 which coincides with  the equation (8.7).
 Moreover, the relation between
 two operators $L_{2,{\rm cov}}$ and  $L_2$ are related through %is given by
$$M_{{\rm cov}}=M+3\partial_xL.$$
The compatibility condition of the Darboux covariant Lax pair (8.2)
exactly gives the  NKdV equation(1.1) in Lax representation
 $$\begin{aligned}
&[M_{{\rm cov}}, L]=
q_{4x,t}+4(v_0+q_{2x})q_{2x,t}+q_{3x}(q_{xt}+w_0)\\
&=v_{xxx}+4vw_x+2v_xw=0.
\end{aligned}$$
$\square$

 In the above repeated procedure, %similar way   step by step,
 we are able to obtain higher order
operators, which are also Darboux {\rm cov}ariant with respect to $T$,
to produce higher order members of the negative order  KdV
hierarchy.
\\[12pt]
%%%%%%%%%%%%%%%%%%%%%%%%%%%%%%%%%%%%%%%%%%%%%%%%%%%%%%%%%%%%%%%%%%%%%%%%%%
{\bf\large  9.  Conservation laws of NKdV equations}\\

 In this section, we will present infinitely many %derive the
conservation laws in a local form for the NKdV equation (1.1) based
on a generalized Miura transformation.

 {\bf Theorem
10.}   The NKdV equation (1.1) possesses the following infinitely many % consequence of
conservation laws
$$F_{n,t}+G_{n,x}=0, \ n=1, 2, \cdots.\eqno(9.1)$$
where  the conversed densities $F_n's$ are recursively given by
 recursion formulas explicitly
$$\begin{aligned}
&F_0=v_{xx}-v^2,\ \ F_1=-v_{xxx}+2vv_x,\\
&
F_{n}=I_{n,xx}-\sum_{k=0}^{n}I_kI_{n-k}+\sum_{k=0}^{n-2}I_kI_{n-2-k,x},
\ \ n=2, 3, \cdots.
\end{aligned}\eqno(9.2)$$
and the  fluxes $G_n's$ are
$$\begin{aligned}
&G_0=2wI_0=2wv,\ \ G_1=2wI_1=-2wv_x,\\
 &G_{n}=2wI_n, \ \ n=2, 3, \cdots.
\end{aligned}\eqno(9.3)$$

{\it Proof.}   For the simplicity, let us select $v_0=w_0=0$ in the
transformation (6.9). We introduce a new potential function
$$q_{2x}=\eta+\varepsilon\eta_x+\varepsilon^2\eta^2,\eqno(9.4)$$
where $\varepsilon$ is a constant parameter.
 Substituting (9.4) into the Lax equation (7.10) leads to
$$\begin{aligned}
&0=[L,
M]=(1+\varepsilon\partial_x+2\varepsilon^2\eta)[-4(\eta+\varepsilon^2\eta^2)\eta_t
-2(q_{x}-\varepsilon\eta)_t\eta_x+\eta_{2x,t}],
\end{aligned}$$
which implies that $v=q_{2x},\ w=q_{xt}$ given  by (9.4) are a
solution of the NKdV equation (1.1) if $\eta$ satisfies the
following equation
$$-4(\eta+\varepsilon^2\eta^2)\eta_t
-2(q_x-\varepsilon\eta)_t\eta_x+\eta_{2x,t}-4\eta_t=0.\eqno(9.5)$$

On the other hand, it follows from (9.5) that
$$[(q_x-\varepsilon\eta)_t]_x=-(\eta+\varepsilon^2\eta^2)_t.$$
Therefore, the equation (9.5) can be rewritten as
$$(\eta_{2x}-\eta^2)_t+[2\eta(\varepsilon^2\eta-q_x)_t]_x=0,$$
or a divergent-type form
$$(\eta_{2x}+2\varepsilon^2\eta\eta_x-\eta^2)_t+(2\eta w)_x=0\eqno(9.6)$$
by replacing $q_{xt}=w$.

%To proceed,
Inserting the expansion
$$\eta=\sum_{n=0}^{\infty} I_n(q,q_x, q_t\cdots)\varepsilon^{n},\eqno(9.7)$$
 into the equation (9.4)
and  comparing %equating
the coefficients for power of $\varepsilon$, we obtain
the recursion relations to calculate $I_n$ in an explicit form
$$\begin{aligned}
&I_0=q_{2x}=v,\ \  I_1=-I_{0,x}=-v_{x},\\
 &I_{n}=-I_{n-1,x}-\sum_{k=0}^{n-2}I_k I_{n-2-k}, \ \ n=2,
3, \cdots.
\end{aligned}\eqno(9.8)$$

Substituting (9.7) into (9.6) and simplifying terms in the power of
$\varepsilon$  provide us infinitely many conservation laws
$$F_{n,t}+G_{n,x}=0, \ n=1, 2, \cdots$$
where  the conversed densities $F_n's$  and the  fluxes $G_n's$ are
by (9.2) and (9.3), respectively. $\square$

Here, we already give recursion formulas (9.7) and (9.8) to show how to generate
conservation laws (9.6) based on the first few explicitly provided.  Apparently, the first
equation in conservation laws  (9.6)
$$v_{xxt}-2vv_t+2wv_x+2w_xv=0$$
is exactly the NKdV equation (1.1)
$$\begin{aligned}
&v_t+w_x=0,\\
 &w_{xxx}+4vw_x+2wv_x=0.
\end{aligned}$$
which is reduced to the NKdV equation (1.2) under the constraints
$v=-u_{xx}/u$ and $w=u^2.$

 In conclusion, the NKdV equation (1.1) is completely
integrable and admits bilinear B\"{a}cklund
transformation, Lax pair and infinitely many local
conservation laws.\\[12pt]
%%%%%%%%%%%%%%%%%%%%%%%%%%%%%%%%%%%%%%%%%%%%%%%%%%%%%%%%%%%%%%%%%%%%%%%%%%%%%%%%%%%%%%%%%%%%%%%%%%%%%%%%%%%
{\bf\large 10. Quasi-periodic solutions of the NKdV equation}\\

In this section, we study  quasi-periodic wave solutions of the NKdV
equation (1.1)  by using  bilinear B\"{a}cklund transformation (7.1)
and  bilinear formulas derived in section 9.

 In fact, a quasi-periodic solution, also called  algebro-geometric solutions or finite gap solutions,  was
  originally studied in the late 1970s by Novikov, Dubrovin, McKean, Lax, Its, and Matveev et
al \cite{ Du, Its, Mc, Nov}, based on the inverse spectral theory
and algebro-geometric method. In recent years, this theory has been
extended to a large class of nonlinear integrable equations
including sine-Gordon equation, Camassa-Holm equation, Thirring
model equation, Kadomtsev-Petviashvili equation, Ablowitz-Ladik
lattice, and Toda lattice
\cite{Cao,Geng1,Geng2,Geng3,Fritz1,Fritz2,Fan5,Qiao-PhD,Qiao1,Qiao-1998,Zam,Zhou-PhD,Zhou}.
  The algebro-geometric theory, however, needs Lax pairs and is also involved in
  complicated analysis procedure on the Riemann surfaces.
 It is  rather difficult to directly determine
the characteristic parameters of waves, such as frequencies and
phase shifts for a function with given wave-numbers and amplitudes.
On the other hand, the bilinear derivative method developed by
Hirota is a powerful approach for constructing exact solution of
nonlinear equations in an explicit form. If a nonlinear equation is
able to be written in a bilinear form by a dependent variable
transformation, then multi-solitary wave solutions are usually
obtained for the equation \cite{Hirota1,Hirota2,Hu,Hu2,Zhang}. Based
on the Hirota forms, Nakamura proposed a convenient way to find a
kind of explicit quasi-periodic  solutions of  nonlinear equations
\cite{Na2}, where the  periodic  wave solutions of the KdV equation
and the Boussinesq equation were obtained. Such a method indeed
displays some advantages over algebro-geometric methods. For
example, it does not need any Lax pair and Riemann surface for the
given nonlinear equation, and is also able to find the explicit
construction of multi-periodic wave solutions. The method relies on
the existence of the Hirota's bilinear form as well as arbitrary
parameters appearing in Riemann matrix \cite{Fan1, Fan2}.
 \\[8pt]
%%%%%%%%%%%%%%%%%%%%%%%%%%%%%%%%%%%%%%%%%%%%%%%%%%%%%%%%%%%%%%
{\bf 10.1. Multi-dimensional Riemann theta functions}

Let us first begin with some preliminary work about
multi-dimensional Riemann theta functions and their
quasi-periodicity.
 The
multi-dimensional Riemann theta function is defined by
$$
\vartheta(\boldsymbol{\zeta},
\boldsymbol{\varepsilon},\boldsymbol{s}|\boldsymbol{\tau})=\sum_{\boldsymbol{n}\in
\mathbb{{Z}}^N}\exp\{2\pi
i\langle\boldsymbol{\zeta}+\boldsymbol{\varepsilon},\boldsymbol{n}+\boldsymbol{s}\rangle-\pi
\langle\boldsymbol{\tau}
(\boldsymbol{n}+\boldsymbol{s}),\boldsymbol{n}+\boldsymbol{s}\rangle\},\eqno(10.1)$$
where  $\boldsymbol{n}=(n_1,\cdots,n_N)^T\in \mathbb{Z}^N$ is an
integer value vector, and $\boldsymbol{s}=(s_1,\cdots,s_N)^T,
\boldsymbol{\varepsilon}=(\varepsilon_1,\cdots,\varepsilon_N)^T\in
\mathbb{{C}}^N$ is a complex parameter vector. $\boldsymbol{\zeta}
=(\zeta_1, \cdots, \zeta_N)^T, \
\zeta_j=\alpha_jx+\beta_jt+\delta_j$, $\ \alpha_j, \beta_j,
\delta_j\in\Lambda_0$, $  j=1, 2, \cdots, N$ are complex phase
variables, where $x, t$ are ordinary real variables and $\theta$ is
a Grassmann variable. The inner product of two vectors
$\boldsymbol{f}=(f_1, \cdots, f_N)^T$ and $\boldsymbol{g}=(g_1,
\cdots, g_N)^T$
 %their  inner product
 is defined by
 $$\langle \boldsymbol{f}, \boldsymbol{g}\rangle=f_1g_1+f_2g_2+\cdots+f_Ng_N.$$
 The matrix $\boldsymbol{\tau}=(\tau_{ij})$ is  a positive
definite and real-valued symmetric $N\times N$ matrix. The entries
$\tau_{ij}$ of the periodic matrix $\boldsymbol{\tau}$ can be
considered as free parameters of the theta function (10.1).

   In this paper, we choose $\tau$ to be purely imaginary matrix to make
the theta function (10.1) real-valued.   In definition (10.1) for  the case of
$\boldsymbol{s}=\boldsymbol{\varepsilon}=\boldsymbol{0}$, %hereafter
we denote
$\vartheta(\boldsymbol{\zeta},{\boldsymbol{\tau}})=\vartheta(\boldsymbol{\zeta},\boldsymbol{0},
\boldsymbol{0}|\boldsymbol{\tau})$ for simplicity.
Therefore, we have  $\vartheta(\boldsymbol{\zeta},\boldsymbol{\varepsilon},
\boldsymbol{0}|\boldsymbol{\tau})=\vartheta(\boldsymbol{\zeta}+\boldsymbol{\varepsilon},
\boldsymbol{\tau})$.

{\bf Remark 4.}  The above periodic matrix $\boldsymbol{\tau}$  is
different from the one in the algebro-geometric approach discussed in
\cite{Nov}-\cite{Fan2}, where it is usually constructed on %through
a
compact Riemann surface $\Gamma$ with genus $N\in\mathbb{N}$. One may
see that the entries in the matrix $\boldsymbol{\tau}$ are not
free and difficult to be explicitly given. $\square$

{\bf Definition 3.}  A function  $g(\boldsymbol{x},t)$ on
$\mathbb{C}^N\times\mathbb{C}$  is said to be quasi-periodic in $t$
with fundamental periods  $T_1, \cdots, T_k\in \mathbb{C}$ if
 $T_1, \cdots, T_k$ are linearly dependent over $\mathbb{Z}$ and there exists a function
  $G(\boldsymbol{x},\boldsymbol{y})\in \mathbb{C}^N\times\mathbb{C}^k$ such that
$$ G(\boldsymbol{x}, y_1,\cdots, y_j+T_j, \cdots, y_k)=G(\boldsymbol{x},y_1,\cdots, y_j, \cdots, y_k ),
 \ \ {\rm for\ all}\ y_j\in \mathbb{C}, \ j=1, \cdots, k.$$
$$ G(\boldsymbol{x}, t,\cdots, t, \cdots, t )=g(\boldsymbol{x}, t). $$
In particular, $g(\boldsymbol{x},t)$ becomes periodic with $T$ if
and only if $T_j=m_jT$. $\square$

Let's first see periodicity of the theta function $\vartheta(\boldsymbol{\zeta}, \boldsymbol{\tau})$.

 {\bf Proposition 10.} \cite{Far} Let $\boldsymbol{e_j}$ be the $j-$th
 column of  $N\times N$  identity matrix $I_N$;
 ${\tau_j}$ be the $j-$th column of $\boldsymbol{\tau}$, and $\tau_{jj}$ the
 $(j,j)$-entry of $\boldsymbol{\tau}$. Then the theta function $\vartheta(\boldsymbol{\zeta}, \boldsymbol{\tau})$ has the periodic properties
$$\begin{aligned}
&\vartheta(\boldsymbol{\zeta}+\boldsymbol{e_j}+i\boldsymbol{\tau_j},\boldsymbol{\tau})=\exp(-2\pi
i\zeta_j+\pi
\tau_{jj})\vartheta(\boldsymbol{\zeta},\boldsymbol{\tau}).
\end{aligned}$$

  The theta function $\vartheta(\boldsymbol{\zeta},\boldsymbol{\tau})$ which satisfies the condition (5.4)  is called
 a multiplicative function. We regard the vectors
$\{\boldsymbol{e_j},\ \ j=1, \cdots, N\}$ and
$\{i\boldsymbol{\tau_j}, \ \ j=1, \cdots, N\}$ as periods of the
theta function $\vartheta(\boldsymbol{\zeta},\boldsymbol{\tau})$
with multipliers $1$ and $\exp({-2\pi i\zeta_j+\pi \tau_{jj}})$,
respectively. Here, only the first $N$ vectors are actually periods
of the theta function $\vartheta(\boldsymbol{\zeta},
\boldsymbol{\tau})$, but  the last $N$ vectors are  the periods of
the functions  $\partial^2_{\zeta_k,\zeta_l}\ln
\vartheta(\boldsymbol{\zeta},\boldsymbol{\tau})$ and $
\partial_{\zeta_k}\ln[\vartheta(\boldsymbol{\zeta}+\boldsymbol{e},
\boldsymbol{\tau})/\vartheta(\boldsymbol{\zeta}+\boldsymbol{h},\boldsymbol{\tau})],
\ k, l=1, \cdots, N$.

{\bf Proposition 11.} Let $\boldsymbol{e_j}$ and
$\boldsymbol{\tau_j}$ be defined as above proposition 2. The
meromorphic functions $f(\boldsymbol{\zeta})$ are as follow
$$\begin{aligned}
&(i) \ \ \ \
 \  f(\boldsymbol{\zeta})=\partial_{\zeta_k\zeta_l}^2\ln\vartheta(\boldsymbol{\zeta},
 \boldsymbol{\tau}),\ \ \boldsymbol{\zeta}\in C^N, \ \ \ k,  l=1, \cdots,
 N,
\end{aligned}$$
$$\begin{aligned}
&(ii) \ \ \ \
f(\boldsymbol{\zeta})=\partial_{\zeta_k}\ln\frac{\vartheta(\boldsymbol{\zeta}+
\boldsymbol{e},\boldsymbol{\tau})}{
\vartheta(\boldsymbol{\zeta}+\boldsymbol{h},\boldsymbol{\tau})},\ \
\boldsymbol{\zeta},\ \boldsymbol{e},\ \boldsymbol{h}\in C^N, \ \
j=1, \cdots, N.
\end{aligned}$$
   then in all two cases (i) and  (ii), it holds that
$$\begin{aligned}
&f(\boldsymbol{\zeta}+\boldsymbol{e_j}+i\boldsymbol{\tau_j})=f(\boldsymbol{\zeta}),
\ \ \ \boldsymbol{\zeta}\in C^N,\ \ \ j=1, \cdots, N,
\end{aligned}$$
which implies that $f(\boldsymbol{\zeta})$ is a  quasi-periodic
function.
\\[8pt]
%%%%%%%%%%%%%%%%%%%%%%%%%%%%%%%%%%%%%%%%%%%%%%%%%%%%%%%%%%%%%%%%%%%%%%%%%%%%
{\bf 10.2.  Bilinear formulae of theta functions}

To construct a kind of explicitly quasi-periodic solutions of the
NKdV equation (1.1),  we propose  some important bilinear formulas
of multi-dimensional Riemann theta functions, whose  derivations are
similar to the case of super bilinear equations \cite{Fan3}, so we
just  list them without proofs.

{\bf Theorem 11.}   Suppose that
$\vartheta(\boldsymbol{\zeta},\boldsymbol{\varepsilon'},
\boldsymbol{0}|\boldsymbol{\tau})$ and $
\vartheta(\boldsymbol{\zeta},\boldsymbol{\varepsilon},
\boldsymbol{0}|\boldsymbol{\tau})$ are two Riemann theta functions,
in which $\boldsymbol{\varepsilon}=(\varepsilon_1, \dots,
\varepsilon_N)$, $\boldsymbol{\varepsilon'}=(\varepsilon_1', \dots,
\varepsilon_N')$, and  $\boldsymbol{\zeta}=(\zeta_1, \cdots,
\zeta_N)$, $\zeta_j=\alpha_jx+\omega_jt+\delta_j, \ \ j=1, 2,
\cdots, N$. Then operators $D_x, D_t$ and
 $S$ exhibit the following perfect
properties when they act on a pair of theta functions
$$\begin{aligned}
&D_x \vartheta(\boldsymbol{\zeta},\boldsymbol{\varepsilon'},
\boldsymbol{0}|\boldsymbol{\tau})\cdot
  \vartheta(\boldsymbol{\zeta},\boldsymbol{\varepsilon}, \boldsymbol{0}|\boldsymbol{\tau})\\
  &=\sum_{\boldsymbol{\mu}}\partial_x\vartheta(2\boldsymbol{\zeta},\boldsymbol{\varepsilon'}-\boldsymbol{\varepsilon},
  -\boldsymbol{\mu}/2|2\boldsymbol{\tau})|_{\boldsymbol{\zeta}=\boldsymbol{0}}
  \vartheta(2\boldsymbol{\zeta},\boldsymbol{\varepsilon'}+\boldsymbol{\varepsilon},\boldsymbol{\mu}/2|2\boldsymbol{\tau}),
\end{aligned}\eqno(10.2)$$
 where $\boldsymbol{\mu}=(\mu_1, \cdots, \mu_N)$,  and  the notation  $\sum_{\boldsymbol{\mu}}$
  represents $2^N$ different transformations corresponding to all possible
  combinations  $\mu_1=0,1;  \cdots;  \mu_N=0,1$.

  In general, for a polynomial operator $H( D_x,
D_t)$ with respect to  $D_x$ and $ D_t$, we have the following
useful formula
$$\begin{aligned}
&H( D_x, D_t)
\vartheta(\boldsymbol{\zeta},\boldsymbol{\varepsilon'},
\boldsymbol{0}|\boldsymbol{\tau})\cdot
  \vartheta(\boldsymbol{\zeta},\boldsymbol{\varepsilon}, \boldsymbol{0}|\boldsymbol{\tau})
  =\sum_{\boldsymbol{\mu}}C(\boldsymbol{\varepsilon'},\boldsymbol{\varepsilon}, \boldsymbol{\mu})
  \vartheta(2\boldsymbol{\zeta},\boldsymbol{\varepsilon'}
  +\boldsymbol{\varepsilon}, \boldsymbol{\mu}/2|2\boldsymbol{\tau}),\end{aligned}\eqno(10.3)$$
in which, explicitly
$$\begin{aligned}
  &C(\boldsymbol{\varepsilon},\boldsymbol{\varepsilon'},\boldsymbol{\mu})=\sum_{\boldsymbol{n}\in \mathbb{Z}^N}
  H(\boldsymbol{\mathcal{M}})\exp\left[-2\pi\langle \boldsymbol{\tau}(\boldsymbol{n}-\boldsymbol{\mu}/2), \boldsymbol{n}
  -\boldsymbol{\mu}/2\rangle-2\pi i
  \langle \boldsymbol{n}-\boldsymbol{\mu}/2, \boldsymbol{\varepsilon'}-\boldsymbol{\varepsilon})\right].
\end{aligned}\eqno(10.4)$$
where we denote $ \boldsymbol{\mathcal{M}}=(4\pi i\langle
\boldsymbol{n}-\boldsymbol{\mu}/2, \boldsymbol{\alpha}\rangle,\ 4\pi
i\langle
  \boldsymbol{n}-\boldsymbol{\mu}/2, \boldsymbol{\omega}\rangle).$

{\bf Remark 6.}  The formulae (10.3) and (10.4) show  that if the
following equations are   satisfied
  $$C(\boldsymbol{\varepsilon},\boldsymbol{\varepsilon'},\boldsymbol{\mu})=0,\eqno(10.5)$$
  for all possible combinations
  $\mu_1=0,1; \mu_2=0,1;  \cdots;  \mu_N=0,1$, in other word, all
  such  combinations  are solutions of equation (10.5),   then
  $\vartheta(\boldsymbol{\zeta},\boldsymbol{\varepsilon'}, \boldsymbol{0}|
  \boldsymbol{\tau})$ and $
  \vartheta(\boldsymbol{\zeta},\boldsymbol{\varepsilon}, \boldsymbol{0}|\boldsymbol{\tau})$
  are  $N$-periodic wave solutions of the bilinear equation
  $$H( D_x, D_t)\vartheta(\boldsymbol{\zeta},\boldsymbol{\varepsilon'}, \boldsymbol{0}|
  \boldsymbol{\tau})\cdot\vartheta(\boldsymbol{\zeta},\boldsymbol{\varepsilon}, \boldsymbol{0}|\boldsymbol{\tau})=0.$$
 We call the formula (10.5) constraint equations, whose number  is  $2^N$.
 This formula  actually provides us an unified approach to construct
  multi-periodic wave  solutions for supersymmetric equations. Once a
  supersymmetric equation is written bilinear forms, then its  multi-periodic wave  solutions
  can be directly obtained by solving system (10.5).

{\bf Theorem 12.} Let
$C(\boldsymbol{\varepsilon},\boldsymbol{\varepsilon'},\boldsymbol{\mu})$
and $H( D_x, D_t)$  be  given in Theorem 10,  and  make a choice
such that $\varepsilon_j'-\varepsilon_j=\pm 1/2, \ j=1, \cdots, N$.
Then

(i) \ If  $H( D_x, D_t)$ is an symmetric operator, i. e.
$$H( -D_x,
-D_t)=H( D_x, D_t),$$
 then $C(\boldsymbol{\varepsilon},\boldsymbol{\varepsilon'},\boldsymbol{\mu})$ vanishes automatically for
 the case when  $\sum_{j=1}^{N}\mu_j$ is
an odd number, namely
  $$C(\boldsymbol{\varepsilon},\boldsymbol{\varepsilon'},\boldsymbol{\mu})|_{\boldsymbol{\mu}}=0, \ \  {\rm for} \  \ \
 \sum_{j=1}^{N}\mu_j=1,\ {\rm mod }\ 2.$$

(ii) \ If $H( D_x, D_t)$ is a skew-symmetric operator,  i.e.
$$H( -D_x,
-D_t)=-H( D_x, D_t),$$ then
$C(\boldsymbol{\varepsilon},\boldsymbol{\varepsilon'},\boldsymbol{\mu})$
vanishes automatically for
 the case when  $\sum_{j=1}^{N}\mu_j$ is
an even number, namely
$$C(\boldsymbol{\varepsilon},\boldsymbol{\varepsilon'},\boldsymbol{\mu})|_{\boldsymbol{\mu}}=0,
\ \ {\rm for} \ \sum_{j=1}^{N}\mu_j=0, \ {\rm mod }\ 2.$$

{\bf Proposition 12. }  Let $\varepsilon_j'-\varepsilon_j=\pm 1/2, \
j=1, \cdots, N$.  Assume $H( D_x, D_t)$ is a linear combination of
even and  odd functions
$$H( D_x, D_t)=H_1(D_x, D_t)+H_2( D_x, D_t),$$
where $H_1$ is  even  and $H_2$ is odd.  In addition,
$C(\boldsymbol{\varepsilon},\boldsymbol{\varepsilon'},\boldsymbol{\mu})$
corresponding (10.8)  is given by
$$C(\boldsymbol{\varepsilon},\boldsymbol{\varepsilon'},\boldsymbol{\mu})
=C_1(\boldsymbol{\varepsilon},\boldsymbol{\varepsilon'},\boldsymbol{\mu})
+C_2(\boldsymbol{\varepsilon},\boldsymbol{\varepsilon'},\boldsymbol{\mu}),$$
where
  $$C_1(\boldsymbol{\varepsilon},\boldsymbol{\varepsilon'},\boldsymbol{\mu})
=\sum_{\boldsymbol{n}\in \mathbb{Z}^N}
  H_1(\boldsymbol{\mathcal{M}})
  \exp\left[-2\pi\langle \boldsymbol{\tau}(\boldsymbol{n}-\boldsymbol{\mu}/2), \boldsymbol{n}
  -\boldsymbol{\mu}/2\rangle-2\pi i
  \langle \boldsymbol{n}-\boldsymbol{\mu}/2,
  \boldsymbol{\varepsilon'}-\boldsymbol{\varepsilon})\right],$$
  $$C_2(\boldsymbol{\varepsilon},\boldsymbol{\varepsilon'},\boldsymbol{\mu})
=\sum_{\boldsymbol{n}\in \mathbb{Z}^N}
 H_2(\boldsymbol{\mathcal{M}})
  \exp\left[-2\pi\langle \boldsymbol{\tau}(\boldsymbol{n}-\boldsymbol{\mu}/2), \boldsymbol{n}
  -\boldsymbol{\mu}/2\rangle-2\pi i
  \langle \boldsymbol{n}-\boldsymbol{\mu}/2,
  \boldsymbol{\varepsilon'}-\boldsymbol{\varepsilon})\right].$$
Then
$$\begin{aligned}
  &C(\boldsymbol{\varepsilon},\boldsymbol{\varepsilon'},\boldsymbol{\mu})
  =C_2(\boldsymbol{\varepsilon},\boldsymbol{\varepsilon'},\boldsymbol{\mu}) \ \  {\rm for} \  \ \
 \sum_{j=1}^{N}\mu_j=1,\ {\rm mod }\ 2,
\end{aligned}$$
$$\begin{aligned}
  &C(\boldsymbol{\varepsilon},\boldsymbol{\varepsilon'},\boldsymbol{\mu})
  =C_1(\boldsymbol{\varepsilon},\boldsymbol{\varepsilon'},\boldsymbol{\mu}), \ \ {\rm for} \ \sum_{j=1}^{N}\mu_j=0,
   \ {\rm mod }\ 2.
\end{aligned}$$

The theorem 2 and corollary 1 are very useful to deal with coupled
super-Hirota's bilinear equations, which will  be  seen  in the
following section 10.

By introducing differential operators
\begin{eqnarray*}
&&\nabla=(\partial_{\zeta_1}, \partial_{\zeta_2}, \cdots, \partial_{\zeta_N}), \\
&&\partial_x=\alpha_1\partial_{\zeta_1}+\alpha_2
\partial_{\zeta_2}+\cdots+\alpha_N
\partial_{\zeta_N}=\boldsymbol{\alpha}\cdot\nabla,\\
&&\partial_t=\beta_1\partial_{\zeta_1}+\beta_2
\partial_{\zeta_2}+\cdots+\beta_N
\partial_{\zeta_N}=\boldsymbol{\beta}\cdot\nabla,
\end{eqnarray*}
then we have
 $$\begin{aligned}
 &\partial_x^k\partial_t^l\vartheta(\boldsymbol{\zeta},\boldsymbol{\tau})
 =(\boldsymbol{\alpha}\cdot\nabla)^k
 (\boldsymbol{\beta}\cdot\nabla)^l\vartheta(\boldsymbol{\zeta},\boldsymbol{\tau}),\ \
 k, l=0, 1, \cdots.
\end{aligned}$$
\\[8pt]
%%%%%%%%%%%%%%%%%%%%%%%%%%%%%%%%%%%%%%%%%%%%%%%%%%%%%%%%%%%%%%%%%%%%%%%%%%%%%%%%%%%%%%%%%%%%%%%%%%%%%%%%%%%%%%%%%%%%%%%%%%%%%%%%%%%%%%
%%%%%%%%%%%%%%%%%%%%%%%%%%%%%%%%%%%%%%%%%%%%%%%%%%%%%%%%%%%%%%%%%%%%%%%%%%%%%%%%%%%%%%%%%%%%%%%%%%%%%%%%%%%%%%%%%%%%%%%%%%%%%%%%%%%%%%
{\bf 10.3. One-periodic waves and  asymptotic analysis}\\

Let us first construct one-periodic wave solutions of the NKdV
equation  (1.1) by using bilinear B\'{a}cklund transformation (7.6).
As a simple case of the theta function
 (10.1)  with $N=1, s=0$,  we choose $F$ and $G$ as follows
$$\begin{aligned}
&F=\vartheta(\zeta,0,0|\tau)=\sum_{n\in\mathbb{Z}}\exp({2\pi
in\zeta-\pi n^2\tau}),\\
&G=\vartheta(\zeta,1/2,0|\tau)=\sum_{n\in\mathbb{Z}}\exp({2\pi
in(\zeta+1/2)-\pi n^2\tau})\\
&\ \ \ =\sum_{n\in\mathbb{Z}}(-1)^n\exp({2\pi in\zeta-\pi
n^2\tau}),\end{aligned}\eqno(10.6)$$
 where $\zeta=\alpha
x+\beta t+\delta$ is the phase variable, and $\tau>0$ is a positive parameter.

By Theorem 6, %in section 9,
the operator $H_1=D_x^2-\lambda$  in bilinear equation (7.6) is
symmetric, and its corresponding constraint equation in the formula
(10.5) automatically vanishes
 for $\mu=1$. Meanwhile,  $H_2=D_tD_x^2-2w_0
D_x+(4v_0+3\lambda)D_t$  are  skew-symmetric, and its corresponding
constraint equation automatically vanishes for $\mu=0$.  Therefore,
the Riemann theta function (10.6)  is a solution of the bilinear
equation (7.6), provided the following equations
 $$\begin{aligned}
&\sum_{n\in\mathbb{Z}}\{[4\pi i(n-\mu/2)]^2\alpha^2 -\lambda\}\exp(-2\pi \tau (n-\mu/2)^2+\pi i(n-\mu/2))|_{\mu=0}=0,\\
&\sum_{n\in\mathbb{Z}}\{[4\pi i(n-\mu/2)]^3\alpha^2\beta
+8\pi i(n-\mu/2)\alpha w_0+ 4\pi i(n-\mu/2)(4v_0+3\lambda)\beta\}\\
&\times\exp(-2\pi \tau (n-\mu/2)^2+\pi i(n-\mu/2))|_{\mu=1}=0.
\end{aligned}\eqno(10.7)$$
hold.

Let %We introduce the notations by
$$\begin{aligned}
&\rho=e^{-\pi\tau/2 },\\
&\vartheta_1(\zeta,\rho)=\vartheta(2\mathbf{\zeta},1/4,-1/2|
2\tau)=\sum_{n\in\mathbb{Z}}\rho^{(2n-1)^2}\exp[4i\pi(n-1/2)
(\zeta+1/4)]
,\\
&\vartheta_2(\zeta,\rho)=\vartheta(2\zeta, 1/4,0 |
2\tau)=\sum_{n\in\mathbb{Z}}\rho^{4n^2}\exp[4i\pi
n(\zeta+1/4)],\end{aligned}$$ then, the equation (10.7) can be
written as a linear system about $\beta$ and $\lambda$
 $$\begin{aligned}
&\vartheta_2''\alpha^2-\vartheta_2\lambda=0,\\
&\vartheta_1'''\alpha^2\beta+2\vartheta_1'\alpha
w_0+(4v_0+3\lambda)\vartheta_1' \beta=0,
\end{aligned}\eqno(10.8)$$
where the derivative value of $\vartheta_j(\zeta,\rho)$ at $\zeta=0$
is denoted by simple notations
$$\vartheta_j'=\vartheta_j'(0,\rho)=\frac{d\vartheta_j(\zeta,\rho)}{d\zeta}|_{\zeta=0}, \ \ j=1,2.$$

It is not hard to see that  the system (10.8)  admits the following
solution for the NKdV equation (1.1)
 $$\begin{aligned}
&\lambda=\frac{\vartheta_2''\alpha^2}{\vartheta_2},\ \ \
\beta=\frac{-2\vartheta_1'\vartheta_2w_0}
{\vartheta_1'''\vartheta_2\alpha^2+4\vartheta_1'\vartheta_2v_0+3\vartheta_1'\vartheta_2''\alpha^2}.
\end{aligned}\eqno(10.9)$$
So, we obtain %In this way,
the following one-periodic wave solution %reads
$$V=v_0+2\partial_{x}^2\ln \vartheta(\zeta,0,0|\tau), \ \ W=w_0+2\partial_x\partial_{t}\ln
\vartheta(\zeta,0,0|\tau),\eqno(10.10)$$
 where $\zeta=\alpha x+\beta t+\delta$ and  parameter $\beta$ is given by (10.9), while
other parameters $\alpha, \tau, v_0, w_0$ are arbitrary. Among the four parameters, the two
%three
ones $\alpha$ and $\tau$ completely dominate a one-periodic wave.

In summary, one-periodic wave (10.10)   is one-dimensional and has
two fundamental periods $1$ and $i\tau$  in phase variable $\zeta$
(see Figure 3).

\input epsf
\begin{figure}
\ \ \ \ \ \  \ \ \ \ \ \ \ \footnotesize $\textbf{(a)}$ \ \ \ \ \ \
\ \ \ \ \ \ \ \ \   \ \ \ \ \ \ \ \ \ \ \ \ \ \footnotesize
$\textbf{(b)}$ \ \ \ \ \ \ \  \ \ \ \ \ \ \ \  \ \ \ \
 \begin{center}
\includegraphics[scale=0.50]{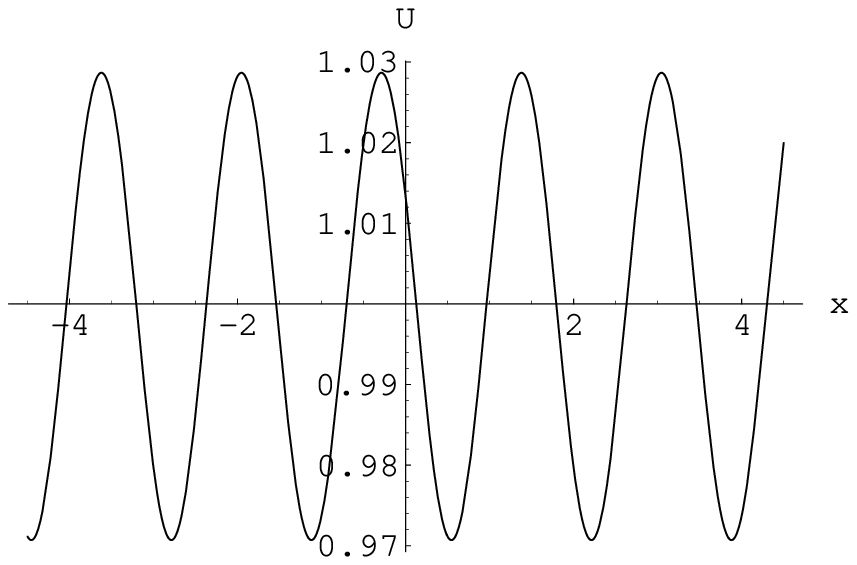}\  \ \ \ \
\includegraphics[scale=0.50]{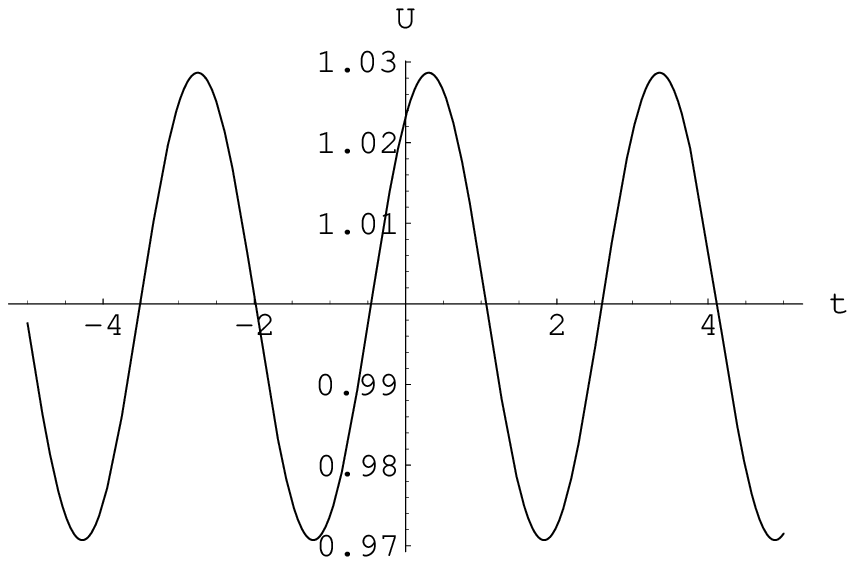}
\\[12pt]
\ \ \ \ \ \  \ \ \ \ \ \ \ \footnotesize $\textbf{(c)}$ \ \ \ \ \ \
\ \ \ \ \ \ \ \ \   \ \ \ \ \ \ \ \ \ \ \ \ \ \footnotesize
$\textbf{(d)}$ \ \ \ \ \ \ \  \ \ \ \ \ \ \ \  \ \ \ \
\\[8pt]
\includegraphics[scale=0.55]{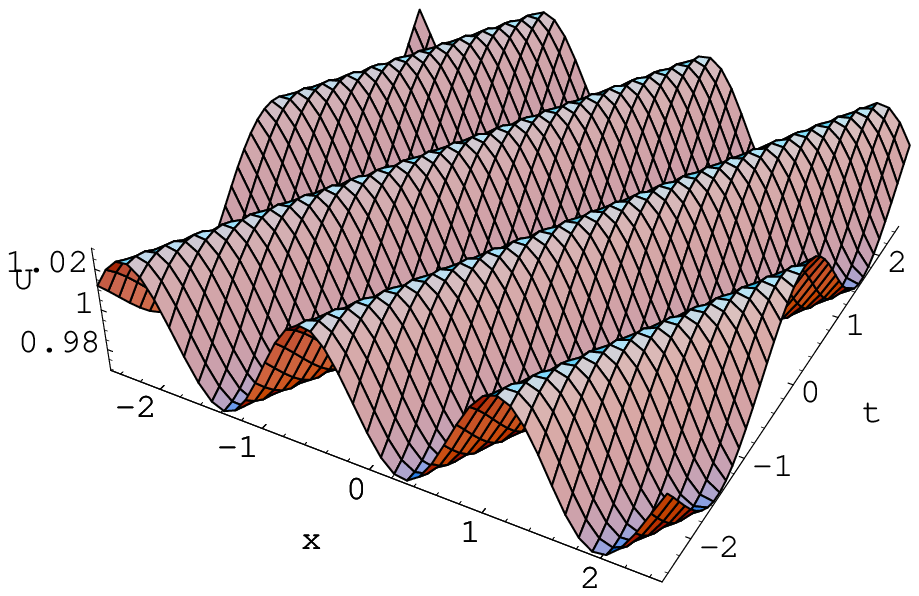} \ \ \ \  \ \ \ \
\includegraphics[scale=0.40]{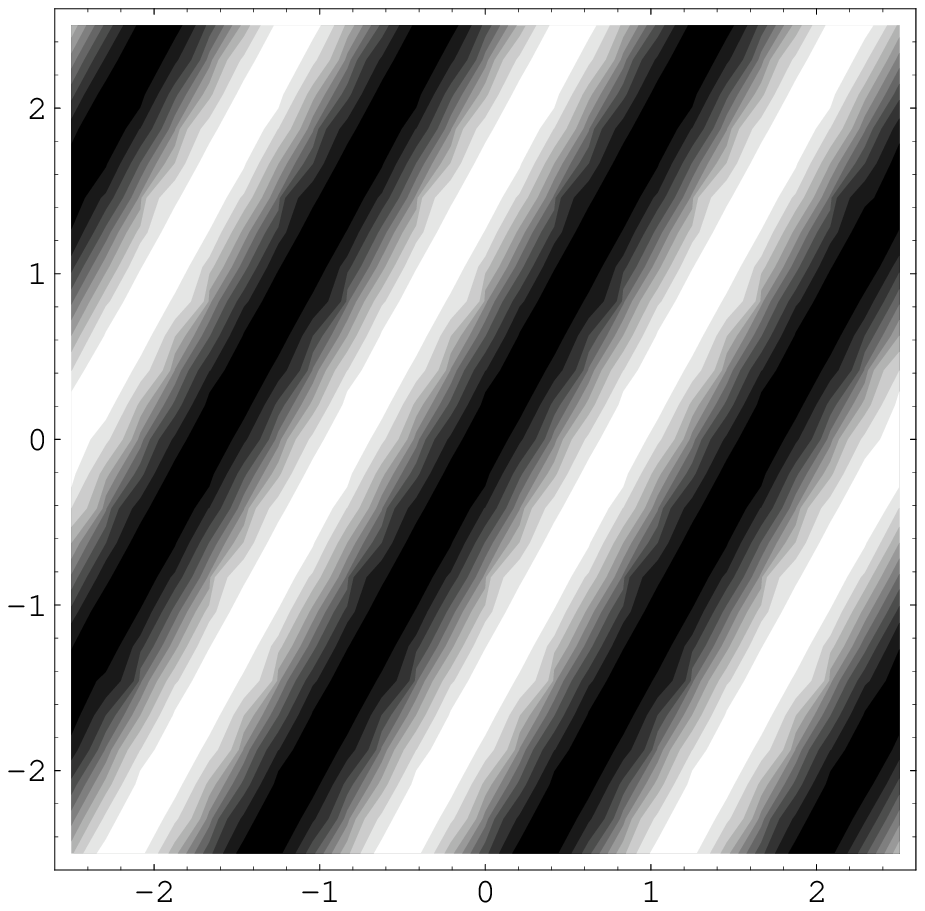}
\end{center}

\caption{\footnotesize\baselineskip=15pt  One-periodic wave for the
NKdV equation (1.1) with parameters: $\alpha=0.6, $ $\tau=2,
v_0=0.5, w_0=1$. (a) and (b) show that every one-periodic wave is
 periodic in both  $x$ and $y$ directions. (c) Perspective view of the
wave.  (d) Overhead view of the wave, with contour plot shown. The
bright hexagons are crests and the dark hexagons are troughs.}
\end{figure}

In the following theorem, we will see that  the one-periodic wave
solution (10.10) can be broken into soliton solution (6.20) under a
long time limit and their relation can be established as follows.

{\bf Theorem 13.}  In the one-periodic wave solution (10.6), the
parameter $\beta$ is given by (10.9), other parameters are chosen
as
$$\alpha=\frac{k}{2\pi i}, \ \ \
\delta=\frac{\gamma+\pi \tau}{2\pi i},\eqno(10.11)$$ where $k_1$ and
$\gamma$ are the same as those  in (6.20).  Then under a small
amplitude limit,  one-periodic wave solution (10.10) can be broken
into the single  soliton solutions (6.20), that is,
$$V\longrightarrow v, \ \ W\longrightarrow w, \ \ {\rm as } \ \
\rho\rightarrow 0.\eqno(10.12)$$

In particular,  in the case of $v_0=0, \ w_0=1$, the one-periodic
solution (10.5) tends to the kink-type soliton solution (5.2), that
is,
$$V\longrightarrow \tilde{v}^{I}, \ \ W\longrightarrow \tilde{w}^{I}, \ \ {\rm as } \ \
\rho\rightarrow 0.\eqno(10.13)$$

{\it Proof.}  Here we use the system (10.8) to analyze asymptotic
properties of the one-periodic solution (10.10).  Let us
%explicitly
expand the coefficients of the system (10.8) as follows
$$\begin{aligned}
&\vartheta_1'=-4\pi\rho+12\pi\rho^9+\cdots,\quad
\vartheta_1'''=16\pi^3\rho+432\pi^3\rho^9+\cdots,\\
&\vartheta_2=1+2\rho^4+\cdots,\quad
\vartheta_2''=32\pi^2\rho^4+\cdots,
\end{aligned}\eqno(10.14)$$
Suppose that the solution of the system (10.8) has the following
form
$$\begin{aligned}
&\lambda=\lambda_0+\lambda_1\rho+\lambda_2\rho^2+\cdots=\lambda_0+o(\rho),\\
&\beta=\beta_0+\beta_1\rho+\beta_2\rho^2+\cdots=\beta_0+o(\rho).
\end{aligned}\eqno(10.15)$$

Substituting the expansions (10.14) and (10.15) into the system
(10.8) and letting $\rho\rightarrow 0$, we immediately obtain
 the following relation
$$
 \begin{aligned}
 &\lambda_0=0,\ \ \beta_0=\frac{-\alpha w_0}{-2\pi^2\alpha^2+2v_0}.
 \end{aligned}\eqno(10.16)$$

Combining  (10.11) and (10.16) leads to
$$\begin{aligned}
&\lambda\longrightarrow 0, \\
& 2\pi i\beta\longrightarrow 2\pi i\beta_0=\frac{-2\pi i\alpha
w_0}{-2\pi^2\alpha^2+2v_0}=\frac{-2kw_0}{k^2+4v_0}, \ \ {\rm as } \
\ \rho\rightarrow 0, \end{aligned}$$
 or equivalently rewritten as
$$\begin{aligned}
&\hat{\zeta}=2\pi i\zeta-\pi \tau=k x+2\pi i\beta t+\gamma\\
&\quad \longrightarrow kx-\frac{2kw_0}{k^2+4v_0}t+\gamma=\xi,\ \
{\rm as}\ \ \rho\rightarrow 0.
\end{aligned}\eqno(10.17)$$

It remains to  verify  that the one-periodic wave  (10.11) has  the
same form as the one-soliton solution (6.20) under the limit
$\rho\rightarrow 0$. Let us expand the function $F$   in the
following form
$$ F=1+\rho^2(e^{2\pi i\zeta}+e^{-2\pi i\zeta})+\rho^8(e^{4\pi i\zeta}+e^{-4\pi i\zeta})
+\cdots .$$
 It follows from (10.11) and (10.17) %,  that
$$\begin{aligned}
&F=1+e^{\hat{\zeta}}+\rho^4(e^{-\hat{\zeta}}+e^{2\hat{\zeta}})+\rho^{12}(e^{-2\hat{\zeta}}+e^{3\hat{\zeta}})
+\cdots\\
&\quad \longrightarrow 1+e^{\hat{\zeta}}\longrightarrow 1+e^{\xi},\
\ {\rm as}\ \ \rho\rightarrow 0.
\end{aligned}\eqno(10.18)$$

So, combining (10.11) and (10.18) yields %leads to
$$\begin{aligned}
&v\longrightarrow v_0+2\partial_{xx} \ln
(1+e^{\xi}),\\
& w\longrightarrow w_0+2\partial_t\partial_{x} \ln (1+e^{\xi}),\ \
{\rm as}\ \ \rho\rightarrow 0.
\end{aligned}$$
Thus, we conclude that the one-periodic solution (10.10) may go to a
bell-type soliton solutions (6.20) as the amplitude $\rho\rightarrow
0$.
$\square$\\[8pt]
%%%%%%%%%%%%%%%%%%%%%%%%%%%%%%%%%%%%%%%%%%%%%%%%%%%%%%%%%%%%%%%%%%%%%%%%%%%%%%%%%%%%%%%%%%%%%%%%%%%%%%%%%%%%%%%%%%%%%%%%%%%%%%%%%%%%%%
%%%%%%%%%%%%%%%%%%%%%%%%%%%%%%%%%%%%%%%%%%%%%%%%%%%%%%%%%%%%%%%%%%%%%%%%%%%%%%%%%%%%%%%%%%%%%%%%%%%%%%%%%%%%%%%%%%%%%%%%%%%%%%%%%%%%%%
{\bf 10.4. Two-periodic waves and  asymptotic properties}\\

Let us now  consider  two-periodic wave solutions to the NKdV
equation (1.1).   For the case of $N=2, \
\boldsymbol{s}=\boldsymbol{0},\
\boldsymbol{\varepsilon}=\boldsymbol{1/2}=(1/2,1/2)$ in the Riemann
theta function (10.1), we choose  $F$ and $G$ as follows
$$\begin{aligned}
&F=\vartheta(\boldsymbol{\zeta},\boldsymbol{0},\boldsymbol{0}|
\boldsymbol{\tau})=\sum_{\boldsymbol{n}\in \mathbb{Z}^2} \exp\{2\pi
i\langle\boldsymbol{\zeta},\boldsymbol{n}\rangle-\pi\langle\boldsymbol{\tau}\boldsymbol{ n},\boldsymbol{n}\rangle\}£¬\\
 &G=\vartheta(\boldsymbol{\zeta},\boldsymbol{1/2},\boldsymbol{0}|
\boldsymbol{\tau})=\sum_{\boldsymbol{n}\in \mathbb{Z}^2} \exp\{2\pi
i\langle\boldsymbol{\zeta}+\boldsymbol{1/2},\boldsymbol{n}\rangle-\pi\langle\boldsymbol{\tau}\boldsymbol{
n},\boldsymbol{n}\rangle\}\\
&\ \ =\sum_{\boldsymbol{n}\in \mathbb{Z}^2}(-1)^{n_1+n_2} \exp\{2\pi
i\langle\boldsymbol{\zeta},\boldsymbol{n}\rangle-\pi\langle\boldsymbol{\tau}\boldsymbol{
n},\boldsymbol{n}\rangle\}
\end{aligned}\eqno(10.19)$$
where  $\boldsymbol{n}=(n_1, n_2)\in Z^2,\ \
\boldsymbol{\zeta}=(\zeta_1, \zeta_2)\in\mathcal{C}^2, \ \
\zeta_i=\alpha_jx+\beta_jt+\delta_j, \ \ j=1, 2$, and $
\boldsymbol{\alpha}=(\alpha_1, \alpha_2),\
\boldsymbol{\beta}=(\beta_1, \beta_2) \in \mathcal{C}^2$.
The matrix
$\boldsymbol{\tau}$ is a positive definite  and real-valued
symmetric ${2\times 2}$ matrix£¬ that is, %which can  take  the form
$$\boldsymbol{\tau}=(\tau_{ij})_{2\times 2}, \ \ \tau_{12}=\tau_{21},\ \ \tau_{11}>0,\ \
 \tau_{22}>0, \ \ \tau_{11}\tau_{22}-\tau_{12}^2>0.$$

 According to Theorem 5, constraint equations associated with  $H_1=D_x^2-\lambda$ and $H_2=D_tD_x^2-2w_0
D_x+(4v_0+3\lambda)D_t$
 automatically vanish for $(\mu_1, \mu_2)=(0,1),
(1, 0)$ and for $(\mu_1, \mu_2)=(0,0), (1, 1)$, respectively. Hence,
making the theta functions (10.19)  satisfy the bilinear equation
(7.6) gives the following  constraint equations
$$\begin{aligned}
&\sum_{n_1,n_2\in \mathbb{Z}}\left[ -16\pi^2\langle
\boldsymbol{n}-\boldsymbol{\mu}/2,
\boldsymbol{\alpha}\rangle^2-\lambda\right]\exp\{-2\pi \langle
\boldsymbol{\tau} (\boldsymbol{n}-\boldsymbol{\mu}/2),
\boldsymbol{n}-\boldsymbol{\mu}/2\rangle\\
&+\pi
i\sum_{j=1}^{2}(n_j-\mu_j/2)\}|_{\boldsymbol{\mu}=(\mu_1,\mu_2)}=0,\
\ {\rm for}\ \ (\mu_1,\mu_2)=(0,0),\ (1,1)=0,\\
 &\sum_{n_1,n_2\in
\mathbb{Z}}\left[ -64\pi^3i\langle
\boldsymbol{n}-\boldsymbol{\mu}/2,
\boldsymbol{\alpha}\rangle^2\langle
\boldsymbol{n}-\boldsymbol{\mu}/2,\boldsymbol{\beta}\rangle+8\pi
i\langle \boldsymbol{n}-\boldsymbol{\mu}/2,
\boldsymbol{\alpha}\rangle w_0+4\pi i\langle
\boldsymbol{n}-\boldsymbol{\mu}/2,\boldsymbol{\beta}\rangle
(4v_0+3\lambda)\right]\\
&\times\exp\{-2\pi \langle \boldsymbol{\tau}
(\boldsymbol{n}-\boldsymbol{\mu}/2),
\boldsymbol{n}-\boldsymbol{\mu}/2\rangle+\pi
i\sum_{j=1}^{2}(n_j-\mu_j/2)\}|_{\boldsymbol{\mu}=(\mu_1,\mu_2)}=0,\\
&{\rm for}\ \ (\mu_1,\mu_2)=(0,1),\ (1,0).
\end{aligned}\eqno(10.20)$$

 Let %Next, let us introduce the following notations
 \begin{eqnarray*}
 &&\rho_{kl}=e^{-\pi\tau_{kl}/2}, k, l=1,2, \boldsymbol{\rho}=(\rho_{11},\rho_{12},
\rho_{22})\\
&&\vartheta_j(\boldsymbol{\zeta},\boldsymbol{\rho})=\vartheta(2\zeta,\boldsymbol{1/4},-\boldsymbol{s}_j/2|2\tau)\\
 && \ \ \ \  \ \ \ \ \ \  \   =\sum_{n_1,n_2\in Z}\exp\{4\pi i\langle\boldsymbol{\zeta+1/4} ,\boldsymbol{n}-\boldsymbol{s_j}/2\rangle\}
 \prod_{k,l=1}^{2}\rho_{kl}^{(2n_k-s_{j,k})(2n_j-s_{j,l})} ,\\
&& \boldsymbol{s_j}=(s_{j,1}, s_{j,2}),\quad j=1, 2,\ \
\boldsymbol{s}_1=(0,1),\ \ \boldsymbol{s_2}=(1,0),\ \
\boldsymbol{s_3}=(0,0),\ \ \boldsymbol{s_4}=(1,1)
\end{eqnarray*}
then the  system (10.20) can be rewritten as a linear system
 $$\begin{aligned}
& (\boldsymbol{\alpha}\cdot\nabla)^2\vartheta_j-\lambda \vartheta_j
=0,\ j=3, 4,
 \end{aligned}\eqno(10.21)$$
 $$\begin{aligned}
&
(\boldsymbol{\beta}\cdot\nabla)(\boldsymbol{\alpha}\cdot\nabla)^2\vartheta_j+
2w_0(\boldsymbol{\alpha}\cdot\nabla)\vartheta_j+(4v_0+3\lambda)(\boldsymbol{\beta}\cdot\nabla)
\vartheta_j =0,\
 j=1, 2,
 \end{aligned}\eqno(10.22)$$
where $\vartheta_j$ represent the derivative values of functions
$\vartheta_j(\boldsymbol{\zeta},\boldsymbol{\rho})$ at $\zeta_1=\zeta_2=0$.

The system (10.22) admits a unique solution
 $$\begin{aligned}
\left(
\begin{matrix}\beta_1\cr
\beta_2\end{matrix}\right)=\left[\frac{\partial(f,
g)}{\partial(\zeta_1, \zeta_2)}\right]^{-1}\left(
\begin{matrix}2w_0(\boldsymbol{\alpha}\cdot\nabla)\vartheta_1\cr
2w_0(\boldsymbol{\alpha}\cdot\nabla)\vartheta_2\end{matrix}\right)
 \end{aligned}\eqno(10.23)$$
where $ \frac{\partial(f, g)}{\partial(\zeta_1, \zeta_2)}$ is the Wronskinan matrix given by
 $$\begin{aligned}
&\frac{\partial(f, g)}{\partial(\zeta_1, \zeta_2)}=\left(
\begin{matrix}\partial_{\zeta_1}f&\partial_{\zeta_2}f\cr\partial_{\zeta_1}g&\partial_{\zeta_2}g\end{matrix}\right),\\
&f=[(\boldsymbol{\alpha}\cdot\nabla)^2+4v_0+3\lambda]\vartheta_1,\ \
g= [(\boldsymbol{\alpha}\cdot\nabla)^2+4v_0+3\lambda]\vartheta_2.
 \end{aligned}$$
With the help of the above $(\beta_1, \beta_2)$, we are able to get
a two-periodic wave solution to the NKdV equation  (1.1)
$$V=v_0+\partial_x^2\ln\vartheta(\boldsymbol{\zeta},\boldsymbol{0},\boldsymbol{0}|\boldsymbol{\tau}),
\ \
W=w_0+\partial_x\partial_t\vartheta(\boldsymbol{\zeta},\boldsymbol{0},\boldsymbol{0}|\boldsymbol{\tau}),\eqno(10.24)$$
where $\alpha_1,$ $ \alpha_2, \tau_{12}, \delta_1$ and $\delta_2$
are arbitrary parameters,  while other parameters $\beta_1, \beta_2$
and  $\tau_{11}$, $\tau_{22}$ are given by (10.23) and (10.21),
respectively.

In summary, the two-periodic wave (10.24) is  a direct
generalization of two one-periodic waves.  Its surface pattern is
two-dimensional with two phase variables $\zeta_1$ and $\zeta_2$.
The two-periodic wave (10.24) has $4$ fundamental periods $\{e_1,
e_2\}$ and $\{i\tau_1, i\tau_2\}$ in $(\zeta_1, \zeta_2)$, and  is
spatially periodic in  two directions $\zeta_1, \zeta_2$. Its real
part is not periodic in $\theta_1$ direction, while its imaginary
part and modulus are all periodic in
 both  $x$ and   $t$   directions.

\input epsf
\begin{figure}
\ \ \ \ \ \  \ \ \ \ \ \ \ \footnotesize $\textbf{(a)}$ \ \ \ \ \ \
\ \ \ \ \ \ \ \ \   \ \ \ \ \ \ \ \ \ \ \ \ \ \footnotesize
$\textbf{(b)}$ \ \ \ \ \ \ \  \ \ \ \ \ \ \ \  \ \ \ \
 \begin{center}
\includegraphics[scale=0.50]{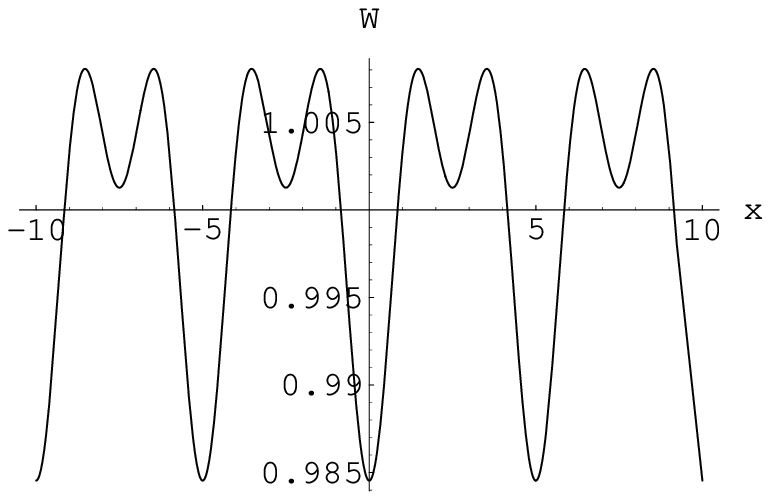}\  \ \ \ \
\includegraphics[scale=0.50]{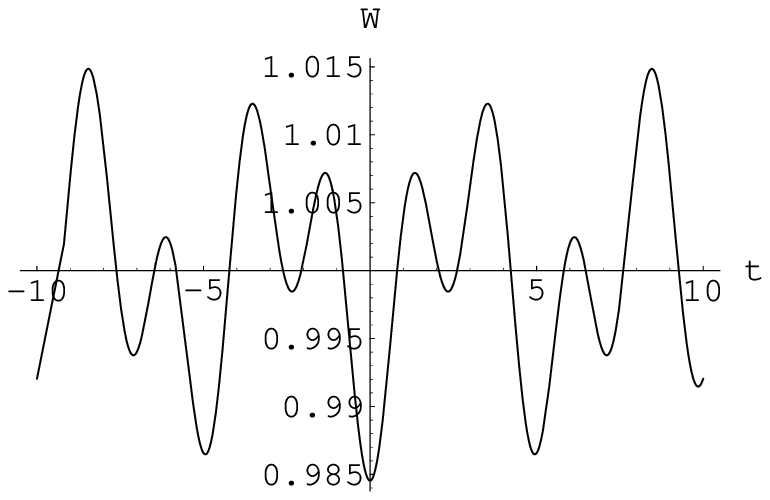}
\\[12pt]
\ \ \ \ \ \  \ \ \ \ \ \ \ \footnotesize $\textbf{(c)}$ \ \ \ \ \ \
\ \ \ \ \ \ \ \ \   \ \ \ \ \ \ \ \ \ \ \ \ \ \footnotesize
$\textbf{(d)}$ \ \ \ \ \ \ \  \ \ \ \ \ \ \ \  \ \ \ \
\\[8pt]
\includegraphics[scale=0.50]{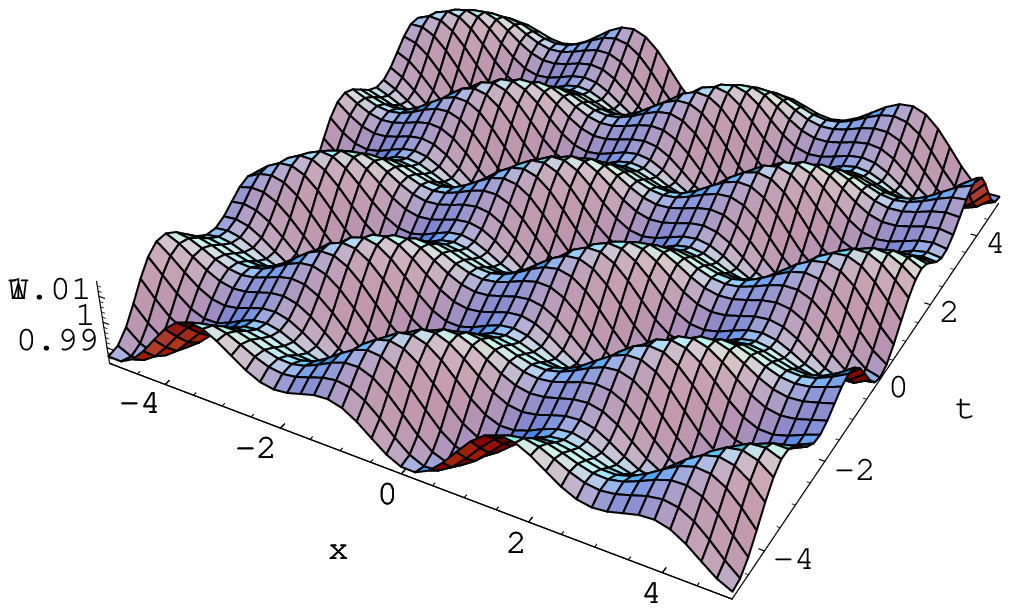} \ \ \ \  \ \ \ \
\includegraphics[scale=0.40]{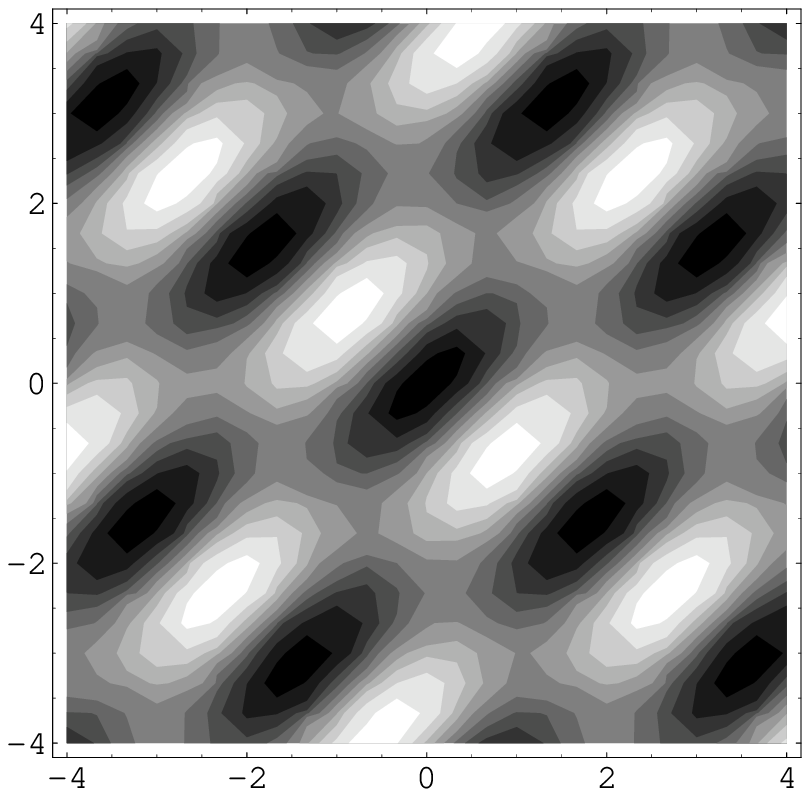}
\end{center}

\caption{\footnotesize\baselineskip=15pt  Two-periodic wave for the
NKdV equation (1.1).   (a) and (b) show that every one-periodic wave
is periodic in both $x$- and $y$-directions. (c) Perspective view of
the wave.  (d) Overhead view of the wave, with contour plot shown.
The bright hexagons are crests and the dark hexagons are troughs.}
\end{figure}

Finally,  we study the asymptotic properties of the two-periodic
solution (10.24).
 In a similar way to Theorem 5, we figure out the relation between
 the two-periodic solution (10.24)  and
the two-soliton solution (6.21) as follows.

{\bf Theorem 14.}   Assume that $(\beta_1, \beta_2)$ is a solution
of the system (10.22), and in the two-periodic wave solution
(10.24), parameters $\alpha_j, \delta_j, \tau_{12}$ are chosen as
$$
\begin{aligned}
&\alpha_j=\frac{k_j}{2\pi i}, \ \
\delta_j=\frac{\gamma_j+\pi\tau_{jj} }{2\pi i},\ \
\tau_{12}=-\frac{A_{12}}{2\pi},\ \ j=1,2,
\end{aligned}\eqno(10.25)$$
 where $k_j,\gamma_j,
j=1, 2$ and $A_{12}$  are those given in (6.21).  Then,   we have
the following asymptotic relations
$$
\begin{aligned}
&\lambda\longrightarrow 0, \ \ \
\zeta_j\longrightarrow\frac{\eta_j+\pi \tau_{jj}}{2\pi i}, \ \
j=1, 2,\\
& F\longrightarrow 1+e^{\eta_1}+e^{\eta_2}+e^{\eta_1+\eta_2+A_{12}},
\  \  {\rm as} \ \ \rho_{11}, \rho_{22}\rightarrow 0.
\end{aligned}\eqno(10.26)$$
So, the two-periodic wave  solution (10.24) just tends to the
two-soliton  solution (6.21) under a limit condition
$$V\longrightarrow v, \ \  W\longrightarrow w, \ \ {\rm as }\ \ \rho_{11}, \rho_{22}\rightarrow 0.$$

{\it Proof.}     Using (10.20), we may expand  the  function $F$
in the following explicit form
\begin{eqnarray*}
&& F=1+(e^{2\pi i\zeta_1}+e^{-2\pi i\zeta_1})e^{-\pi \tau_{11}}
+(e^{2\pi i\zeta_2}+e^{-2\pi i\zeta_2})e^{-\pi
\tau_{22}}\\
&&\ \ \ \ \ \ +(e^{2\pi i(\zeta_1+\zeta_2)}+e^{-2\pi
i(\zeta_1+\zeta_2)})e^{-\pi (\tau_{11}+2\tau_{12}+\tau_{22})}+\cdots
\end{eqnarray*}
Furthermore,  adopting (10.25) and making a transformation we infer
that
\begin{eqnarray*}
&&F=1+e^{\hat{\zeta}_1}+e^{\hat{\zeta}_2}+e^{\hat{\zeta}_1+\hat{\zeta}_2-2\pi
\tau_{12} }+\rho_{11}^4e^{-\hat{\zeta}_1}
 +\rho_{22}^4e^{-\hat{\zeta}_2}+\rho_{11}^4\rho_{22}^4e^{-\hat{\zeta}_1-\hat{\zeta}_2-2\pi \tau_{12}}+\cdots\\
&& \ \ \ \ \ \longrightarrow
1+e^{\hat{\zeta}_1}+e^{\hat{\zeta}_2}+e^{\hat{\zeta}_1+\hat{\zeta}_2+A_{12}},\
\ {\rm as}\ \ \rho_{11}, \rho_{22} \rightarrow 0,
 \end{eqnarray*}
where $\hat{\zeta}_j=\alpha_jx+\hat{\beta}_jt+\delta_j, \ \ j=1,2,$ and $\hat{\beta}_j=2\pi i \beta_j, j=1, 2$.

Now, we need to prove
$$
\begin{aligned}
& \hat{\beta}_j\longrightarrow \frac{-2k_jw_0}{k_j^2+4v_0}, \ \
\hat{\zeta}_j\longrightarrow \xi_j,\ \ j=1, 2, \ \ \ {\rm as} \ \
\rho_{11}, \rho_{22}\rightarrow 0.\end{aligned}\eqno(10.27)$$

As in the case of $N=1$,  the solution of the system (10.23) has the
following form
$$\begin{aligned}
&\beta_1=\beta_{1,0}+\beta_{1,1}\rho_{11}+\beta_{2,2}\rho_{22}+o(\rho_{11},\rho_{22}),\\
&\beta_2=\beta_{2,0}+\beta_{2,1}\rho_{11}+\beta_{2,2}\rho_{22}+o(\rho_{11},\rho_{22}),\\
&\lambda=\lambda_0+\lambda_1\rho_{11}+\lambda_2\rho_{22}+o(\rho_{11},\rho_{22}).
\end{aligned}\eqno(10.28)$$

Expanding  functions  $\vartheta_{j}, \ j=1, 2, 3, 4$ in equations
(10.21) and (10.22) with substitution of assumption (10.28), and
letting $\rho_{11},\rho_{22}\longrightarrow 0$ , we will obtain
$$\begin{aligned}
&\lambda_0=0, \\
&16\pi i(-\pi^2\alpha_1^2+v_0)\beta_{1,0}-8\pi i w_0\alpha_1=0,\\
&16\pi i(-\pi^2\alpha_2^2+v_0)\beta_{2,0}-8\pi i w_0\alpha_2=0.
\end{aligned}\eqno(10.29)$$

Using  (10.28) and (10.29),  we conclude  that
$$\begin{aligned}
&\lambda=o(\rho_{11}, \rho_{22})\longrightarrow 0,
\\
 &\beta_j=\frac{-2k_jw_0}{k_j^2+4v_0}+o(\rho_{11}, \rho_{22})\longrightarrow
\frac{-2k_jw_0}{k_j^2+4v_0},  \ \ \ {\rm as } \ \ \rho_{11},
\rho_{22}\rightarrow 0,
\end{aligned}$$
and therefore  we have  (10.26).  So,  the two-periodic wave
solution (10.24) tends to the two-soliton solution (6.21) as
$\rho_{11},
\rho_{22}\rightarrow 0$. $\square$\\[8pt]
%%%%%%%%%%%%%%%%%%%%%%%%%%%%%%%%%%%%%%%%%%%%%%%%%%%%%%%%%%%%%%%%%%%%%%%%%%%%%%%%%%%%%%%%%%%%%%%%%%%%%%%%%%%%%%%%%%%%%%%%
{\bf  10.5. Multi-periodic wave solutions}

 The system (10.5) indicates that constructing
multi-periodic wave solutions depends on the solvability of the
system (10.5). Obviously, the number of constraint  equations of the
type (10.5) is $2^{N-1}+1$. On the other hand, we have
$\frac{1}{2}N(N+1)+3N+3$ parameters $\tau_{ii}, \tau_{ij}, \alpha_i,
\omega_i, \lambda, u_0, v_0$. Among them, $2N$ parameters
$\tau_{ii}, \alpha_i$ may be the given parameters relate to the
amplitudes and wave numbers of $N$-periodic waves.  Therefore,  the
number of the unknown parameters is $\frac{1}{2}N(N+1)+N+3$ while
$\frac{1}{2}N(N+1)$ parameters $\tau_{ij}$, implicitly appearing in
the series form, can not to be solved explicitly in general. So, the
number of the explicit unknown parameters is only $N+3$, and the
number of equations is larger than the unknown parameters in the
case of $N> 4$.   This fact means that if equation (10.5) is
satisfied, then we have at least $N$-periodic wave solutions ($N\leq
4$).  In this paper, we only consider one- and two-periodic wave
solutions of the NKdV equation (1.1). There are still  certain
computation difficulties in the calculation for the case of $N>2$,
which will be studied in the future.\\[12pt]
  %%%%%%%%%%%%%%%%%%%%%%%%%%%%%%%%%%%%%%%%%%%%%%%%%%%%%%%%%%%%%%%%%%%%%%%%%%%%%%%%%%%%%%%%%%%%%%%%%%%%%%%%%%%%%%%%%%%%%%%
%%%%%%%%%%%%%%%%%%%%%%%%%%%%%%%%%%%%%%%%%%%%%%%%%%%%%%%%%%%%%%%%%%%%%%%%%%%%%%%%%%%%%%%%%%%%%%%%%%%%%%%%%%%%%%%%%%%%%%
{\bf\large  Acknowledgment}

This work is  supported by the U. S. Army Research Office under
contract/grant No. W911NF- 08-1-0511 and by the Texas Norman
Hackerman Advanced Research Program under Grant No. 003599-
0001-2009,  the National Science Foundation of China (No. 10971031)
and Shanghai Shuguang Tracking Project (No. 08GG01).


\begin{thebibliography}{99} %% The number "99" means that this list has more than nine items.

\small
 \baselineskip=20pt

\bibitem{Ab1}   Ablowitz  M J and  Clarkson P A,  {\it  Solitons, Nonlinear
Evolution Equations and Inverse Scattering}, (  Cambridge University
Press, Cambridge, 1991).

\bibitem{ath}  Athorne C, Stability and periodicity in coupled Pinney equations, \emph{J. Diff. Eqns.}
\textbf{100}, 82-94 (1992).
%%%%%%%%%%%%%%%%%%%%%%%%%%%%%%%%%%%%%%%%%%%%%%%%%%%%%%%%%%%%%%%%%%%%%%%%%%%%%%%%%%%%%%%%%%%%%%%%%%%%%%

\bibitem{Ba1}   Battig D,  Kappeler T, and  Mityagin B,   On the Korteweg-de
Vries equation: frequencies and initial value problem. Pacific J.
Math., {\bf 181}, 1-55 (1997).

\bibitem{Belo}  Belokolos E., Bobenko A.,  Enol'skij V.,  Its  A. and   Matveev V. B.,
            {\it Algebro-Geometrical Approach to Nonlinear Integrable
              Equations} ( Springer, Berlin, 1994).

\bibitem{Bullough} Bullough  R K and
Caudrey P J,   Solitons and the Korteweg-de Vries Equation:
Integrable Systems in 1834-1995. Acta Appl. Math.,  {\bf 39},
193-228 (1995).
%%%%%%%%%%%%%%%%%%%%%%%%%%%%%%%%%%%%%%%%%%%%%%%%%%%%%%%%%%%%%%%%%%%%%%%%%%%%%%%%%%%%%%%%%%%%%%%%%%%%%%

\bibitem{Camassa} Camassa R and Holm D D, An integrable shallow water equation with
peaked solitons. Phys. Rev. Lett. {\bf 71},  1661-1665
(1993).

\bibitem{Cao}   Cao C W,   Wu  Y T and  Geng X G, Relation
between the Kadometsev-Petviashvili equation
         and the confocal involutive system,  J. Math. Phys. {\bf 40},  3948-3970 (1999).


\bibitem{Che1}  Cherednik I. V.,   {\it Basic Methods of Soliton Theory}, ( World Scientific, Singapore,
 1996).
%%%%%%%%%%%%%%%%%%%%%%%%%%%%%%%%%%%%%%%%%%%%%%%%%%%%%%%%%%%%%%%%%%%%%%%%%%%%%%%%%%%%%%%%%%%%%%%%%%%%%%

\bibitem{DP[1999]}  Degasperis A and Procesi M, Asymptotic integrability, in symmetry
and perturbation theory,pp.23-37: edited by A. Degasperis and G.
Gaeta, World Scientific, 1999.


\bibitem{DHH2002}  Degasperis A, Holm D D,  Hone A N W, A new integrable equation with
peakon solutions, Theor. and Math. Phys., {\bf 133},
1463-1474(2002).

\bibitem{Du}    Dubrovin B. A., Periodic problem for the KdV equation in the class of finite band potentials.
             Funct. Anal. Appl. {\bf 9}, 265-277 (1975).
%%%%%%%%%%%%%%%%%%%%%%%%%%%%%%%%%%%%%%%%%%%%%%%%%%%%%%%%%%%%%%%%%%%%%%%%%%%%%%%%%%%%%%%%%%%%%%%%%%%%%%

 \bibitem{VE}  Ermakov V, Second-order differential equations. Conditions of complete integrability, Univ. Izv. Kiev,
 {\bf 20}, 1-25 (1880).
%%%%%%%%%%%%%%%%%%%%%%%%%%%%%%%%%%%%%%%%%%%%%%%%%%%%%%%%%%%%%%%%%%%%%%%%%%%%%%%%%%%%%%%%%%%%%%%%%%%%%%

 \bibitem{Fan1}   Fan E G and  Hon Y C, Quasi-periodic waves and   asymptotic
behavior  for  the  (2+1)-dimensional  Bogoyavlenskii's breaking
soliton equation, Phys Rev E,  {\bf 78}, 036607-13 (2008).

\bibitem{Fan2}  Fan E G, Quasi-periodic waves and   asymptotic
property  for  the  asymmetrical Nizhnik-Novikov-Veselov  equation,
J. Phys A {\bf 42},  095206.1-11  (2009).


 \bibitem{Fan3}    Fan   E G and  Hon Y C, Quasi-periodic wave solutions of  $\mathcal{N}=2$
supersymmetric KdV equation in superspace, Stud. Appl. Math. {\bf
125}, 343-371 (2010).

\bibitem{Fine}  Finelli F,  Vacca G P, and  Venturi G, Chaotic inflation from a
scalar field in nonclassical states, Phys. Rev. D {\bf 58},
103514.1-14 (1998)

\bibitem{Fuchssteiner} Fuchssteiner  B., Some tricks from the symmetry-toolbox for nonlinear equations:
Generalizations of the Camassa-Holm equation, Physica D 95, 229-243
(1996)
%%%%%%%%%%%%%%%%%%%%%%%%%%%%%%%%%%%%%%%%%%%%%%%%%%%%%%%%%%%%%%%%%%%%%%%%%%%%%%%%%%%%%%%%%%%%%%%%%%%%%%

\bibitem{Gardner1}   Gardner C S, Greene J M,   Kruskal M D and  Miura R M,
Method for solving the Korteweg-de Vries equation. Phys. Rev. Lett.,
{\bf 19}, 1095-1097 (1967).

\bibitem{Gardner2}  Gardner C S,   Korteweg-de Vries equation and
generalizations. IV. The Korteweg-de Vries equation as a Hamiltonian
system. J. Math. Phys., {\bf 12}, 1548-1551 (1971).


\bibitem{Dikii}  Gel'fand  I M and   Dikii L A,   Asymptotic behaviour of the
resolvent of Sturm-Liouville equations and the algebra of the
Korteweg-de Vries equations. Russian Math. Surveys, {\bf 30}, 77-113
(1975).

\bibitem{Geng1}   Geng X G,   Wu Y T and Cao  C W,  Quasi-periodic solutions of the modified Kadomtsev-Petviashvili equation,
     J.  Phys. A  {\bf 32},  3733-3742 (1999).

\bibitem{Geng2}   Geng X G, and  Cao C W, Decomposition of the (2+1)-dimensional Gardner equation and
              its quasi-periodic solutions,  Nonlinearity,  {\bf 14},  1433-1452 (2001).

\bibitem{Geng3}  Geng X G,   Dai H H,  Zhu  J Y and   Wang H Y,
Decomposition of the discrete Ablowitz-Ladik hierarchy,  Stud. Appl.
Math.   {\bf 118},  281-312 (2007).

\bibitem{Fritz2}   Gesztesy F and   Holden H, Real-valued algebro-geometric
solutions of the Camassa-Holm hierarchy  Phil Trans R Soc A  {\bf
 366},  1025-1054 (2008).

\bibitem{Fritz1} Gesztesy F,  and Holden H, {\it Soliton Equations and Their Algebro-Geometric Solutions}
(Cambridge University Press, New York, 2003).

\bibitem{Gilson} Gilson C,  Lambert F,  Nimmo J and  Willox R, On the combinatorics of the Hirota D-operators,
 Proc. R. Soc. Lond. A {\bf 452},  223-234  (1996).

 \bibitem{Guha}  Guha P., Nonholonomic deformation of generalized KdV-type
equations, J. Phys. A,  {\bf 42},  345201.1-17 (2009).
%%%%%%%%%%%%%%%%%%%%%%%%%%%%%%%%%%%%%%%%%%%%%%%%%%%%%%%%%%%%%%%%%%%%%%%%%%%%%%%%%%%%%%%%%%%%%%%%%%%%%%

\bibitem{Hirota}  Hirota R,  Exact solution of the Korteweg-de Vries equation for multiple
collisions of solitons, Phys. Rev. Lett. {\bf 27}, 1192-1194 (1971).



\bibitem{Hirota1}    Hirota R and  Satsuma J, Nonlinear Evolution Equations Generated from the
 B\"{a}cklund Transformation for the Boussinesq Equation,
 Prog. Theor. Phys.  {\bf 57},  797-807 (1977).

\bibitem{Hirota2}   Hirota R,  {\it Direct methods in soliton theory} (Springer-verlag, Berlin, 2004)

\bibitem{Fan5} Hon Y C and  Fan E G,  An algebro-geometric solution for a Hamiltonian
system with application to dispersive long wave equation, J.
 Math. Phys.    {\bf 46},  032701-21 (2005).

\bibitem{H1} Hone A N W, The associated Camassa-Holm equation and the KdV
equation, J. Phys. A, {\bf 27},  L307-L314 (1999).

 \bibitem{HW1} Hone  A N W and  Wang J P, Integrable peakon
equations with cubic nonlinearity,   J. Phys. A, {\bf 41},
372002.1-11 (2008).

\bibitem{Hu}    Hu X B and  Clarkson P A,  Rational solutions of a differential-difference KdV equation,
 the Toda equation and the discrete KdV equation,
J. Phys. A  {\bf 28},  5009-5016 (1995).

\bibitem{Hu2}    Hu X B,  Li C X,  Nimmo J J C and   Yu G F,
An integrable symmetric (2+1)-dimensional Lotka-Volterra equation
and a family of its solutions, J. Phys. A {\bf 38}, 195-204(2005).

\bibitem{Its}   Its A,  and Matveev V B, Hill's operator with finitely many gaps. Funct. Anal. Appl. {\bf 9},
 65-66 (1975).
%%%%%%%%%%%%%%%%%%%%%%%%%%%%%%%%%%%%%%%%%%%%%%%%%%%%%%%%%%%%%%%%%%%%%%%%%%%%%%%%%%%%%%%%%%%%%%%%%%%%%%

\bibitem{Kersten}  KerstenP. H. M.,  Krasil¡äshchik I.S., Verbovetsky  A.M.,  Vitolo R.£¬
Integrability of Kupershmidt deformations, Acta Appl Math, {\bf
109}, 75-86 (2010)

\bibitem{Korteweg}  Korteweg D J,  and   de Vries G,   On the change of form of long
waves advancing in a rectangular canal, and on a new type of long
stationary waves. Phil. Mag.,  {\bf 39}, 422-443 (1895).

\bibitem{Kupershmidt}  KupershmidtB. A., KdV6: An integrable system, Phys. Lett. A, {\bf
372}, 2634-2639 (2008).
%%%%%%%%%%%%%%%%%%%%%%%%%%%%%%%%%%%%%%%%%%%%%%%%%%%%%%%%%%%%%%%%%%%%%%%%%%%%%%%%%%%%%%%%%%%%%%%%%%%%%%

\bibitem{Lambert1}   Lambert F,  Loris I and Springael J, Classical Darboux transformation and the Kp hierarchy,
 Inverse Probl. {\bf 17},  1067-174  (2001).

\bibitem{Lambert2}   Lambert F and  Springael J, Soliton equations and Simple combinatorics,
Acta Appl.  Math.  {\bf 102},  147-178 (2008).

\bibitem{Lax}   Lax P D,  Periodic solutions of the KdV equation, Comm. Pure Appl. {\bf 28}, 141-188 (1975).

\bibitem{Lax1}  Lax P D,   Integrals of nonlinear equations of evolution and
solitary waves. Comm. Pure Appl. Math., {\bf 21}, 467-490 (1968).

\bibitem{Leon}  Leon J and  Latifi A, Solution of an initial-boundary value
problem for coupled nonlinear waves, J. Phys. A, 23,  1385-1403
(1990)


\bibitem{Liu4}    Liu Q P, Hu X B and  Zhang M X, Supersymmetric modified Korteweg¨Cde Vries
       equation: bilinear approach,  Nonlinearity   {\bf 18}, 1597-1603 (2005).

\bibitem{Lax3}  Lax P D,  Outline of a theory of the KdV equation. Pages
70-102 of:  T. Ruggeri,  G. Bolliat, C. M. Dafermos, P. D. Lax, and
T. P. Liu,  Recent Mathematical Methods in Nonlinear Wave
Propagation. Lecture Notes in Mathematics, vol. 1640. Berlin:
Springer  1996.

\bibitem{Lou}  Lou S Y,    Symmetries of the KdV equation and four hierarchies
of the integrodifferential KdV equations,   J. Math. Phys. {\bf 35},
2390-2396 (1994).

\bibitem{Lundmark-Szm} Lundmark H. and   Szmigielski J., Multi-peakon solutions of the Degasperis-Procesi equation,
    Inverse Problems, {\bf 19}, 1241-46 (2003).
%%%%%%%%%%%%%%%%%%%%%%%%%%%%%%%%%%%%%%%%%%%%%%%%%%%%%%%%%%%%%%%%%%%%%%%%%%%%%%%%%%%%%%%%%%%%%%%%%%%%%%

\bibitem{Mc}    Mckean H P, and  Moerbeke P., The spectrum of Hill's equation, Invent. Math. {\bf 30}, 217-274 (1975).


\bibitem{Me1} Mel'nikov V K, Exact solutions of the Korteweg-de Vries equation
with a self-consistent source, Phys. Lett. A, {\bf 128},  488-492
(1988).

\bibitem{Me2} Mel'nikov V K, Interaction of solitary waves in the system described by the
Kadomtsev-Petviashvili equation with a self-consistent source,
Commun. Math. Phys. {\bf 126}, 201-215 (1989).

\bibitem{Me3} Mel'nikov V K,  Integration of the Korteweg-de Vries equation with a source, Inverse Problem,  {\bf 6},
233-246 (1990).



\bibitem{Miura1}  Miura R M,  Korteweg-de Vries equation and generalizations.
I. A remarkable explicit nonlinear transformation. J. Math. Phys.,
{\bf 9},  1202-1204 (1968).

\bibitem{Miura2}  Miura R M,  Gardner C. S. and   Kruskal M. D.,  Korteweg-de
Vries equation and generalizations. II. Existence of conservation
laws and constants of motion. J. Math. Phys., {\bf 9}, 1204-1209
(1968).

\bibitem{Far}  Mumford D,  {\it Tata Lectures on Theta II},  Progress in Mathmatics,
Vol. 43 ( Boston: Birkh\"{a}user,  1984).
%%%%%%%%%%%%%%%%%%%%%%%%%%%%%%%%%%%%%%%%%%%%%%%%%%%%%%%%%%%%%%%%%%%%%%%%%%%%%%%%%%%%%%%%%%%%%%%%%%%%%%


\bibitem{Na2}   Nakamura A,  A direct method of calculating periodic wave solutions to nonlinear evolution
equations. I. exact two-periodic wave solutions, J. Phys. Soc. Jpn.
 {\bf 47}, 1701-1705 (1979).

 \bibitem{Nov}    Novikov S P, Periodic problem for the KdV equation,  Funct. Anal. Appl. {\bf 8},  236-246 (1974).

\bibitem{Novikov}  Novikov V, Generalizations of the Camassa-Holm equation,
J. Phys. A, {\bf 42} 342002.1-15 (2009)
%%%%%%%%%%%%%%%%%%%%%%%%%%%%%%%%%%%%%%%%%%%%%%%%%%%%%%%%%%%%%%%%%%%%%%%%%%%%%%%%%%%%%%%%%%%%%%%%%%%%%%


\bibitem{Palais}  Palais R S,  The symmetry of solitons. Bull. Amer. Math. Soc., {\bf 34}, 339-403 (1997).

 \bibitem{Pinney}  Pinney E, The nonlinear equation $y"+p(x)y+cy^{-3}=0$,Proc. Amer.
Math. Soc, 1, 681 (1950).
%%%%%%%%%%%%%%%%%%%%%%%%%%%%%%%%%%%%%%%%%%%%%%%%%%%%%%%%%%%%%%%%%%%%%%%%%%%%%%%%%%%%%%%%%%%%%%%%%%%%%%

\bibitem{Qiao}  Qiao Z J  and  Li J B,
Negative order KdV equation with solitons, periodic solitary wave
and kink wave solutions,
  Eur. Phs. Lett. {\bf 94},  50003.1-6 (2011).

\bibitem{Qiao1}  Qiao Z J, The Camassa-Holm hierarchy, N-dimensional integrable systems, and algebro-geometric
 solution on a symplectic submanifold, Comm Math Phys  {\bf  239},  309-341 (2003).

\bibitem{Qiao2}   Qiao Z J,   A new integrable equation with cuspons and W/M-shape
  peakon solitons, J. Math Phys. {\bf 47}, 11270.1-9 (2006).

\bibitem{Qiao-AAM} Qiao Z J, Integrable Hierarchy, $3\times 3$ Constrained Systems, and Parametric
Solutions, Acta Appl. Math., {\bf 83}, 199-220 (2004).

\bibitem{Qiao-JMP2} Qiao Z J,
New integrable hierarchy,cuspons,one-peak solitons,and
M/W-shape-peak solutions, J. Math. Phys. {\bf  48},
082701.1-20(2007).

\bibitem{Qiao-book} Qiao Z J,  {\it Finite-dimensional Integrable System and Nonlinear
Evolution Equations}, (Chinese National Higher Education Press,
Beijing, 2002).

\bibitem{Qiao-PhD} Qiao Z J, ``Generalized Lax algebra, $r$-matrix and algebro-geometric
soultion for the integrable system", preprint 1996, Ph D Thesis
(10246/950011), Fudan University, PR China, 1997.


\bibitem{Qiao-1998} Qiao Z J,  Non-dynamical r-matrix and algebraic-geometric solution for a discrete system,
preprint 1996, Chin. Sci. Bull. {\bf 43}, 1149-1153 (1998).
%%%%%%%%%%%%%%%%%%%%%%%%%%%%%%%%%%%%%%%%%%%%%%%%%%%%%%%%%%%%%%%%%%%%%%%%%%%%%%%%%%%%%%%%%%%%%%%%%%%%%%


\bibitem{HL}   Rachael M H and James E L, Ermakov-Pinney equation in
scalar field cosmologies,Phys. Rev. D {\bf 66}, 023523.1-8 (2002)

\bibitem{RR80}  Reid J, Ray J., Ermakov systems, nonlinear superposition and solution of nonlinear equations of motion,
 J. Math. Phys. {\bf 21}, 1583-1587 (1980).

 \bibitem{CHR}  Rogers C,  Hoenselaers C. and  Ray J.R., On 2+1-dimensional Ermakov systems, \emph{J. Phys. A}
  \textbf{26}, 2623-2633 (1996).

 \bibitem{CBKA} Rogers C, Malomed B, Chow K and  An H, Ermakov-Ray-Reid systems in nonlinear optics,
 \emph{J. Phys. A}, {\bf 43}, 455214-30  (2010).

\bibitem{CRHA}  Rogers C and  An H, Ermakov-Ray-Reid systems in 2+1-dimensional rotating shallow water theory,
 {\em Stud. Appl. Math.}, {\bf 125}, 275-99 (2010).


\bibitem{RS96}  Rogers C and  Schief W K, Multi-component Ermakov systems: structure and linearization,
{\em J. Math. Anal. Appl.} {\bf 198}, 194-220 (1996).

\bibitem{Rosu}  Rosu H C,  Espinoza P,  Reyes M, Ermakov approach for $Q=0$ empty FRW
minisuperspace oscillators, Nuovo Cimento B, {\bf 114}, 1435-1440
(1999).

\bibitem{Russell}  Russell J S  Report of the committee on waves. Pages
417-496 (1837). of: Report of the 7th Meeting of the British
Association for the Advancement of Science, Liverpool. London: John
Murray.
%%%%%%%%%%%%%%%%%%%%%%%%%%%%%%%%%%%%%%%%%%%%%%%%%%%%%%%%%%%%%%%%%%%%%%%%%%%%%%%%%%%%%%%%%%%%%%%%%%%%%%

\bibitem{Tu1}  Tu G Z, The trace identity, a powerful tool for constructing Hamiltonian structure of
  integrable systems.  J Math Phys,  {\bf 30},   330-338   (1989).

\bibitem{Tu2}  Tu G Z,  A new hierarchy of integrable system and its Hamiltonian
 structure. Sci China Ser  A-Math, {\bf 32}, 142-153
 (1989).

\bibitem{Shahinpoor}   Shahinpoor M and  Nowinski J L, Exact solution to the problem of forced large amplitude radial
oscillations of a thin hyperelastic tube, Int. J. Nonlinear Mech.
{\bf 6}, 193-208 (1971)
%%%%%%%%%%%%%%%%%%%%%%%%%%%%%%%%%%%%%%%%%%%%%%%%%%%%%%%%%%%%%%%%%%%%%%%%%%%%%%%%%%%%%%%%%%%%%%%%%%%%%%

\bibitem{Verosky} Verosky J M, Negative powers of Olver recursion operators, J. Math. Phys. {\bf 32}, 1733-1736 (1991).



\bibitem{Zakharov}  Zakharov V E  and  Faddeev L. D.,   Korteweg-de Vries equation:
A completely integrable Hamiltonian system, Functional Anal. Appl.,
{\bf 5}, 280-287 (1971).

\bibitem{Zam}  Zampogni L,  On
algebro-geometric solutions of the Camassa-Holm hierarchy,  Advanced
Nonl Studies  {\bf 3},  345-380 (2007).

\bibitem{Zhang-qiao}  Zhang G P  and  Qiao Z J,
Cuspons and Smooth Solitons of the Degasperis-Procesi Equation under
Inhomogeneous Boundary Condition, Mathematical Physics, Analysis and
Geometry {\bf 10}, 205-225 (2007).

\bibitem{Zhang}  Zhang D J, The N-soliton solutions for the modified KdV equation with self-consistent sources,
 J. Phys. Soc. Jpn. {\bf 71},  2649-2656 (2002).


\bibitem{Zhou1}  Zhou  R. G.,   Mixed hierarchy of soliton equations, J. Math. Phys. {\bf 50}, 123502.1-12 (2009).

\bibitem{Zhou}   Zhou R G, The finite-band solution of the Jaulent-Miodek equation,
  J. Math. Phys. {\bf 38},  2535-2346 (1997).

\bibitem{Zhou-PhD}  Zhou R G, ``The
finite dimensional integrable systems related to the soliton
equations", preprint 1996, Ph D Thesis, Fudan University, PR China
1997.

\end{thebibliography}
\end{document}